\documentclass[a4paper,11pt]{article}
\pdfoutput=1

\usepackage{jcappub} 
\usepackage[T1]{fontenc} 
\usepackage{multirow}
\usepackage{array, makecell}
\usepackage{float} 
\usepackage{tabu}

\newcommand{\be}{\begin{eqnarray}}
\newcommand{\ee}{\end{eqnarray}}

\title{\boldmath Baryonic effects for weak lensing. Part I. Power spectrum and covariance matrix}

\author[a,b]{Aurel Schneider,}
\author[b]{Nicola Stoira,}
\author[b]{Alexandre Refregier,}
\author[c]{Andreas J. Weiss,}
\author[a]{Mischa Knabenhans,}
\author[a]{Joachim Stadel,}
\author[a]{and Romain Teyssier\,}

\affiliation[a]{Institute for Computational Science, University of Zurich, Winterthurerstrasse 190, 8057 Zurich, Switzerland.}
\affiliation[b]{Institute for Particle Physics and Astrophysics, ETH Zurich, Wolfgang Pauli Strasse 27, 8093 Zurich, Switzerland.}
\affiliation[c]{Space Research Institute, Austrian Academy of Sciences, Schmiedlstrasse 6, 8042 Graz, Austria}

\emailAdd{aurel.schneider@uzh.ch}

\abstract{Baryonic feedback effects lead to a suppression of the weak lensing angular power spectrum on small scales. The poorly constrained shape and amplitude of this suppression is an important source of uncertainties for upcoming cosmological weak lensing surveys such as Euclid or LSST. In this first paper in a series of two, we use simulations to build a Euclid-like tomographic mock data-set for the cosmic shear power spectrum and the corresponding covariance matrix, which are both corrected for baryonic effects following the \emph{baryonification} method of \citet{Schneider:2018pfw}. In addition, we develop an emulator to obtain fast predictions of the baryonic power suppression, allowing us to perform a likelihood inference analysis for a standard $\Lambda$CDM cosmology with both cosmological and astrophysical parameters. Our main findings are the following: (i) ignoring baryonic effects leads to a greater than 5$\sigma$ bias on the cosmological parameters $\Omega_m$ and $\sigma_8$; (ii) restricting the analysis to the largest scales, that are mostly unaffected by baryons, makes the bias disappear, but results in a blow-up of the $\Omega_m$-$\sigma_8$ contour area by more than a factor of 10; (iii) ignoring baryonic effects on the covariance matrix does not significantly affect cosmological parameter estimates; (iv) while the baryonic suppression is mildly cosmology dependent, this effect does not noticeably modify the posterior contours. Overall, we conclude that including baryonic uncertainties in terms of nuisance parameters results in unbiased and surprisingly tight constraints on cosmology.}

\begin{document}
\maketitle

\section{Introduction}
Within the next few years, large-scale weak lensing surveys will provide unprecedented data to study the cosmological model ($\Lambda$CDM) and its unknown cold dark matter and dark energy components. The range of scales probed by future surveys like Euclid\footnote{\texttt{https://www.euclid-ec.org/}}, LSST\footnote{\texttt{https://www.lsst.org/}}, and WFIRST\footnote{\texttt{https://wfirst.gsfc.nasa.gov/}} will include the highly nonlinear regime of structure formation that has to be modelled with numerical simulations.

Over the last decades, gravity-only $N$-body simulations have been continuously developed towards higher particle numbers and improved accuracy. As a result, they are now close to approaching percent-level precision regarding the non-linear clustering signal \citep[e.g. Refs.][]{Heitmann:2007hr,Schneider:2015yka,Garrison:2018juw}. However, $N$-body simulations are time-consuming and cosmological parameter estimation requires fast predictors that can be used to scan the high-dimensional parameter space of $\Lambda$CDM. The current approach of cosmological inference therefore relies on simulations to adjust fitting functions and regression routines based for example on the halo model \citep{Takahashi:2012em,Mead:2015yca,Smith:2018zcj,Cataneo:2018cic}, on direct emulation techniques \citep{Heitmann:2013bra,Knabenhans:2018cng,DeRose:2018xdj}, or on neural networks \citep{Ravanbakhsh:2017bbi,Fluri:2018hoy,Alsing:2019xrx,Manrique-Yus:2019hqc}.

However, it has become increasingly evident over the last few years that baryonic processes (which are ignored in gravity-only $N$-body simulations and associated regression techniques) have a significant effect on current and future weak-lensing observations. Several hydrodynamical simulations which include feedback effects from active galactic nuclei (AGN) and supernova explosions have shown that the clustering signal is affected by 10-30 percent at the nonlinear scales \citep{vanDaalen:2011xb,Hellwing:2016ucy,Mummery:2017lcn,Springel:2017tpz,Chisari:2018prw,vanDaalen:2019pst,Foreman:2019ahr,Barreira:2019ckp}. Similar effects have been found using the more analytical and physically intuitive approach of the halo model \citep{Semboloni:2011aaa,Mohammed:2014mba,Fedeli:2014gja,Debackere:2019cec}. A short review about the baryonic effects and their connection to cosmology is provided in \citet{Chisari:2019tus}.

An alternative model halfway between the halo model and simulations has recently been introduced in the Refs.~\citep{Schneider:2015wta,Schneider:2018pfw}. This \emph{baryonification} approach relies on an empirical parametrisation of halo profiles (including gas, stars, and dark matter) and modifies the output of a gravity-only $N$-body simulation by slightly displacing particles around halo centres. The advantage of \emph{baryonifying} $N$-body simulations is that the baryonic effects are empirically parametrised, making it possible to perform many fast realisations of the nonlinear cosmic density field with varying baryonic parameters. As a result it is possible to go beyond the power spectrum, studying for example weak-lensing maps via peak statistics \citep{Weiss:2019jfx} or a deep-learning approach \citep{Fluri:2019qtp}.

The present paper is the first in a series of two, where we build upon the \emph{baryonification} model of Ref.~\citep{Schneider:2018pfw} to perform a cosmological forecast analysis for the cosmic shear power spectrum of a stage-IV weak-lensing survey. We construct mock observations based on a Euclid-like survey configuration with a covariance matrix obtained from a suite of \emph{baryonified} $N$-body simulations. This allows us to perform a number of Monte-Carlo Markov Chain runs to study the effects of baryons on the posterior contours of cosmological and baryonic parameters.

There are several questions we want to address with this paper: (i) To what accuracy will it be possible to measure cosmological parameters of the $\Lambda$CDM model if we marginalise over baryonic uncertainties? (ii) By how much will the errors decrease if we fix the baryonic parameters to their true values, implying that they can be constrained with other observations? (iii) How much constraining power do we lose if we only rely on data from large scales that remain unaffected by baryons? (iv) What is the bias introduced if we analyse all the data but completely ignore the effects of baryons in the prediction pipeline? Next to these main questions, we also check the validity of several simplifying assumptions, such as ignoring baryons in the covariance matrix and keeping baryonic effects decoupled from cosmology. Finally, we also investigate the effects of additional freedom in the redshift dependence of the baryonic suppression signal.

All the results from the present paper assume a five-parameter $\Lambda$CDM cosmology where neutrinos are set to be massless. A more realistic setup, including massive neutrinos and several extensions beyond the $\Lambda$CDM framework, will be discussed in the second paper of this series \citep[see Ref.][]{Schneider:2019bbb}.

The present paper is structured as follows: In Sec.~\ref{sec:BCM} we revisit the \emph{baryonification} approach and we present the \emph{baryonic emulator} that allows us to speed up the calculations, making it fit for cosmological inference. In Sec.~\ref{sec:mock} we present our mock data-set including the covariance matrix which is based on a suite of simulations. Sec.~\ref{sec:likelihoodanalysis} describes the results of our parameter inference for both the cosmological and the baryonic parameters. Finally, we discuss simplifying model assumptions in Sec.~\ref{sec:assumptions} and we conclude in Sec.~\ref{sec:conclusions}. The Appendices~\ref{sec:ref_sim} and \ref{sec:emu_perform} provide further details on the construction and the testing of the baryonic emulator.

\section{Baryonic effects on the matter power spectrum}\label{sec:BCM}
In this section, we review the \emph{baryonic correction model} developed in \citet[henceforth S19]{Schneider:2018pfw} and we present the \emph{baryonic emulator} built upon this model. The latter is required to speed up the prediction pipeline in order to perform Markov-Chain Monte-Carlo (MCMC) sampling for cosmological parameter estimates.

\subsection{Summarising the baryonic correction model}\label{sec:BCmodel}
The baryonic correction (BC) model consists of a numerical routine which aims to perturb gravity-only $N$-body simulations in order to account for the effects of baryonic feedback on the large-scale structure of the universe. The idea of the model is to displace simulation particles around each halo so that the original NFW profile ($\rho_{\rm nfw}$) is transformed into a final baryon-dark matter profile ($\rho_{\rm bcm}$), i.e.,
\begin{equation}
\rho_{\rm nfw}(r) \hspace{0.5cm}\longrightarrow\hspace{0.5cm}\rho_{\rm bcm}(r) =  \rho_{\rm clm}(r) + \rho_{\rm gas}(r) + \rho_{\rm cga}(r),
\end{equation}
where the latter consists of a central galactic (cga), a gas, and a collisionless matter (clm) component. Note that $\rho_{\rm clm}$ is dominated by dark matter but also contains satellite galaxies and intracluster stars. We now summarise the different components that go into the above relation. More details about the parametrisation and its comparison to observations can be found in S19 \citep{Schneider:2018pfw}.
\begin{itemize}

\item The stellar profile of the central galaxy ($\rho_{\rm cga}$) is described by the truncated power law
\begin{equation}
\rho_{\rm cga}(r) \propto \frac{f_{\rm cga}(M)}{r^2}\exp\left[-\left(\frac{r}{2 R_{h}}\right)^2\right],\hspace{0.5cm}R_{h}=0.015\times r_{200}.
\end{equation}
The above relation only provides an approximate fit to realistic galaxies. Note, however, that the shape of the stellar profile has no influence on the clustering at scales relevant for weak lensing. The fractions of stars in the central galaxy ($f_{\rm cga}$) and the total stellar fraction ($f_{\rm star}$), including satellite galaxies and stars corresponding to the intra-cluster light ($f_{\rm sga}$) are defined as
\begin{equation}
f_{\rm cga}=0.09\left(\frac{M}{M_s}\right)^{-\eta_{\rm cga}},\hspace{0.5cm}f_{\rm star}=f_{\rm cga}+f_{\rm sga}=0.09\left(\frac{M}{M_s}\right)^{-\eta_{\rm star}},
\end{equation}
where $M_s= 2.5\times 10^{11}$ M$_{\odot}$/h and where $\eta_{\rm star}$ and $\eta_{\rm cga}$ are free model parameters. For consistency reasons, we impose $\eta_{\rm cga}>\eta_{\rm star}$ and $M>2.5\times 10^{11}$ M$_{\odot}$/h. In principle, it would be straight forward to extend these relations to smaller halo masses by introducing a break in the stellar fractions below $M_s$ \citep[see e.g. Ref.][]{Moster:2012fv}. Note, however, that the cosmological clustering signal of stage-IV weak lensing surveys is only affected by haloes above $M\sim10^{12}$ M$_{\odot}$/h \citep[]{Schneider:2015yka}.

\item The collisionless matter profile ($\rho_{\rm clm}$), consisting of the dominant dark matter component plus satellite galaxies (including intra-cluster stars), is given by
\begin{equation}
\rho_{\rm clm}(r)=\left(\Omega_{\rm dm}/\Omega_{m}+ f_{\rm sga}\right)\rho_{\rm nfw}(r).
\end{equation}
This profile is furthermore allowed to react to the other components via the adiabatic relaxation prescription of Ref.~\citep{Abadi:2009ve}. A detailed description of this procedure is given in S19.

\item Finally, the gas profile ($\rho_{\rm gas}$) is described by the relation
\begin{equation}\label{rhogas}
\rho_{\rm gas} \propto \frac{f_{\rm gas}}{(1+r/r_{\rm co})^{\beta}[1+(r/r_{\rm ej})^2]^{(7-\beta)/2}},\hspace{1cm}\beta=3-\left(\frac{M_c}{M}\right)^{\mu}
\end{equation}
where $r_{\rm co}=0.1\times r_{200}$ and
\begin{equation}\label{thetaej}
r_{\rm ej}\equiv\theta_{\rm ej}\times r_{200}.
\end{equation}
The gas fraction is obtained by subtracting the stellar fraction from the cosmic baryon fraction $f_{b}=\Omega_b/\Omega_{m}$, i.e.,
\begin{equation}\label{fractions}
f_{\rm gas}=f_{b} - f_{\rm star}.
\end{equation}
Note that the slope of the gas profile (described by the parameter $\beta$) is assumed to be mass-dependent with a shallower slope for galaxy-groups compared to clusters. Eq.~(\ref{rhogas}) has been shown in S19 to be in good agreement with X-ray observations. 
\end{itemize}
The parametrisation of the BC model summarised above has five free parameters, three related to the gas profile ($M_c$, $\mu$, $\theta_{\rm ej}$) and two related to the stellar fractions ($\eta_{\rm cga}$,  $\eta_{\rm star}$). The model is furthermore sensitive to the cosmic baryon fraction $f_b$, which we will show later on to affect the baryonic suppression signal of the matter power spectrum.

\begin{figure}[tbp]
\centering
\includegraphics[width=\textwidth]{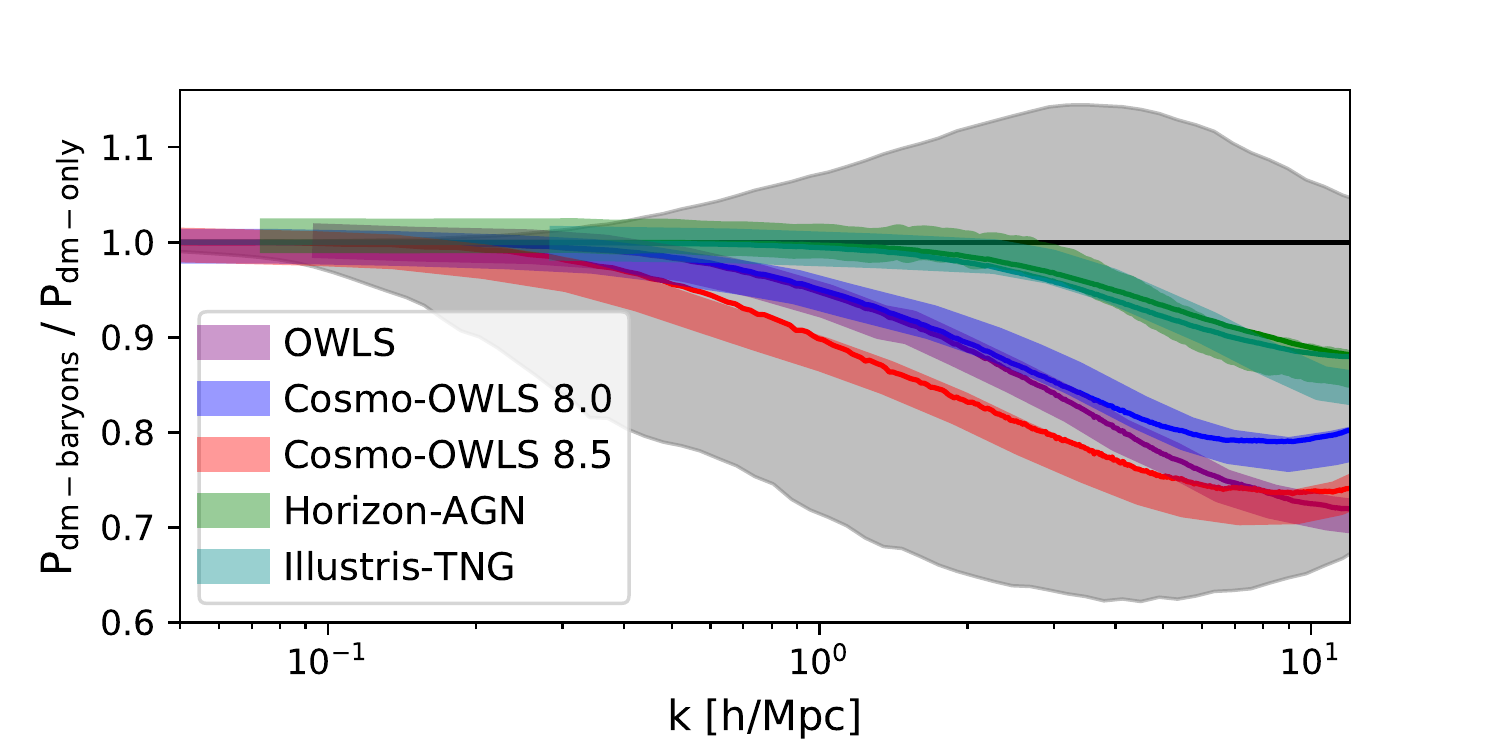}
\caption{Comparison between the baryonic correction (BC) model (coloured lines) and various hydrodynamical simulations (coloured bands) from the literature \citep{Semboloni:2011aaa,Mummery:2017lcn,Chisari:2018prw,Springel:2017tpz}. Note that the BC model parameters are not fitted to the power spectrum but to the gas and stellar fractions of each simulation. See S19 \citep{Schneider:2018pfw} for more details. The grey area in the background shows the prior range assumed in this paper.}
\label{fig:hydrocomparison}
\end{figure}

While the baryon correction method consists of an approximative approach based on outputs of gravity-only $N$-body simulations, it is in good agreement with full hydrodynamical simulations. In S19 it is shown that if the BC model parameters are tuned to reproduce the gas and stellar fraction of a given hydrodynamical simulation, then the baryonic suppression effects on the matter power spectrum can be predicted with an accuracy of 2 percent or better up to $k=10$ h/Mpc. 

Fig.~\ref{fig:hydrocomparison} shows a comparison of the power spectra from the baryonic correction model and the hydrodynamical simulations OWLS \citep{Semboloni:2011aaa}, Cosmo-OWLS \citep{Mummery:2017lcn}, Horizon-AGN \citep{Chisari:2018prw}, and Illustris-TNG \citep{Springel:2017tpz}. As explained above, the BC model parameters $M_c$, $\beta$, $\eta_{\rm cga}$, and $\eta_{\rm sga}$ are not fitted to reproduce the power suppression, but they are selected to match the gas and stellar fractions within $r_{500}$ of the corresponding simulation. As the remaining parameter $\theta_{\rm ej}$ cannot be well constrained with gas fractions at only one enclosed mass scale, it is set to the fiducial value of $\theta_{\rm ej}=4$. The predictions from the BC model are shown as coloured lines, while the results from the hydrodynamical simulations are plotted as broad coloured bands. The good agreement between model and simulations is highly non-trivial and illustrates how well the BC model is able to capture the baryonic effects on the large-scale structure of the universe\footnote{It is of course also possible to directly fit the BC model parameters to the relative power spectrum shown in Fig.~\ref{fig:hydrocomparison}. In this case we would obtain an even better agreement between the BC model and simulations. As a consequence, however, the baryonic parameter would lose their physical meaning. It would still be possible to use them as nuisance parameters in a weak-lensing cosmological inference analysis, but the direct connection to X-ray gas fraction (as shown in S19) would not be valid anymore.}.

The grey area shown in Fig.~\ref{fig:hydrocomparison} will be discussed in more detail in Sec.~\ref{sec:likelihoodanalysis}. It represents the selected prior-range of the baryonic model parameters. This prior is conservative enough to include all known results from hydrodynamical simulations and describes the current uncertainties of baryonic effects on the cosmological clustering signal.

\subsection{Cosmology-dependence of the baryonic effects}\label{cosmo_dependence}
So far, all weak-lensing cosmological studies that included baryonic effects have implicitly assumed them to be independent of cosmology \citep[see e.g. Refs.][]{Joudaki:2016mvz,Joudaki:2016kym,Hildebrandt:2016iqg,Abbott:2017wau,Hikage:2018qbn}. For the case of the cosmic shear power spectrum, this means for example that a cosmology-independent baryonic correction term can be simply multiplied to the dark-matter-only prediction of the matter power spectrum, considerably simplifying the analysis. In this section, we check the validity of this assumption by investigating potential correlations between the baryonic suppression effect and individual cosmological parameters.

We start by running one cosmological gravity-only N-body simulation assuming a standard 5-parameter $\Lambda$CDM cosmology (with massless neutrinos) with $\Omega_m=0.315$, $\Omega_b=0.049$, $h_0=0.674$, $n_s=0.96$, $\sigma_8=0.811$ before calculating the matter power spectrum at $z=0$. We then apply the baryonification method (with default parameter values of $M_c=10^{14}$ M$_\odot$/h, $\mu=0.4$, $\theta_{\rm ej}=4$, $\eta_{\rm cga}=0.3$, and $\eta_{\rm cga}=0.6$) and determine the ratio of the power spectrum with and without baryonic correction. As a next step, we run more $N$-body simulations where we vary each if the cosmological parameters individually while keeping the other parameters unchanged. For each of these simulations, we again determine the ratio of the power spectrum. We investigate the parameter ranges of $\Omega_m\in[0.24,0.35]$, $\Omega_b\in[0.038,0.055]$, $h\in[0.62,0.77]$, $n_s\in[0.92,1.00]$, and $\sigma_8\in[0.78,0.84]$.

\begin{figure}[tbp]
\includegraphics[width=\textwidth]{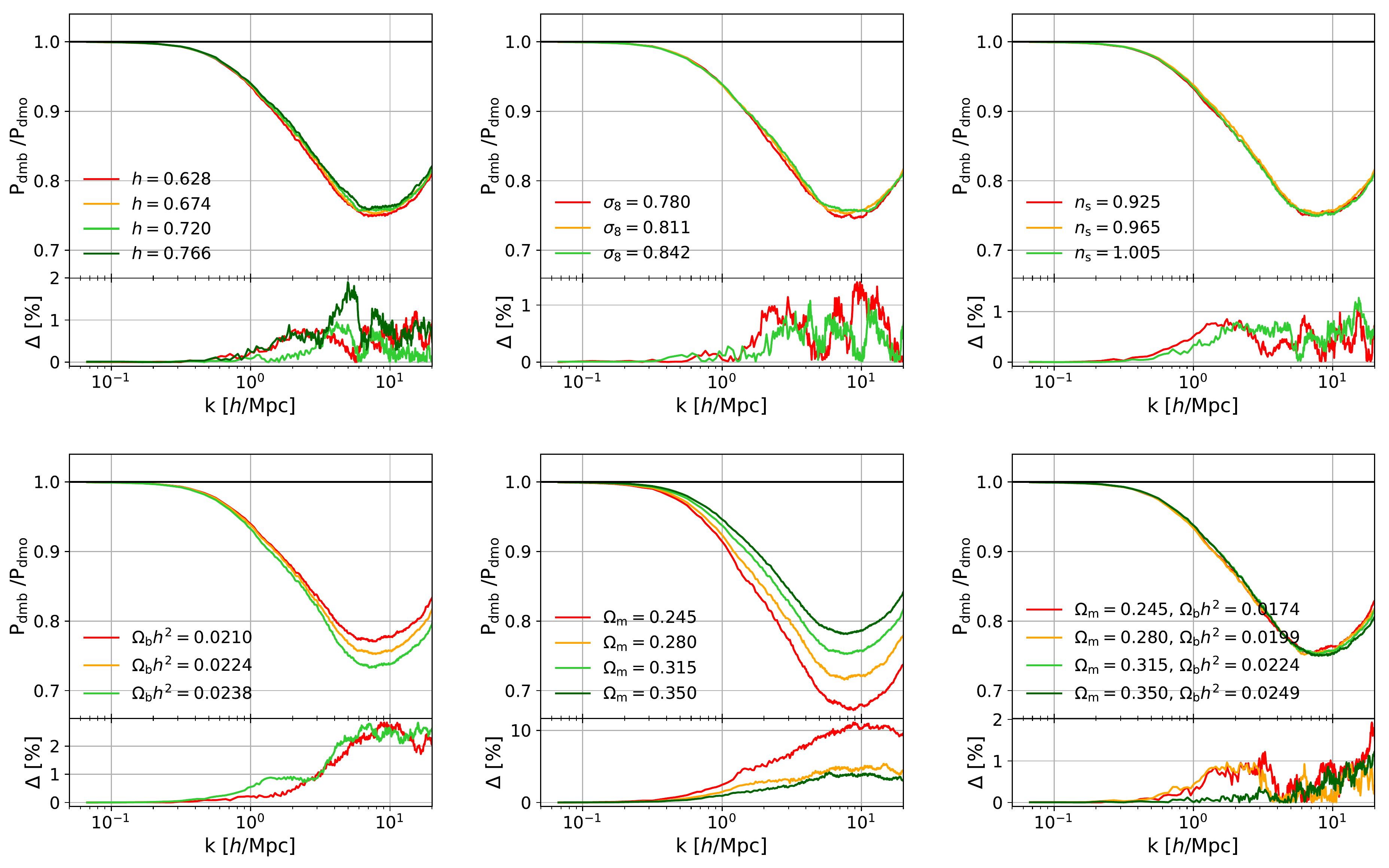}
\caption{Cosmology-dependence of the baryonic suppression effect on the power spectrum, assuming a baryonic correction model with fixed parameters $M_{\rm c}=10^{14}$ M$_{\odot}$/h, $\mu=0.4$, $\theta_{\rm ej}=4.0$, $\eta_{\rm star}=0.3$, and $\eta_{\rm cga}=0.6$. Values of individual cosmological parameters are labelled, all the other parameters are kept at their default values of  $\Omega_m=0.315$, $\Omega_b=0.049$, $h_0=0.674$, $n_s=0.96$, $\sigma_8=0.811$. Fractional differences to the default cosmology ($\Delta$ in percent) are shown in the corresponding sub-panels.}
\label{fig:cosmo_dependence}
\end{figure}

In Fig.~\ref{fig:cosmo_dependence} we plot the resulting ratios of the power spectra with and without baryon correction for different values of the individual cosmological parameters. The top panels show that neither varying $h_0$, $\sigma_8$, nor $n_s$ does significantly affect the spoon-like baryonic signal. The bottom-left and centre panels, on the other hand, show that varying $\Omega_b$ or $\Omega_m$ has a direct influence on the amplitude of the baryonic suppression. Finally, the bottom-right panel reveals that simultaneously changing $\Omega_b$ and $\Omega_m$ while keeping the baryon fraction $f_b=\Omega_b/\Omega_m$ fixed has no visible effect on the ratio of the power spectrum. We therefore conclude that the amplitude of the baryon effect is directly dependent on the comic baryon fraction $f_b$, but largely independent of other cosmological parameters. 

The relation between baryonic feedback effects and the cosmic baryon fraction is not surprising. It is a simple consequence of the fact that feedback effects can have a larger influence on cosmological structure formation, if there is more gas compared to the total amount of matter. Regarding the baryonic correction model, the direct influence of the cosmic baryon fraction is visible in Eq.~(\ref{rhogas}) and Eq.~(\ref{fractions}). A larger $f_b$ automatically leads to a stronger influence of the gas profile and therefore potentially stronger baryonic effects. 

The dependence between the baryonic feedback effects and the cosmic baryon fraction is a generic feature that is not restricted to the baryonic correction model. For example, it is also visible when comparing runs from the BAHAMAS simulations \citep{McCarthy:2017csu} that are based on the same AGN feedback implementation but different cosmologies. The change of the baryonic suppression effect on the BAHAMAS power spectra when going from {\tt WMAP9} \citep{Hinshaw:2012aka} to {\tt Planck13} \citep{Ade:2013zuv} can indeed be directly related to the corresponding shifts in $\Omega_m$ and $\Omega_b$ \citep[see e.g. Fig. 6 in Ref.][]{vanDaalen:2019aaa}. Note that the same direct relation between cosmology and baryonic suppression effect can also be observed when comparing the power spectra from OWLS and cosmo-OWLS (which are based on {\tt WMAP3} \citep{Spergel:2006hy} and {\tt WMAP7} \citep{Komatsu:2010fb} cosmologies, respectively).

All $N$-body simulations described in this section were run with the code {\tt Pkdgrav3} \citep{Stadel:2001aaa,Potter:2016ttn} assuming a box length of $L=256$ Mpc/h and a particle number of $N=512^3$. This box-length and resolution have been shown to be sufficient for converged ratios of the power spectrum at the percent level \citep[see Ref.][]{Schneider:2015wta}. For the halo finding (which is a necessary step of the baryonification procedure) we used the {\tt AHF} code described in Ref.~\citep{Knollmann:2009aaa}.

\begin{table}[tbp]
\centering
\small
\begin{tabular}{c  c  c }
Parameter Description & Acronym  & Emulator Range\\
\hline
Gas parameter 1 (related to the slope of the gas profile) & $\log M_c$ & 12.7, 16.7 \\
Gas parameter 2 (related to the slope of the gas profile) & $\mu$ & 0.1, 1 \\
Gas parameter 3 (related to the maximum gas ejection) & $\theta_{\rm ej}$ & 2, 8 \\
\hline
Stellar parameter 1 (related to the total stellar fraction) & $\eta_{\rm star}$ & 0.2, 0.4\\
Stellar parameter 2 (related to the central galactic stellar fraction) &  $\eta_{\rm cga}$ & 0.5, 0.7\\
\hline
Cosmic baryon fraction & $f_b$ & 0.13, 0.21\\
\hline
Redshift & $z$ & 0, 2\\
\end{tabular}
\caption{Descriptions and ranges of the model parameters from the \emph{baryonic emulator}. The cosmic baryon fraction is emulated between $f_b=0.13-0.21$. Outside of this range and within $f_b=0.05-0.5$, it is approximated using Eq.~(\ref{emu_approx}).}
\label{tab:emurange}
\end{table}

\subsection{Emulation of model parameters}\label{sec:emulator}
In order to perform cosmological parameter inference, we require a fast pipeline to predict the tomographic cosmic shear power spectrum. This is only possible if we have a regression routine for the power suppression caused by baryonic effects that can be included into a MCMC sampling routine. The method should allow us to vary both the free baryonic parameters and the cosmological baryon fraction which affects the signal as established in the previous section. 

Following the example of the {\tt EuclidEmulator} \citep{Knabenhans:2018cng}, we construct an emulator for the baryonic suppression signal 
\begin{equation}\label{barsuppression}
S_{\rm BCM}(k,z)\equiv\frac{P_{\rm dmb}(k,z)}{P_{\rm dmo}(k,z)},
\end{equation}
where $P_{\rm dmo}(k,z)$ and $P_{\rm dmb}(k,z)$ are the absolute power spectra of the uncorrected (dark-matter-only) and the \emph{baryonified} (dark-matter-baryon) density field. We thereby include the cosmological scales $k\leq10$ h/Mpc and the redshift range $z\in [0,2]$. We start by building a training set (i.e. the experimental design) using a optimised latin hypercube sampling (LHS) strategy with 1000 points in the 6 dimensional parameter space and for five fixed redshift values $z=\{0,0.5,1,1.5,2\}$.

At each sample point of the experimental design, we then apply the baryonic correction model, generating a vector in $k$ for $S_{\rm BCM}$. We do not run separate $N$-body simulations for this step, which considerably speeds up the process and allows us to go to a large experimental design. This is an approximation that is only accurate if changing the cosmology of the underlying $N$-body simulation has no (or very little) effect on the suppression signal. In Fig.~\ref{fig:cosmo_dependence}  we have shown that this is indeed the case regarding the parameters $h_0$, $\sigma_8$, and $n_s$. For the remaining parameters $\Omega_m$ and $\Omega_b$, the situation is more complicated. While Fig.~\ref{fig:cosmo_dependence} clearly shows that they do have an effect on $S_{\rm BCM}$, it is unclear wether this is entirely due to the parametrisation of the BC model or also due to the underlying simulation. In Appendix \ref{sec:ref_sim}, we further investigate the provenance of the cosmology dependence on $S_{\rm BCM}$, showing that using simulations with different values of $\Omega_m$ while keeping the BC model parameters fixed, does not lead to substantial changes of $S_{\rm bcm}$. This confirms that using one single $N$-body simulation with fixed cosmology for the experimental design only introduces sub-percent errors for parameter values within the range of the emulator.

Once the experimental design is established, we construct the baryonic emulator using the uncertainty quantification software {\tt UQLab} \citep{Marelli:2014aaa} which follows a spectral decomposition method called polynomial chaos expansion. The algorithm generates a surrogate model which can be evaluated for an arbitrary point in the parameter space. More details on the emulation technique can be found in Ref.~\citep{Knabenhans:2018cng}.

The ranges for each of the six emulated parameters are given in Table~\ref{tab:emurange}. Note that we do not emulate the redshift parameter but we use direct interpolation at each $k$-mode instead. In the Appendix~\ref{sec:emu_perform} we show that this interpolation strategy results gives accurate results at all redshifts between $z=0-2$. The $f_b$ parameter is emulated between the values 0.13 and 0.21. Beyond this range we use the approximate relations
\begin{eqnarray}\label{emu_approx}
S_{\rm BCM}(k,z|f_b) = S_{\rm BCM}(k,z|f_b=0.13)^{(f_b/0.13)^{1.2}},\hspace{0.5cm}{\rm{for}\,\,f_b<0.13}\\
S_{\rm BCM}(k,z|f_b) = S_{\rm BCM}(k,z|f_b=0.21)^{(f_b/0.21)^{1.2}},\hspace{0.5cm}{\rm{for}\,\,f_b>0.21}\nonumber
\end{eqnarray}
We have checked that this provides a good fit to the simulation-based results. Note furthermore, that the posteriors of the resulting inference analysis shown in Sec.~\ref{sec:likelihoodanalysis} (assuming either free or fixed baryonic parameters and including all scales of the mock sample) lie within the range of the emulator where the above approximation does not apply.

\begin{figure}[tbp]
\centering
\includegraphics[width=0.49\textwidth]{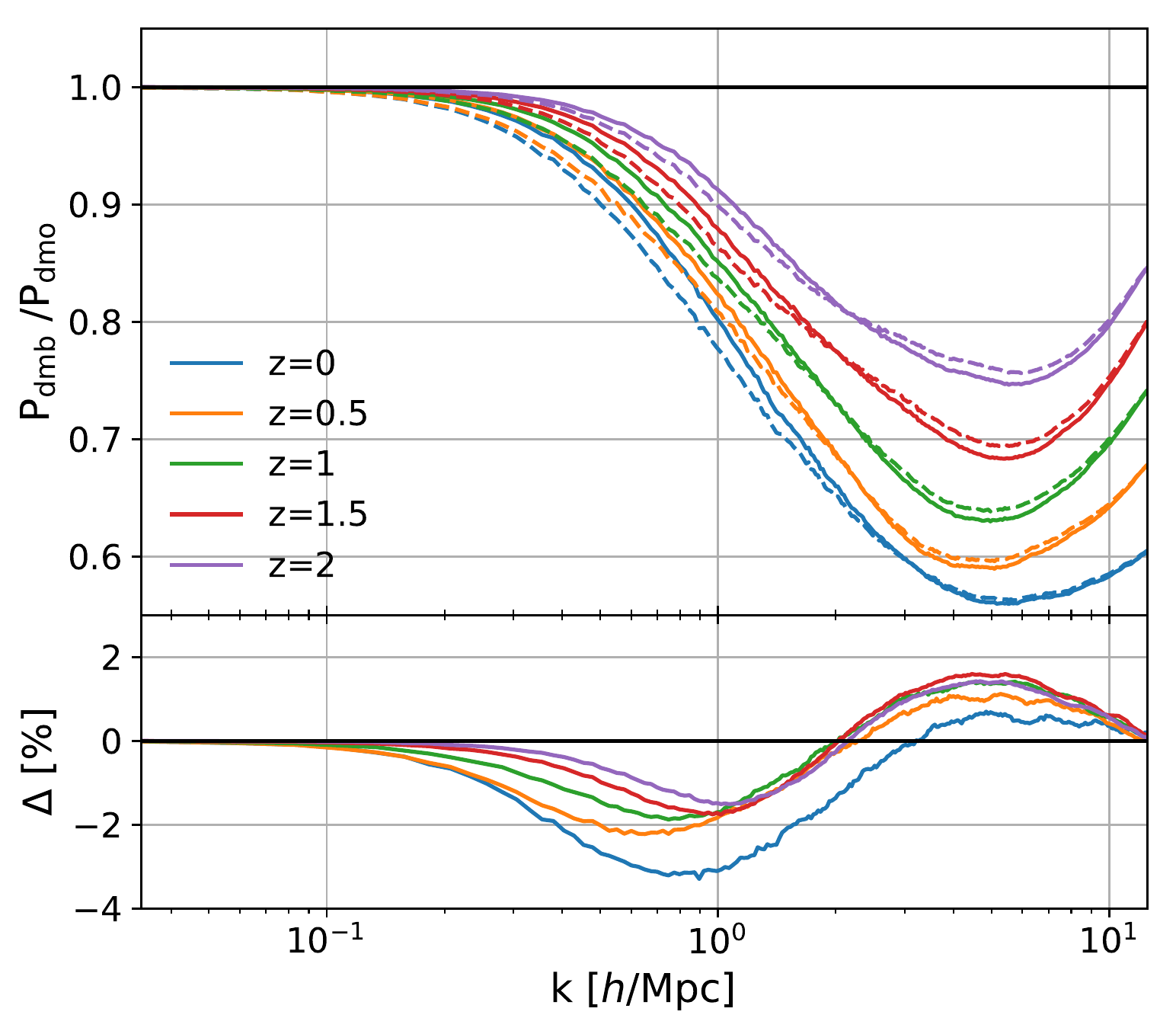}
\includegraphics[width=0.49\textwidth]{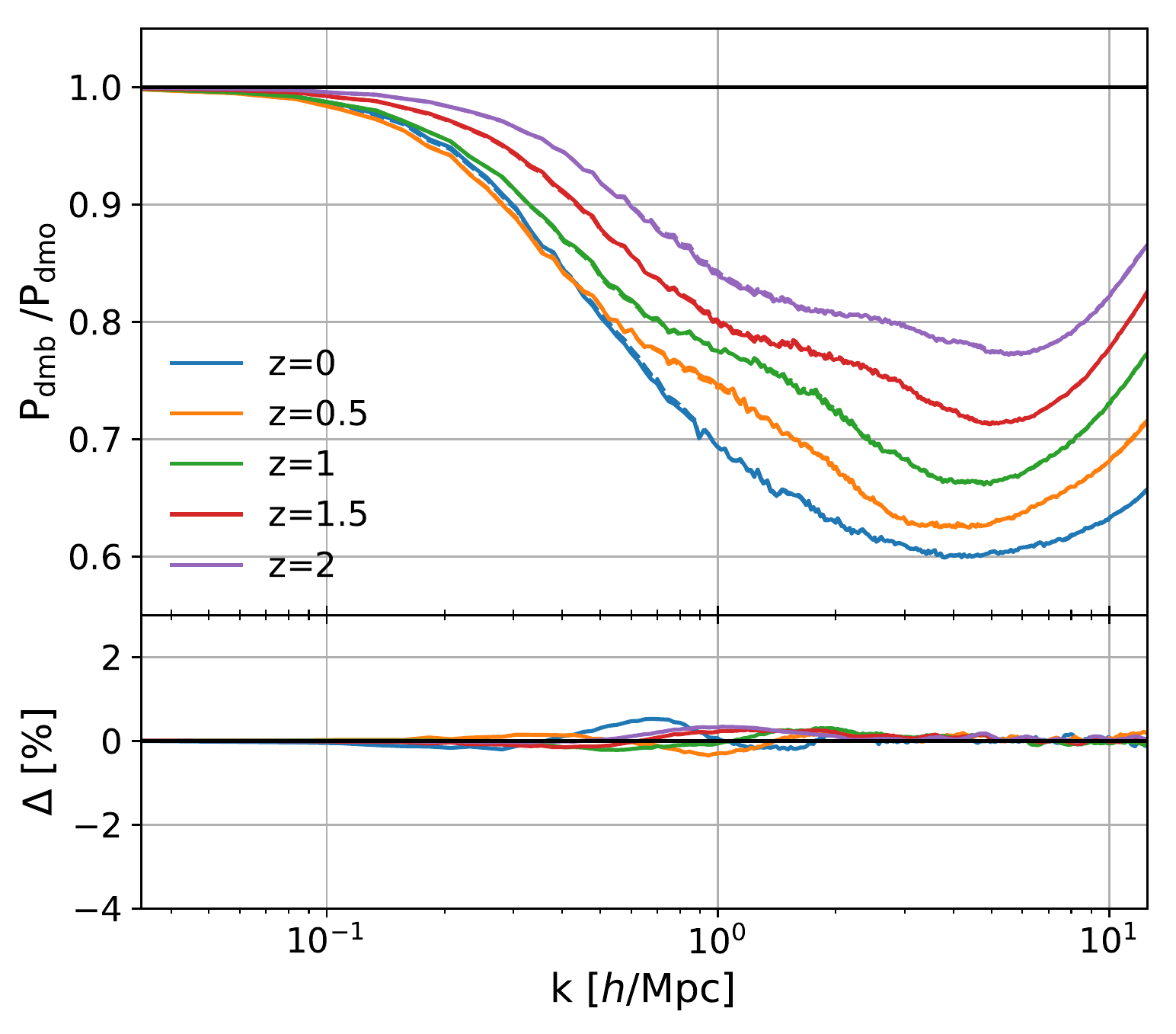}
\caption{Baryonic power suppression from the \emph{baryonic emulator} (dashed lines) and directly from the \emph{baryonic correction model} (solid) for two test sample points in the 6-dimensional parameter space. The left-hand and right-hand panels show sample point 0 and 1, respectively. The parameter values of the two sample points together with the remaining points of the test sample are listed in Table~\ref{tab:emutestsample} of Appendix \ref{sec:emu_perform}. More comparison results from the test sample are shown in Fig.~\ref{fig:SampleALL}.}
\label{fig:Samples0_1}
\end{figure}

In order to test the baryonic emulator, we again use the optimised LHS method to assign six sample points in the parameter space, comparing the power suppression from the emulator with the one directly obtained from the \emph{baryonification} method. In Fig. \ref{fig:Samples0_1} we show a comparison from two of the six sampling points, the other four are plotted in Fig.~\ref{fig:SampleALL} of Appendix \ref{sec:emu_perform}. The upper part of the figure represents the expected (continuous line) and emulated (dashed line) power suppression of the matter power spectrum, whereas the lower part represents the deviations between the two, i.e. the emulation error. While the emulator is very precise at sample point 1 it shows some small deviations at sample point 0, which, however, stay below $\sim 3$ percent at all $k$-modes. The other test sample points show a similar behaviour (see Fig.~\ref{fig:SampleALL}).

\begin{figure}[tbp]
\centering
\includegraphics[width=\textwidth]{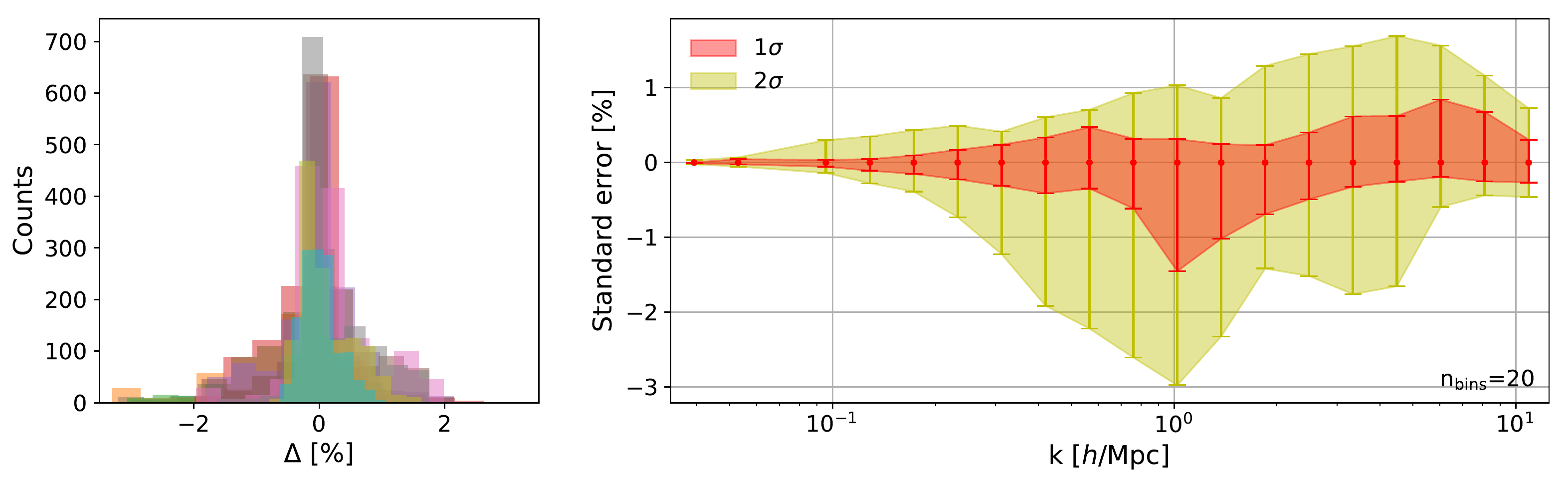}
\caption{\textit{Left}: histograms of the emulation errors $\Delta=\mathrm{P}_{\rm emu}/\mathrm{P}_{\rm true}-1$ for each $k$-bin between $z=0$ and 2. Different colours correspond to different $k$-bins. \textit{Right}: 1-$\sigma$ and 2-$\sigma$ errors of the emulator as a function of $k$-modes.}
\label{fig:emu_error_std}
\end{figure}

In order to study the accuracy of the emulator in a more quantitative way, we follow the approach presented in Fig. \ref{fig:emu_error_std}. Using the test-set with six sample points at 9 different redshift values, we study the error distribution as a function of $k$-modes (assuming 20 logarithmically spaced $k$-bins). The distribution of the relative error $\Delta=\mathrm{P}_{\rm emu}/\mathrm{P}_{\rm true}-1$ is plotted in the left-hand panel of Fig. \ref{fig:emu_error_std}, where the different colours refer to the different $k$-bins. Based on these distributions, we define 1-$\sigma$ and 2-$\sigma$ errors that are plotted as a function of $k$-modes in the right-hand panel of Fig. \ref{fig:emu_error_std}. Not surprisingly, the emulator works best at small values of $k$. The poorest performance is reached on scales around $k=1$ Mpc$/h$, while the precision improves again towards the smallest scales. In total the error of the emulator remains within 1.5 percent at the 1$\sigma$ and within 3 percent at 2$\sigma$ confidence level.

\section{Weak-lensing predictions and mock observations}\label{sec:mock}
In this section, we describe both our prediction pipeline of the weak-lensing angular power spectrum and the construction of our mock observations. We thereby provide information on the modelling of matter power spectrum, the tomographic binning, the map generation, and the construction of the covariance matrix.  

\subsection{Predicting the tomographic shear power spectrum}\label{sec:predictionpipeline}
The prediction of the tomographic cosmic shear power spectrum is based on the Limber approximation \citep{Limber:1953aaa}. This means that the spherical harmonics power spectra are calculated via the integral
\begin{equation}\label{Cofl}
C_{ij}(\ell)=\int_0^{\chi_H}\frac{g_i(\chi)g_j(\chi)}{\chi^2}P_{\rm dmb}\left(\frac{\ell}{\chi},z(\chi)\right)d\chi,
\end{equation}
where the co-moving distance $\chi$ goes from 0 to the horizon $\chi_H$. The lensing weights $g_i$ are given by
\begin{equation}
g_i(\chi)=\frac{3\Omega_m}{2}\left(\frac{H_0}{c}\right)^2 \frac{\chi}{a} \int_{\chi(z)}^{\chi_H}n_i(z)\frac{\chi(z')-\chi(z)}{\chi(z')}dz',
\end{equation}
and $n_i(z)$ stands for the galaxy distribution at redshift-bin $i$. For the full galaxy distribution, we assume
\begin{equation}\label{galdistr}
n(z)\propto z^2\exp(-z/0.24)
\end{equation}
which is a reasonable functional form for future galaxy survey such as Euclid. In our analysis, we assume three redshift bins between $z=0.1$ and $1.5$ with sharp bin-edges and equal galaxy counts, leading to the bin-sizes $\Delta z_1=[0.1,\,0.478]$, $\Delta z_2=[0.478,\,0.785]$, and $\Delta z_3=[0.785,\,1.5]$.

In order to mimic the effect of intrinsic alignment, we furthermore include the \emph{nonlinear linear alignment model} of Ref.~\citep{Hirata:2004aaa,Bridle:2007ft}. The model accounts for intrinsic-intrinsic and intrinsic-shear correlations by adding one free model parameter $A_{\rm IA}$ describing the amplitude of the effect. For simplicity and in accordance with previous work \citep[e.g. Refs.][]{Hildebrandt:2016iqg,Kacprzak:2019tzh}, we ignore potential redshift and luminosity dependence of the model.

The nonlinear baryon-corrected power spectrum appearing in Eq.~(\ref{Cofl}) is given by 
\begin{equation}\label{Pdmb}
P_{\rm dmb}(k,z)=P_{\rm dmo}(k,z)\times S_{\rm BCM}(k,z),
\end{equation}
where $S_{\rm BCM}$ is the emulated baryonic suppression term defined in Eq.~(\ref{barsuppression}) and $P_{\rm dmo}(k,z)$ is the nonlinear dark-matter-only power spectrum. For the latter, we use the revised {\tt halofit} model of \citet{Takahashi:2012em}. The underlying transfer function is based on the \citet{Eisenstein:1998aaa} fitting function. 

The revised {\tt halofit} procedure has an accuracy of about five percent compared to high-resolution $N$-body simulations \citep{Takahashi:2012em, Schneider:2015yka}. Note that this will not be sufficient for future weak-lensing surveys such as Euclid or LSST, where more elaborate emulators will have to be used \citep[see e.g. Ref.][]{Knabenhans:2018cng}. However, the approach is sufficient for a forecast study, especially since we use the same method for the predictions and the mock data set, thereby eliminating all systematic biases introduced by the {\tt halofit} method.

The calculations of the auto and cross angular power spectra including the weak-lensing and the intrinsic alignment contributions are performed using the python package {\tt PyCosmo} of \citet{Refregier:2017seh}. {\tt PyCosmo} applies efficient numerical integration routines, allowing to obtain fast predictions necessary for high-dimensional parameter inference.

\begin{figure}[tbp]
\centering
\includegraphics[width=0.99\textwidth,trim=0.7cm 0.2cm 1.6cm 0.8cm,clip]{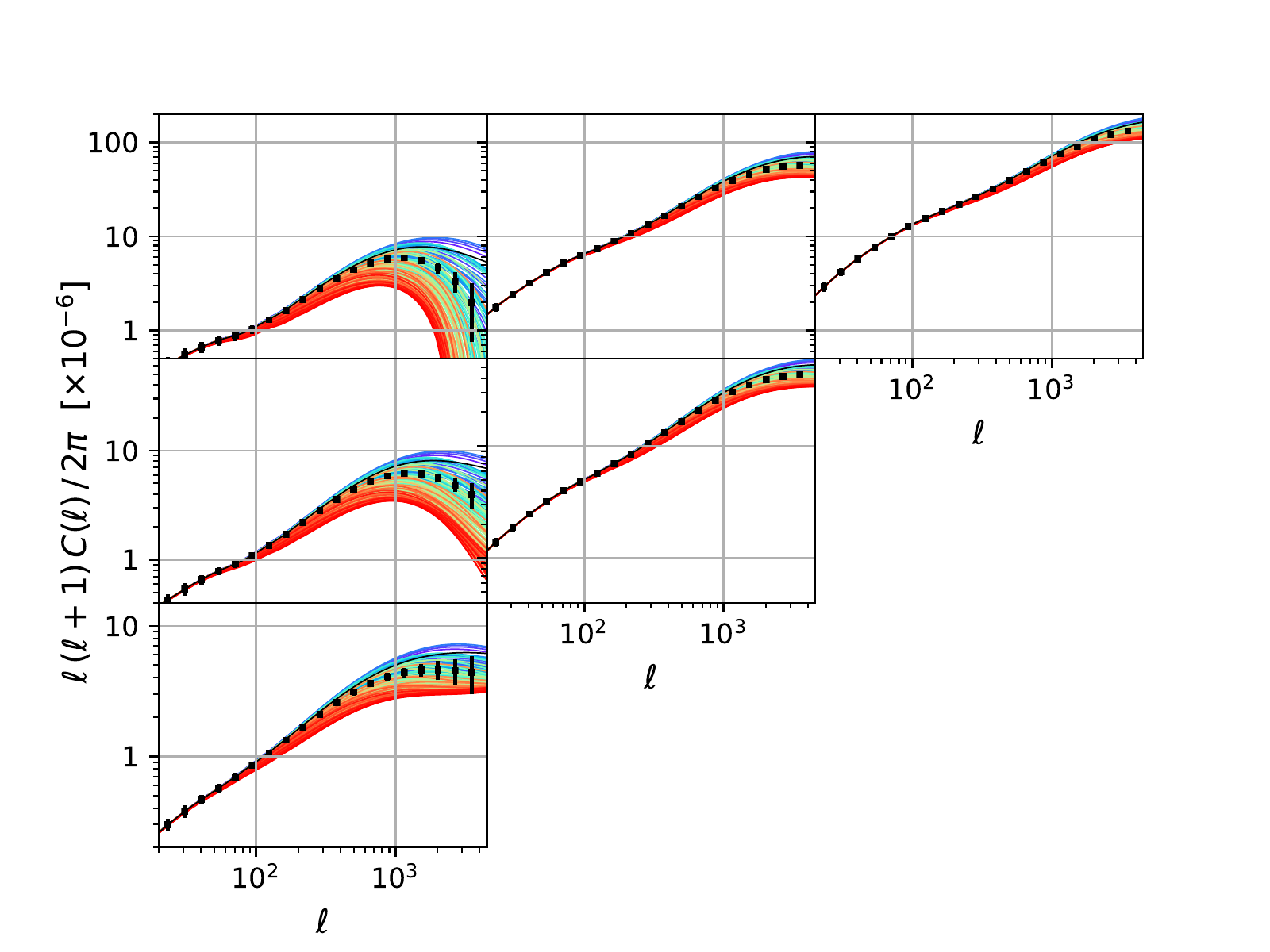}
\caption{Mock auto and cross power spectra of the three redshift bins with error-bars that include both cosmic variance and Gaussian shape noise expected from a Euclid-like survey (black data points). The auto spectra are shown by the diagonal, the cross-spectra by the off-diagonal panels (with increasing redshift from bottom-left to top-right). The coloured lines correspond to the theoretical predictions assuming baryonic parameters $\log M_c=13-16$ (mass in M$_{\odot}$/h), $\mu=0.1-0.7$ and $\theta_{\rm ej}=2-8$. The range covered by the lines illustrates the current level of uncertainty due to the baryonic effects.}
\label{fig:Cofl}
\end{figure}

In Fig.~\ref{fig:Cofl} we plot the resulting angular power spectra of the cosmic shear. The 3 diagonal panels refer to the auto spectra, whereas the off-diagonal ones show the cross-spectra of the three tomographic bins. The coloured lines correspond to the predictions assuming a default cosmology from {\tt Planck} and baryonic parameters varying within the ranges $\log M_c=[13,16]$, $\mu=[0.1,0.7]$ and $\theta_{\rm ej}=[2,8]$. This corresponds to the prior ranges given in Table~\ref{tab:prior0} of Sec.~\ref{sec:likelihoodanalysis}. The black data points of Fig.~\ref{fig:Cofl} show the Euclid-like mock observations which we will discuss in the following section.

\subsection{Mock observations and covariance matrix}\label{sec:mockobs}
We build weak-lensing convergence maps based on simulated light-cones in order to construct realistic mock observations of a Euclid-like survey. The light-cones are generated from $N$-body simulations presented in \citet{Weiss:2019jfx} that have been run using the gravity-only $N$-body code {\tt Pkdgrav3} \citep{Stadel:2001aaa,Potter:2016ttn}. The initial conditions were created with the {\tt MUSIC} code \citep{Hahn:2011aaa} assuming an \citet{Eisenstein:1998aaa} transfer function with a standard Planck cosmology \citep{Ade:2015fva}.

Each simulation contains $N=512^3$ particles and has a box-length of $L=512$ Mpc/h. The simulation boxes are replicated up to 1728 times to obtain a full-sky light-cone between redshift $z=0.1-1.5$. During the replication process, all boxes are randomly shifted and rotated. The light-cone is constructed using 78 concentric shells at different redshifts. The setup used here allows for an accuracy of about five percent in terms of the angular power spectra \citep{Weiss:2019jfx}. While this level of accuracy is not quite sufficient for stage-IV weak lensing surveys \citep[see e.g. Ref.][]{Schneider:2015yka}, it is good enough for the forecast purpose of this paper.

For the weak-lensing map construction, we project each shell onto a HEALPix\footnote{\texttt{https://healpix.sourceforge.io}} map (of $N_{\rm side}=2048$ resolution) and weight them according to the galaxy distributions of Eq.~(\ref{galdistr}) for each of the three redshift bins defined in Sec.~\ref{sec:predictionpipeline}. The weighted HEALPix maps are then combined assuming the Born approximation. This leads to tomographic full-sky maps of the weak-lensing convergence.

In order to obtain realistic mock observations of a stage-IV survey, we cut out a galaxy footprint of 20000 deg$^2$  and add a Gaussian noise component of
\begin{equation}
\langle\sigma^2\rangle=\frac{\sigma_e^2}{A_{\rm pix}n_{\rm gal}}
\end{equation}
to the map, where $\sigma_e=0.3$ is the root-mean-square of the shear ellipticity dispersion, $n_{\rm gal}=10$ arcmin$^{-2}$ is the galaxy density for each redshift bin (summing up to a total of 30 arcmin$^{-2}$ over the full redshift range), and $A_{\rm pix}$ is the area of a single pixel of the map. More details about the $N$-body simulations, the light-cone, and the convergence maps (including potential systematics) can be found in \citet{Weiss:2019jfx}.

In total, we rely on ten $N$-body simulations with different random seeds. During the light-cone construction, we furthermore apply the replication process in five different ways, which means that we are left with 50 independent light cones. Each of these full-sky light cones can account for two survey footprints leading to 100 statistically independent maps. For every map we finally assume three different noise configurations ending up with 300 different maps for every redshift bin. 

\begin{figure}[tbp]
\centering
\includegraphics[height=0.48\textwidth,trim=1.2cm 0.1cm 4.1cm 0.8cm,clip]{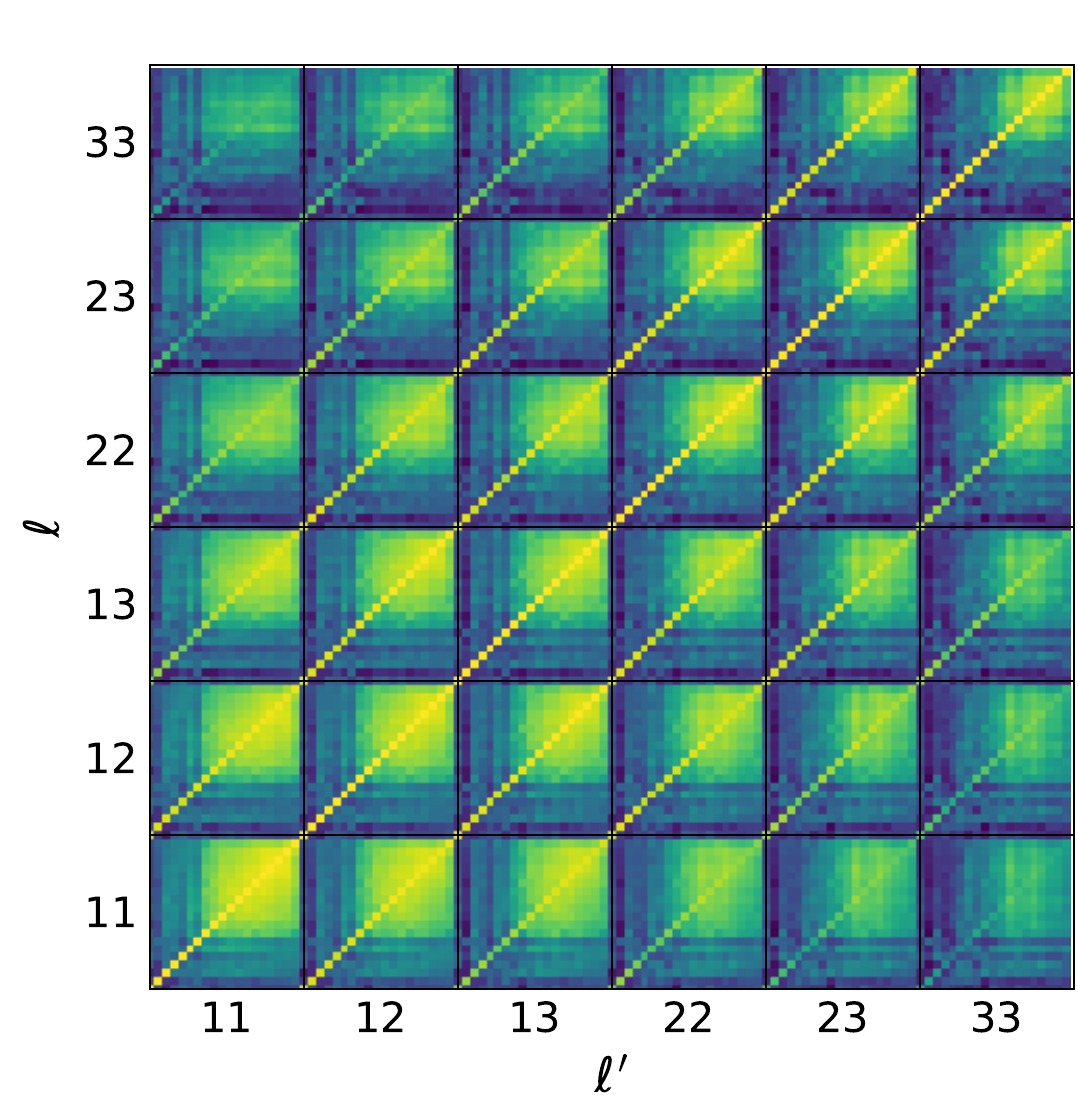}
\includegraphics[height=0.48\textwidth,trim=2.7cm 0.1cm 1.8cm 0.8cm,clip]{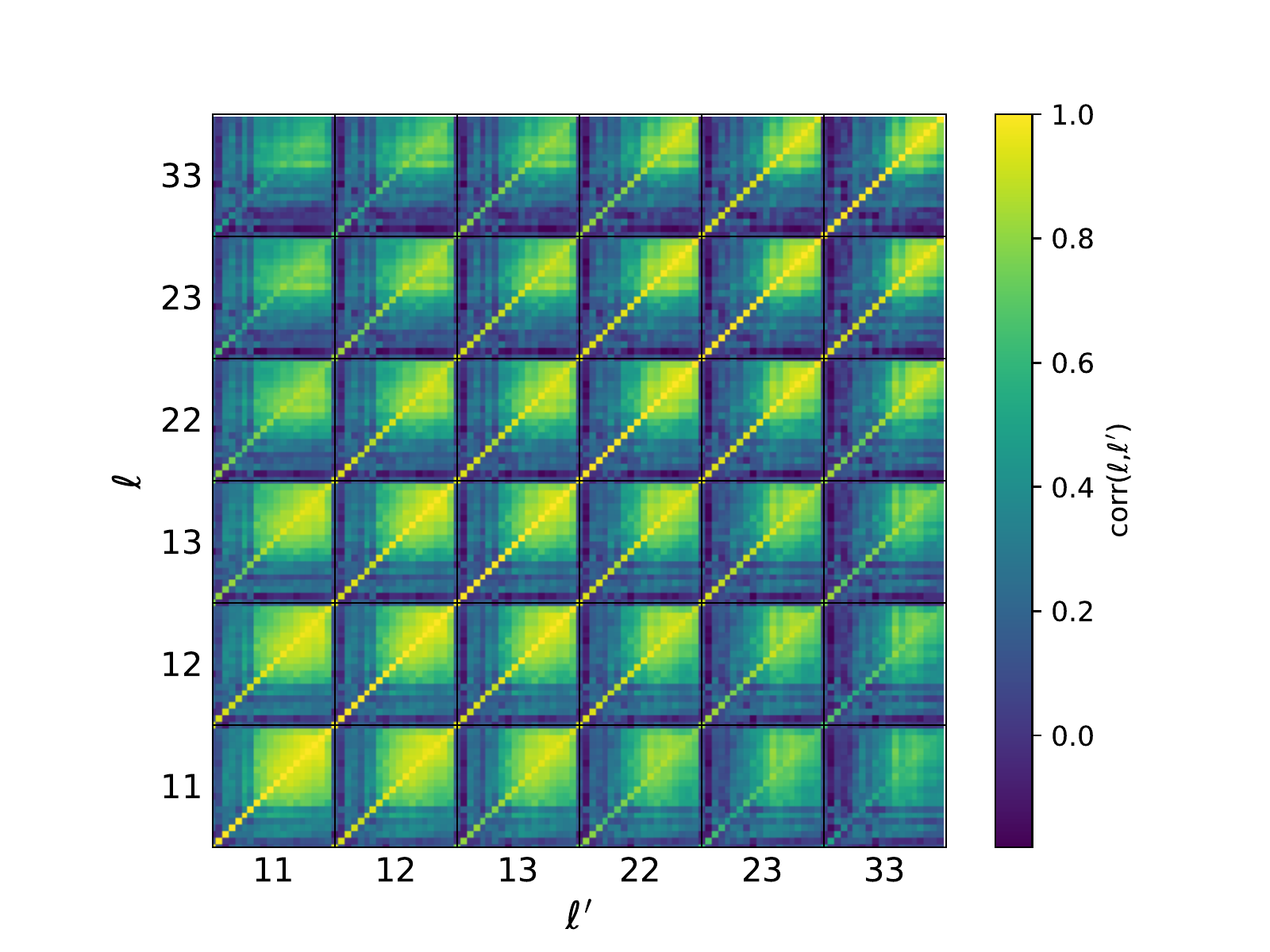}\\
\includegraphics[height=0.48\textwidth,trim=1.2cm 0.1cm 4.1cm 0.8cm,clip]{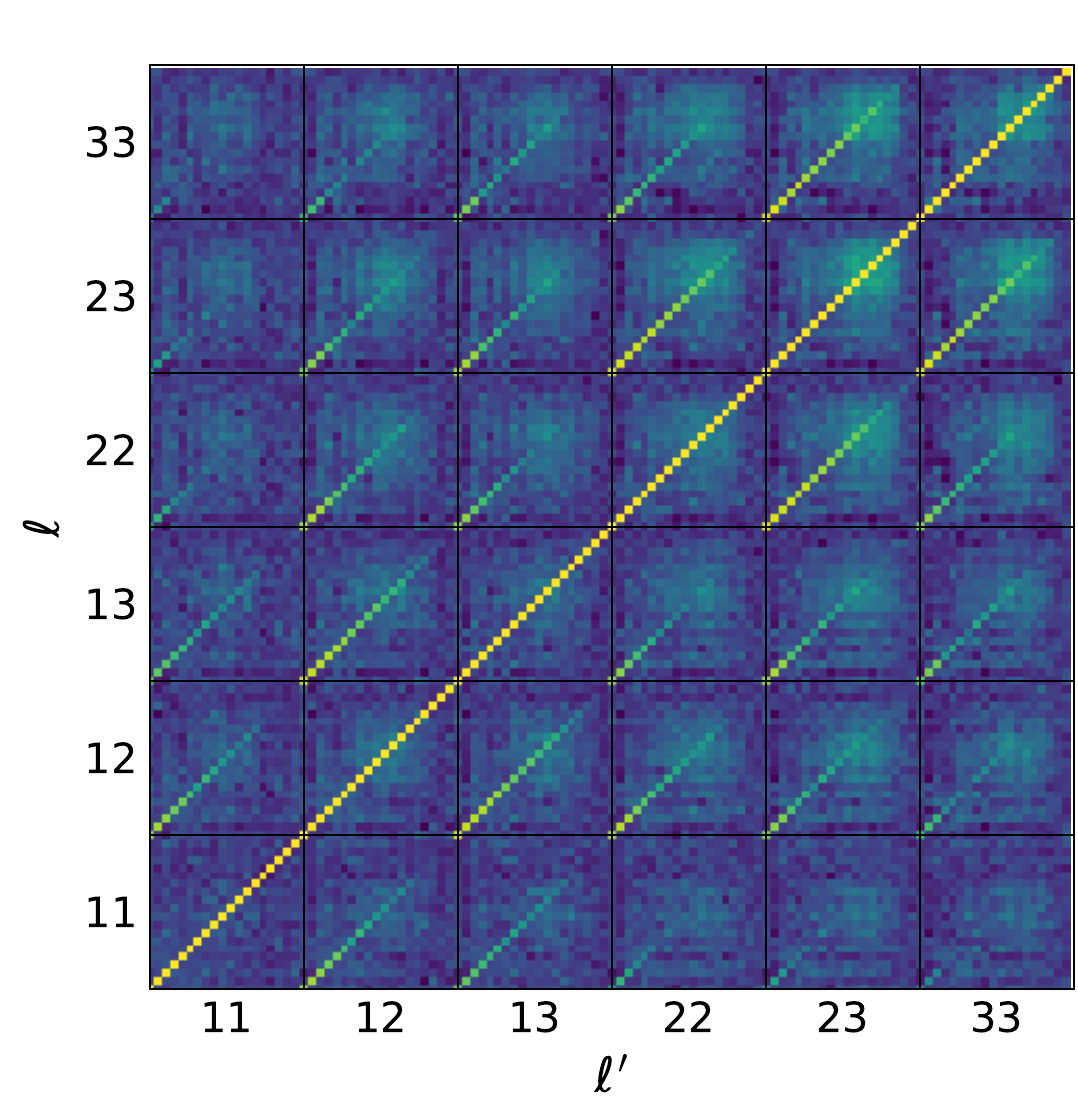}
\includegraphics[height=0.48\textwidth,trim=2.7cm 0.1cm 1.8cm 0.8cm,clip]{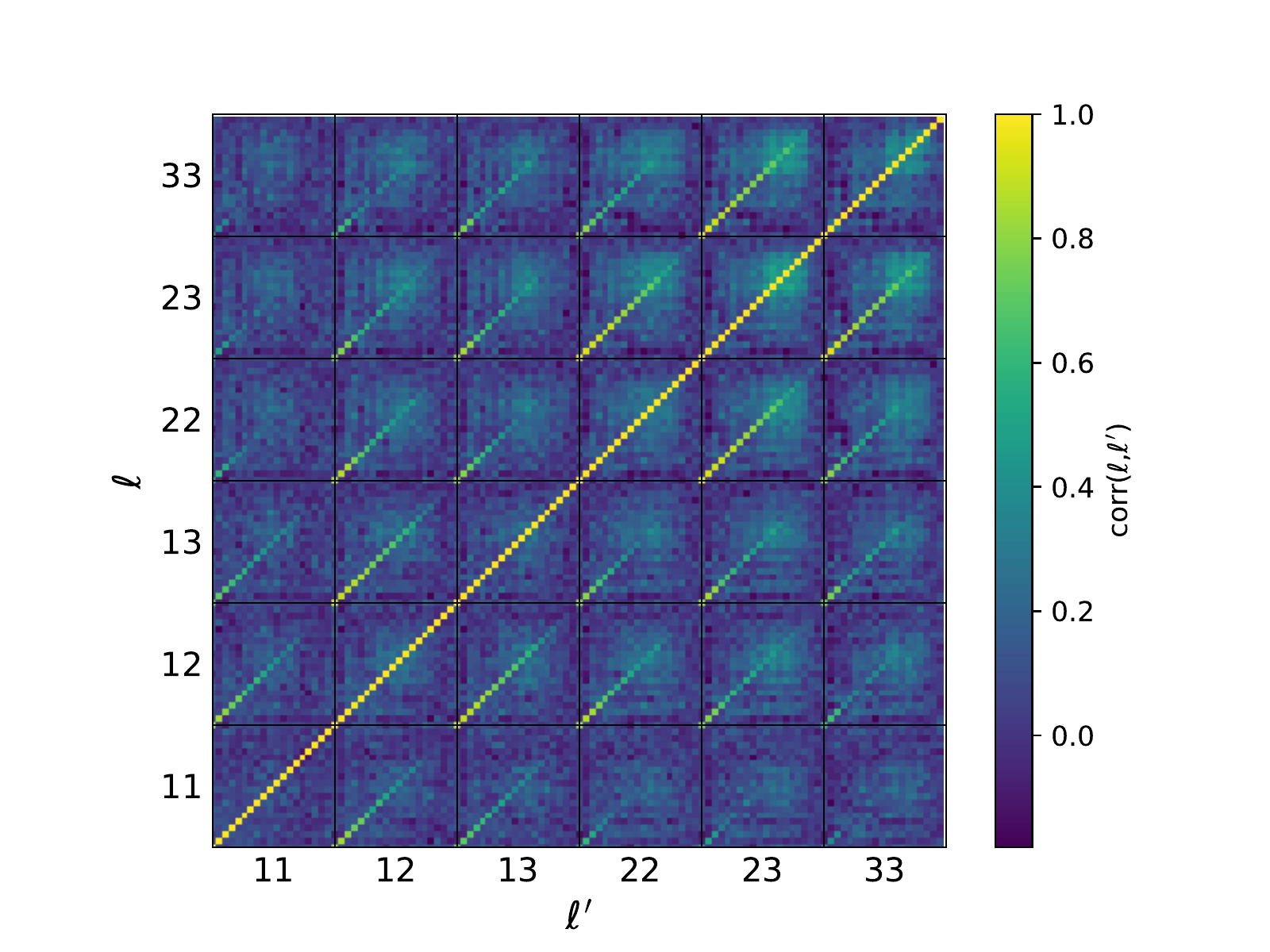}
\caption{Normalised covariance matrices of the tomographic cosmic shear power spectrum without and with baryonic effects (left and right). For the baryonic effects, we assume a realistic model that has been shown in S19 to agree well with X-ray observations (see benchmark model B-avrg in S19 \citep{Schneider:2018pfw}). The top-panels are noise free, whereas the bottom panels include Gaussian shape noise expected for a Euclid-like survey (see text).}
\label{fig:cov}
\end{figure}

We create two different mock observations one for the dark-matter-only case and one including baryonic corrections. The latter is constructed by applying the baryonification method on each of the ten $N$-body simulations at every output in redshift. We assume the baryon parameters $\log M_c=13.8$ in M$_{\odot}$/h, $\mu=0.21$, $\theta_{\rm ej}=4$, $\eta_{\rm star}=0.32$, and $\eta_{\rm sga}=0.6$, which corresponds to the benchmark model B of \citet{Schneider:2018pfw} and consists of a realistic model that agrees both with X-ray observations and hydrodynamical simulations.

The covariance matrix of the tomographic shear auto and cross power spectra is defined as 
\begin{equation}\label{cov}
{\rm cov}\left[C_{ij}(\ell),C_{ij}(\ell')\right]=\langle\, ( C_{ij}(\ell)-\langle C_{ij}(\ell)\rangle)\,( C_{ij}(\ell')-\langle C_{ij}(\ell')\rangle)\, \rangle,
\end{equation}
where $\langle...\rangle$ represents the average over all independent weak-lensing maps. The angular power spectra for the redshift bins $i$ and $j$ are given by $C_{ij}(\ell)$, where $i=j$ corresponds to the auto and $i\neq j$ to the cross power spectra. The variable $\ell$ describing the angular scale has to be discretised in bins. For illustration purposes, it is often more convenient to define the correlation matrix
\begin{equation}
{\rm corr}\left[C_{ij}(\ell),C_{ij}(\ell')\right]=\frac{{\rm cov}\left[C_{ij}(\ell),C_{ij}(\ell')\right]}{\sqrt{{\rm cov}\left[C_{ij}(\ell),C_{ij}(\ell)\right]{\rm cov}\left[C_{ij}(\ell'),C_{ij}(\ell')\right]}},
\end{equation}
which is nothing else than the covariance matrix normalised to its diagonal values.

We construct the covariance matrix based on the 300 baryonified and dark-matter-only weak-lensing maps at each of the three redshift bins. Individual angular power spectra are calculated using the routine anafast of the HEALPix software package. We thereby define 19 $\ell$-bins going from $\ell=20$ to 4000. The bins are equally separated in logarithmic space.

In Fig.~\ref{fig:cov} we plot the correlation matrices obtained from both the dark-matter-only (left) and the baryon-corrected maps (right). They include the three auto (11, 22, 33) and three cross correlation terms (12, 13, 23) for all 19 $\ell$-bins. The top-panels show the correlation matrices without shape noise. They illustrate the well-known fact that small-scale modes are correlated, which is a direct consequence of the non-linear nature of structure formation. The bottom-panels, on the other hand, show the full correlation matrices including shape-noise errors. The latter strongly suppress any correlations, leading to subdominant off-diagonal terms over all scales. A closer look at Fig.~\ref{fig:cov} reveals that the differences between dark-matter-only and baryon-corrected covariances remain very small. This is not only true for the full covariance matrix but also for the noise-free case where potential deviations should be more visible.

Finally, with the covariance matrix at hand, we can build our mock observations. In principle this would be possible using the spherical harmonic auto and cross power spectra directly measured on our simulations. However, the maps suffer from a few-percent power depletion at the largest scales due to the limited box size and the randomisation procedure \citep[see also Ref.][]{Weiss:2019jfx}. Furthermore, we use the revised {\tt halofit} framework of \citet{Takahashi:2012em} for the predictions. The latter is only about 5 percent accurate in the non-linear regime, leading to a shift in the posterior, which becomes visible when stage-IV lensing survey configuration are assumed. We therefore only use the simulated maps for the covariance matrix and directly use the revised {\tt halofit} results for the mock angular power spectrum. Note, however, that this is primarily an aesthetic choice which does not affect the conclusions of the paper.

The resulting mock power spectra are plotted as black symbols in Fig.~\ref{fig:Cofl}. The error bars correspond to the (square-root of the) diagonal values of the covariance matrix and are the result of our light-cone simulations and weak-lensing map generation procedure. They consist of a combination of Gaussian noise and statistical errors assuming a Euclid-like galaxy resolution and survey footprint. The default cosmological, baryonic, and intrinsic alignment parameters used for the mock data are summarised in Table.~\ref{tab:prior0}. 


\section{Parameter forecast analysis}\label{sec:likelihoodanalysis}
Based on the mock angular power spectrum shown in Fig.~\ref{fig:Cofl}, we now perform a number of likelihood analyses to estimate the constraining power of a stage-IV weak lensing survey, assuming a standard $\Lambda$CDM cosmology. This means we simultaneously vary five cosmological parameters ($\Omega_b$, $\Omega_m$, $\sigma_8$, $n_s$, $h_0$), one intrinsic-alignment parameter ($A_{\rm IA}$), and three gas parameters ($M_c$, $\mu$, $\theta_{\rm ej}$). The remaining baryonic parameters describing the stellar components ($\eta_{\rm star}$, $\eta_{\rm cga}$) are kept fixed for simplicity. This choice can be justified by the fact that the stellar components are relatively well known and have a comparable minor effect on the power spectrum (see Fig.~2 in S19). However, we explicitly investigate the role of stellar parameters in Appendix~\ref{sec:53barpars}, showing that they have no noticeable effects on cosmological parameter estimates of stage-IV weak lensing surveys. 

\begin{table}[tbp]
\centering
\small
\begin{tabular}{c  c c  c }
Parameter description & Acronym & True value & Prior range\\
\hline
Cosmic baryon abundance & $\Omega_b$ & 0.049 & $0.04-0.06$ \\
Cosmic matter abundance & $\Omega_m$ & 0.315 & $0.15-0.42$ \\
Clustering amplitude & $\sigma_8$ &0.811 & $0.66-0.9$ \\

Spectral index & $n_s$ & 0.966 & $0.9-1.0$ \\
Reduced Hubble parameter & $h_0$ & 0.673 & $0.6-0.9$ \\
\hline
Baryonic parameter 1 (related to slope of gas profile) & $\log M_c$ & 13.8 & $13-16$ \\
Baryonic parameter 2 (related to slope of gas profile) & $\mu$ & 0.21 & $0.1-0.7$ \\
Baryonic parameter 3 (related to maximum gas ejection) & $\theta_{\rm ej}$ & 4 & $2-8$ \\
\hline
Amplitude of intrinsic alignment & $A_{\rm IA}$ & 1 & $0-2$ \\
\hline
\end{tabular}
\caption{Parameter descriptions, acronyms, true value (used in the mock data set), and prior ranges for the cosmological, baryonic, and intrinsic alignment parameters. All priors are assumed to be flat.}
\label{tab:prior0}
\end{table}

All model parameters including their prior ranges are summarised in Table~\ref{tab:prior0}. The priors on the cosmological and intrinsic-alignment parameters are selected to be wide enough to not affect the posteriors. Only the baryon-abundance $\Omega_b$, which is not very sensitive to the weak-lensing signal, has been set to a range motivated by (but significantly broader than) results from the CMB \citep{Aghanim:2018eyx} and nucleosynthesis \citep[e.g. Ref.][]{Steigman:2007xt}. The priors on the baryonic parameters are selected so that they comfortably include the predictions of all known hydrodynamical simulations. This can be verified in Fig.~\ref{fig:hydrocomparison} where the grey area indicates the resulting spread in the matter power spectrum for modifications of baryonic parameters within the above prior ranges. In the following we will see that the priors on $M_c$ and $\theta_{\rm ej}$ are wide enough so that these parameters are uniquely constrained by the weak-lensing shear power spectrum. This is not the case for the third parameter $\mu$, which, however, has a much weaker effect on the cosmological signal. All priors are assumed to be flat.

\subsection{Cosmological parameters}
The goal of this section is to investigate and compare different strategies for dealing with baryonic effects on the tomographic weak-lensing shear power spectrum. This allows us to assess potential biases from simplifying assumptions. The strategies are the following:
\begin{itemize}
\item[(i)] First, we consider the case where baryonic effects are completely ignored in the prediction pipeline. Although we know this assumption to be inaccurate, we want to quantify the introduced bias on the cosmological parameters.
\item[(ii)]Second, we ignore baryonic effects but consider only data from the largest cosmological scales with $\ell\leq100$ (where baryonic effects are subdominant). While this should reduce the biases on cosmological parameters, it will also increase the size of their contours.
\item[(iii)] As a third step, we carry out a parameter inference analysis including baryons in the prediction pipeline and and marginalising over all baryonic parameters to determine the cosmology. This should give us a realistic estimate of the expected accuracy from the weak-lensing shear power spectrum alone (assuming a Euclid-like survey). 
\item[(iv)] Finally, we perform the same exercise, however, this time keeping all baryonic parameters at a fixed value. This corresponds to an ideal situation where baryonic parameters are perfectly determined with external data from gas observations.
\end{itemize}
For each of these four cases, we run a cosmological inference sampling and we analyse the size of the resulting parameter contours as well as potential biases with respect to the true cosmology of the mock data set. The inference is performed using the Markov-Chain Monte Carlo (MCMC) sampler {\tt UHAMMER} \citep{Akeret:2012aaa} which is based on the {\tt emcee} code \citep{Foreman-Mackey:2013aaa}. 

\begin{figure}[tbp]
\centering
\includegraphics[width=0.99\textwidth,trim=0.3cm 0.2cm 0.7cm 0.3cm,clip]{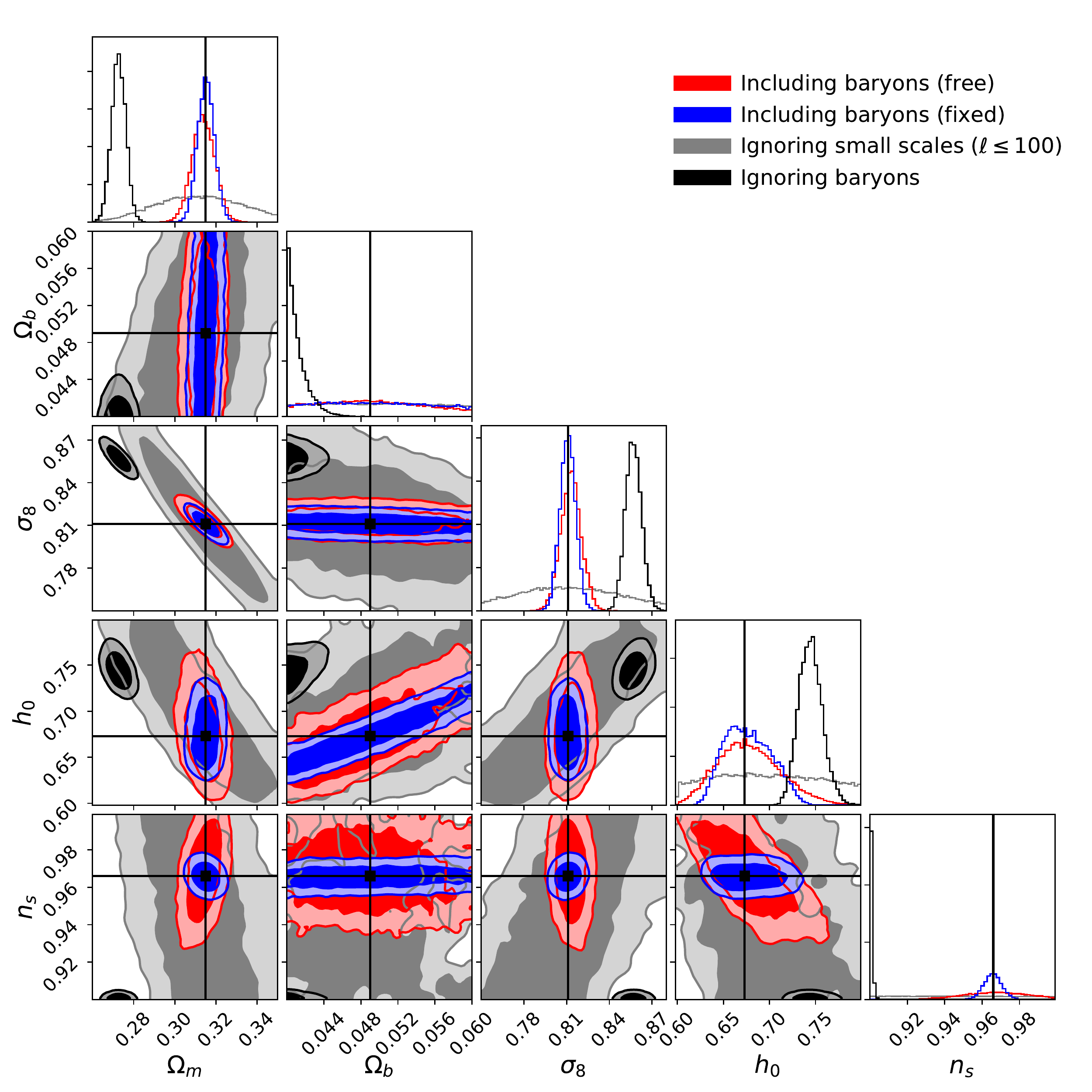}
\caption{Posterior contours of all cosmological parameters assuming a Euclid-like setup for the cases when baryonic effects are either included (red, blue) or ignored (black, grey). The true parameter values of the mock data are indicated by black lines. Ignoring baryonic effects leads to strong biases (black) except if small scales are ignored (grey). Accounting for baryonic effects while including all scales (up to $\ell=4000$) makes the posterior contours shrink significantly, especially if the baryonic parameters are fixed (blue) instead of marginalised over (red).}
\label{fig:contourDMOvsBCM}
\end{figure}

Fig.~\ref{fig:contourDMOvsBCM} highlights the results for the cosmological parameters ($\Omega_m$, $\Omega_b$, $\sigma_8$, $h_0$, and $n_s$) while the baryonic and intrinsic-alignment parameters are marginalised over. Scenario (i) of the above list is shown in black, illustrating what happens if the baryonic effects are ignored in the analysis pipeline. While the resulting contours are very tight, they are strongly biased with respect to the true cosmology. Depending on the cosmological parameter, the bias is typically between about 5 and 10 standard deviations. This is a much stronger effect than in original work \citep{Jing:2005gm,Rudd:2007zx} but in qualitative agreement with more recent findings from Refs.~\citep{Semboloni:2011aaa,Huang:2018wpy} that included the effects of AGN feedback. 
We conclude that ignoring baryonic effects would lead to very wrong conclusions about cosmology.

The grey contours represent the scenario (ii), where all scales significantly affected by baryons are ignored. According to Fig.~\ref{fig:Cofl}, this requires a cut at $\ell=100$, which means that only about one third of the available data points are used in the MCMC analysis. While such a strategy of cutting small-scale information does not lead to a noticeable bias (the grey contours are well centred around the true cosmology indicated by the black cross), the constraining power remains poor for all parameters.

The red contours of Fig.~\ref{fig:contourDMOvsBCM} illustrate the scenario (iii), where all scales are included in the analysis and baryonic effects are modelled according to the baryonic correction model of Sec.~\ref{sec:BCM}. The baryonic model parameters are allowed to vary freely within the prior ranges provided in Table~\ref{tab:prior0}. They are marginalised over together with the intrinsic-alignment parameter. The resulting constraints on $\Omega_m$ and $\sigma_8$ are each about a factor of four tighter than the contours from large-scales only. This highlights the importance of including small scales in the analysis of future weak-lensing observations. Other parameters such as $\Omega_b$ and $n_s$ remain rather poorly constrained.

Finally, the blue contours of Fig.~\ref{fig:contourDMOvsBCM} illustrate the scenario (iv), where baryonic parameters are fixed to their correct values. This corresponds to the ideal situation where all baryonic effects are fully determined. Compared to the case where baryonic effects are marginalised over, the error on $n_s$ shrink by more than a factor of 2, while the errors on $h_0$, $\sigma_8$, and $\Omega_m$ shrink by bout 50 percent or less. While this is a noteworthy improvement, it also tells us that weak-lensing shear alone (without additional data from gas observations) results in surprisingly tight constraints of cosmological parameters. It is unclear whether this will also be the case for extensions of the cosmological model, which could affect the power spectrum in ways that are degenerate with the baryonic effects.

The posterior contours shown in Fig.~\ref{fig:contourDMOvsBCM} make it very clear that future weak lensing surveys such as Euclid or LSST require a proper parametrisation of baryonic effects. In terms of the key cosmological parameters $\Omega_m$ and $\sigma_8$, parametrising and marginalising over baryonic parameters leads to more than 3 snd 5 times tighter error bars compared to the conservative case where only modes that are unaffected by baryons are considered. In terms of the contour area, which is often used as a measure of the Figure of Merit, the improvement is more than a factor of 15.

The above analysis also shows that external prior information on the baryonic parameters has the potential to further reduce the errors on cosmological parameters. However, for $\Omega_m$ and $\sigma_8$, the improvement on the individual errors is about 50 percent at best, for the contour area it is not more than a factor of two. This is a noticeable, but not a dramatic improvement. However, it is possible that adding gas information becomes more important when investigating extensions of the minimal cosmological model. In Paper II \citep{Schneider:2019bbb}, we will combine the angular power spectra with mock data of X-ray gas fractions and we will investigate the models including massive neutrinos and extensions of the $\Lambda$CDM model.

\begin{figure}[tbp]
\centering
\includegraphics[width=0.98\textwidth,trim=0.1cm 0.1cm 0.5cm 0.4cm,clip]{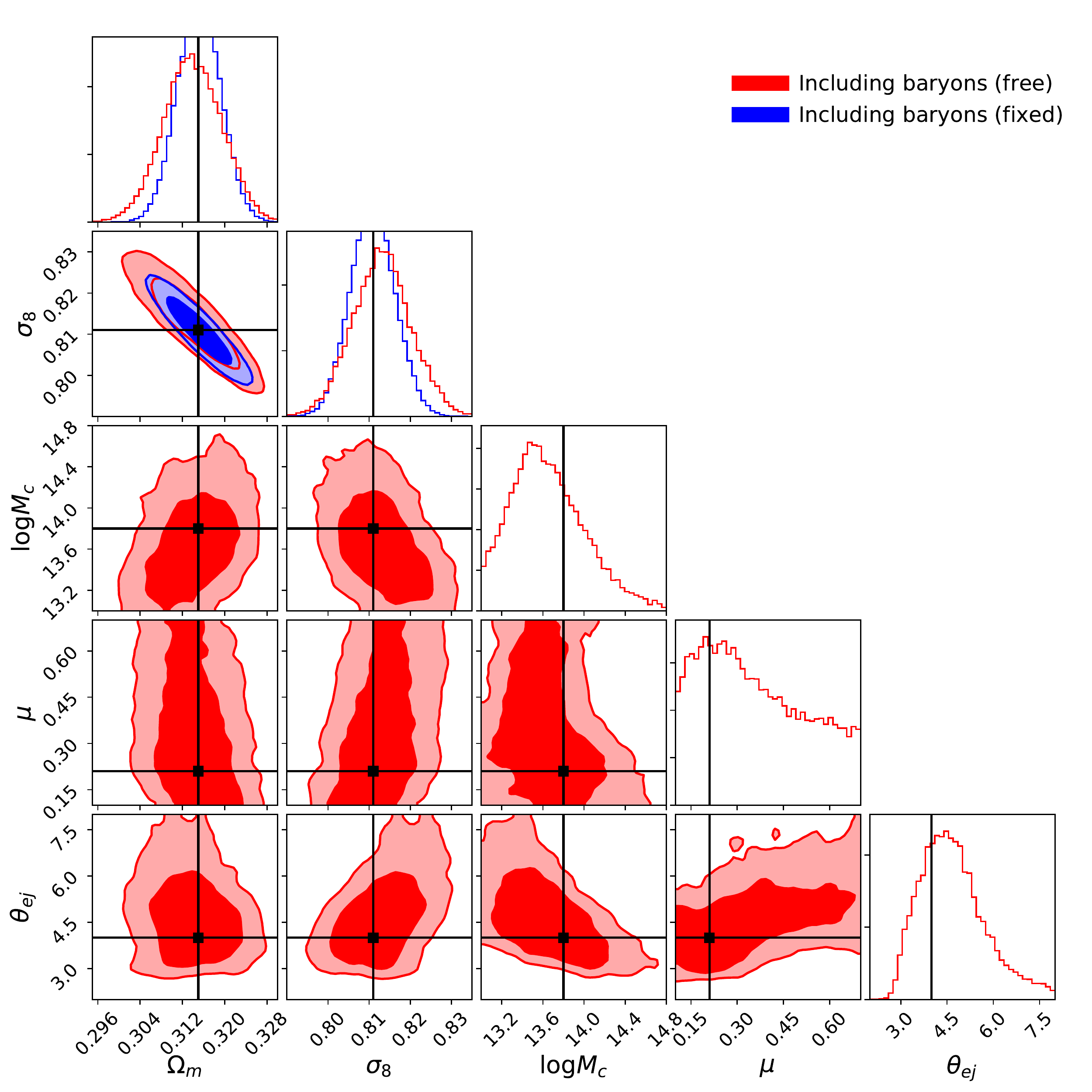}
\caption{Posterior contours of the baryonic parameters $\log M_c$, $\mu$, and $\theta_{\rm ej}$ together with $\Omega_m$ and $\sigma_8$. All other cosmological and intrinsic-alignment parameters are marginalised over. The red and blue contours correspond to the cases (iii) and (iv) in the text, where the three baryonic parameters are either left free to vary or fixed to the values assumed in the mock (indicated by black lines). The red contours show that a Euclid-like shear power spectrum alone is able to put limits on the parameters $\log M_c$ and $\theta_{\rm ej}$ while $\mu$ remains largely unconstrained. Additional observations from X-ray will be able put much more stringent limits on baryonic parameters (see Paper II \citep{Schneider:2019bbb}).}
\label{fig:contourBARPARAMS}
\end{figure}

\subsection{Baryonic parameters}
We now turn our focus towards the baryonic parameters $M_c$, $\mu$, and $\theta_{\rm ej}$. They describe the gas profiles of galaxy groups and clusters (see Eq.~\ref{rhogas}) thereby affecting the weak-lensing signal. While the baryonic parameters are best constrained using direct gas observations, we will now investigate whether weak-lensing shear data alone can constrain baryonic feedback parameters as well.

Fig.~\ref{fig:contourBARPARAMS} shows the posterior contours of the baryonic parameters together with the $\Omega_m$ and $\sigma_8$ parameters. The values of the parameters assumed for the mock data are indicated with black lines (they correspond to the beat-fitting values from current X-ray data, see S19). The red and blue contours correspond to the scenarios (iii) and (iv) from the previous section. The latter does only appear in the top panel, since for this case all baryonic parameters are fixed to their true values. Fig.~\ref{fig:contourBARPARAMS} shows that both the $M_c$ and $\theta_{\rm ej}$ parameters (describing the slope of the maximum extend of the gas profile) are constrained well beyond the original prior-range. The third parameter $\mu$, on the other hand, shows a rather flat posterior distribution over the full prior range. This is not surprising, since $\mu$ has been shown in S19~\citep{Schneider:2018pfw} to be the least sensitive of the baryonic parameters regarding changes of the matter power spectrum.

In summary, Fig.~\ref{fig:contourBARPARAMS} shows that future weak-lensing data will be able to not only constrain cosmology but astrophysical effects such as gas ejection from AGN feedback and its dependence on halo mass. Note, however, that with additional data from gas observations it will be possible to pin down baryonic parameters at much higher precision. In Paper II \citep{Schneider:2019bbb} we will show this using gas fractions from the upcoming X-ray survey eROSITA.

\subsection{Allowing for additional freedom in the redshift evolution}
The baryonic correction (BC) model does not account for any explicit redshift dependence of the gas, stellar, or dark-matter density profiles. The apparent redshift-evolution of the power spectrum (visible for example in Fig.~\ref{fig:Samples0_1}) is a result of the fact that at different redshift the signal is dominated by haloes of different mass. Although this implicit redshift evolution is in good agreement with some hydrodynamical simulations, the true redshift evolution of the baryonic suppression effect remains uncertain \citep[see S19][]{Schneider:2018pfw}.

\begin{figure}[tbp]
\centering
\includegraphics[width=0.49\textwidth,trim=0.0cm 0.1cm 1.5cm 0.4cm,clip]{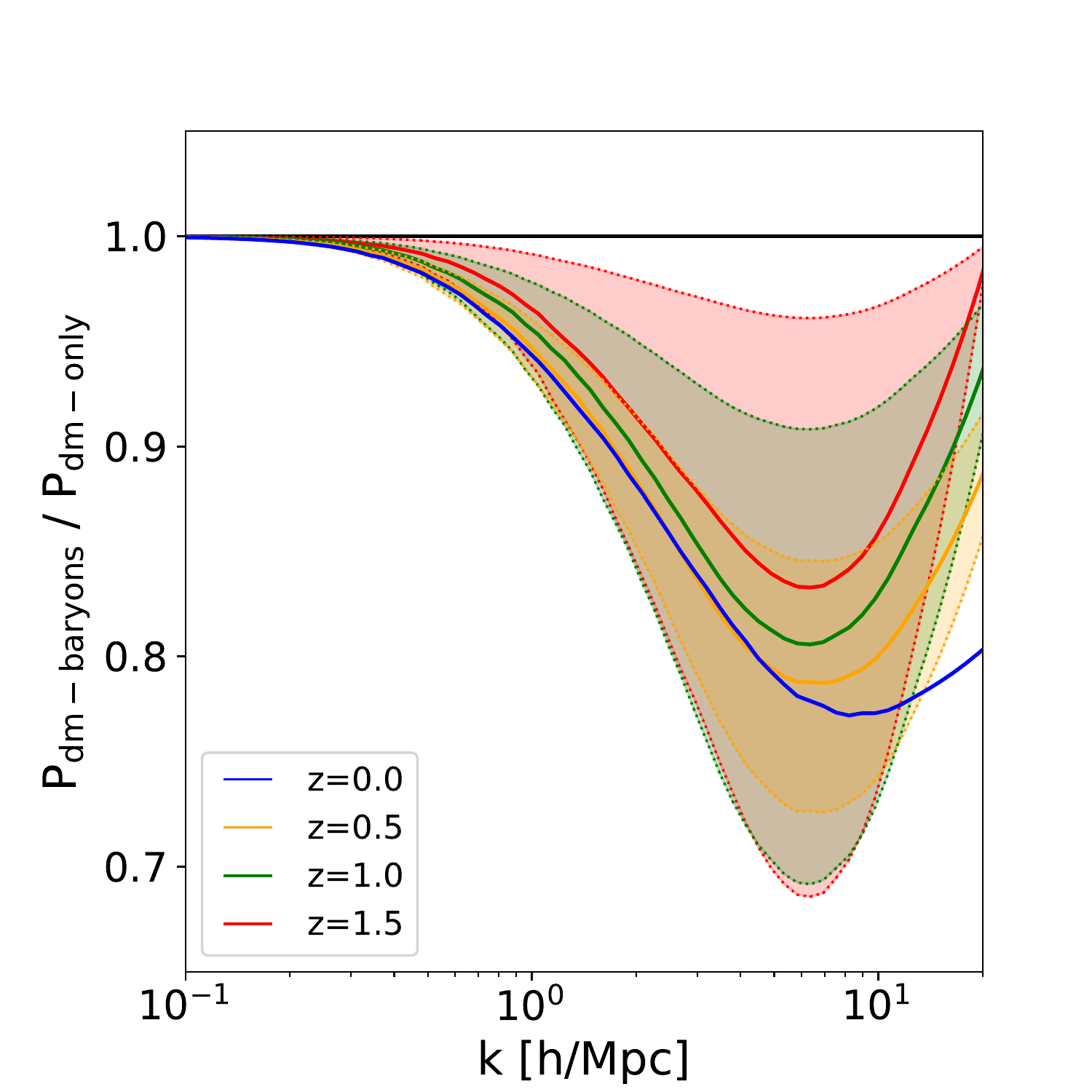}
\includegraphics[width=0.49\textwidth,trim=0.0cm 0.1cm 1.5cm 0.4cm,clip]{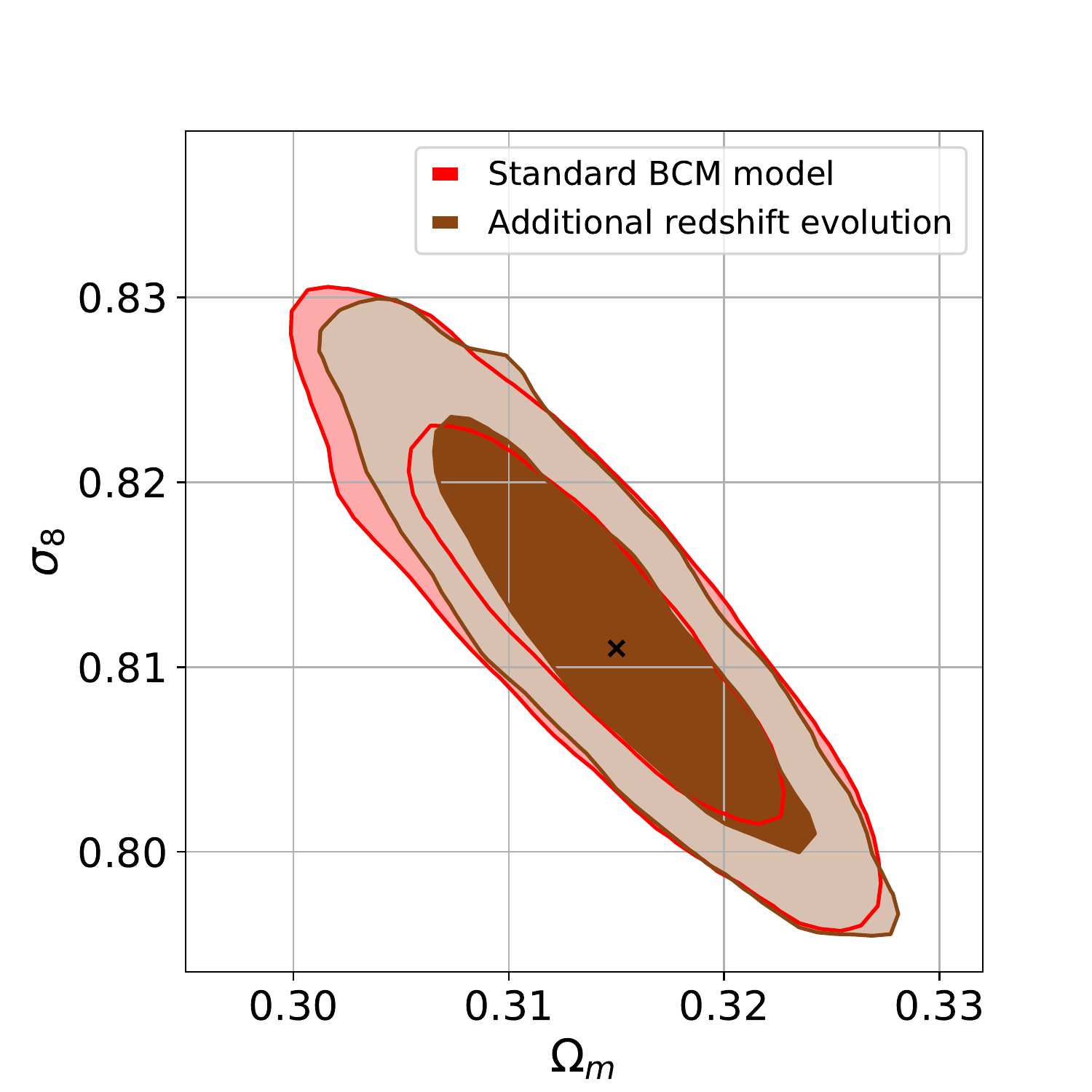}
\caption{Effects of an additional free parameter ($\gamma$) quantifying deviations from the standard redshift evolution of the baryonic suppression signal. \emph{Left:} Baryonic suppression on the matter power spectrum for a {\tt Planck} cosmology and default baryonic parameters. The coloured solid lines show the original redshift evolution of the BC model, the shaded regions (delimited by dotted lines) indicate the additional freedom provided by the $\gamma$-parameter of Eq.~(\ref{zevol}) assuming a prior range of $\gamma \in [-0.5,0.5]$. \emph{Right:} resulting posterior contours (at 0.68 and 0.95 confidence level) of the original case (red) and the case with additional marginalised redshift parameter $\gamma$. We have checked that the agreement between the two cases is similar for the remaining parameters.}
\label{fig:zevol}
\end{figure}

In this section, we investigate whether further freedom in the redshift evolution has a significant effect on the posterior contours of cosmological parameters. We therefore introduce an explicit redshift parameter at the level of the matter power spectrum. This is done by replacing the baryonic suppression signal (defined in Eq.~\ref{barsuppression}) with the relation
\begin{equation}\label{zevol}
S_{\rm BCM}(k,z)\, \rightarrow\, S_{\rm BCM}(k,z) + \left[1-S_{\rm BCM}(k,z)^{\gamma z}\right]
\end{equation}
in Eq.~(\ref{Pdmb}). The new parameter $\gamma$ is allowed to vary freely within the prior ranges $\gamma=[-0.5,0.5]$. If $\gamma>0$ ($\gamma<0$), the baryonic suppression gets reduced (increased) towards higher redshifts, whereas $\gamma=0$ means no change with respect to the previous model.

Note that the above redshift evolution does not allow for the same amount of freedom as adding explicit redshift dependencies into the gas, stellar, and dark matter profiles of the BC model (defined in Sec.~\ref{sec:BCmodel}). However, we have checked that Eq.~(\ref{zevol}) allows us to mimic the redshift evolution of hydrodynamical simulations to reasonable precision. The redshift dependency of the Cosmo-OWLS runs, for example, can be reproduced at the 2 percent level up to $k\sim7$ h/Mpc with $\gamma=0.17$. Similar or better agreement is found for the OWLS, Horizon-AGN, and Illustris-TNG simulations.

In the left panel of Fig.~\ref{fig:zevol}, we illustrate the effect of the additional $\gamma$-parameter on the relative power spectrum ($S_{\rm BCM}$). At each redshift, the shaded region shows the additional freedom of $S_{\rm BCM}$ if $\gamma$ is varied within the prior ranges between $\gamma=-0.5$ (lower limit) and $\gamma=0.5$ (upper limit). The solid lines in the centre of the shaded regions indicate the case of $\gamma=0$. The added variability due to the additional redshift parameter grows from zero at $z=0$ to nearly the size of the baryon suppression signal at $z=1.5$.

The right panel of Fig.~\ref{fig:zevol} shows the $\Omega_8$-$\sigma_8$ contours resulting from MCMC chain including all parameters of Table~\ref{tab:prior0} plus the additional redshift parameter $\gamma$ (brown).The contour shows no significant change in size compared to the standard case with fixed $\gamma=0$ (red). We have checked that this is also the case for the remaining parameter contours not shown in this plot.

Based on Fig.~\ref{fig:zevol} we conclude that, although the BC model might not be general enough to include the full range of potential redshift dependence, such an effect is unlikely to significantly affect cosmological parameter estimates. This confirms the validity of the results shown in the present paper.

\subsection{Potential biases from the estimated covariance}
Before we move on and investigate possible simplifications of the baryonification method, let us discuss the potential presence of biases caused by our estimate of the covariance matrix. It has been argued by \citet{Hartlap:2006kj} that, in order to not underestimate the posterior contours, the inverse of the covariance matrix has to be corrected by the factor $\alpha=(N-p-2)/(N-1)$, where $p$ is the length of the data vector and $N$ is the number of realisations used to construct the covariance matrix. More recently, \citet{Sellentin:2015waz} showed that correcting for the uncertainties in the covariance matrix furthermore requires the Gaussian likelihood to be replaced by a multivariate t-distribution.

Note that all posterior contours from this paper have been estimated using a Gaussian likelihood without recalibrating the covariance matrix. In order to test the validity of this approach, we have performed an additional MCMC run based on the prescription of \citet{Sellentin:2015waz} (i.e. replacing the multivariate Gaussian by a t-distribution and adding the pre-factor $\alpha$ to the inverse of the covariance matrix). As a result, we found posterior contours that are effectively indistinguishable compared to the ones from the uncorrected analysis. We therefore conclude that our results are unlikely to be plagued by significant biases related to the construction of the covariance matrix.

\section{Testing simplifying model assumptions}\label{sec:assumptions}
Current weak-lensing studies generally rely on simplifying assumptions for baryonic effects in their analysis pipeline. For example, baryonic effects are usually not included in the covariance matrix and any potential cosmology dependence of baryonic parameters is ignored. In this section, we evaluate whether it will be acceptable to ignore these effects in the analysis pipeline of future weak-lensing surveys such as Euclid or LSST.

\subsection{Cosmology dependence of the baryonic suppression}
In the first part of this paper, we have established that the baryonic suppression of the power spectrum is sensitive to the mean baryon fraction $f_b=\Omega_b/\Omega_m$. While this dependency is fully accounted for in our analysis, this is not the case for other studies where any cosmology dependence of baryonic effects is usually ignored. We now investigate whether this simplifying assumption has any effect on cosmological parameter estimates.

\begin{figure}[tbp]
\centering
\includegraphics[width=0.49\textwidth,trim=0.0cm 0.1cm 1.5cm 0.4cm,clip]{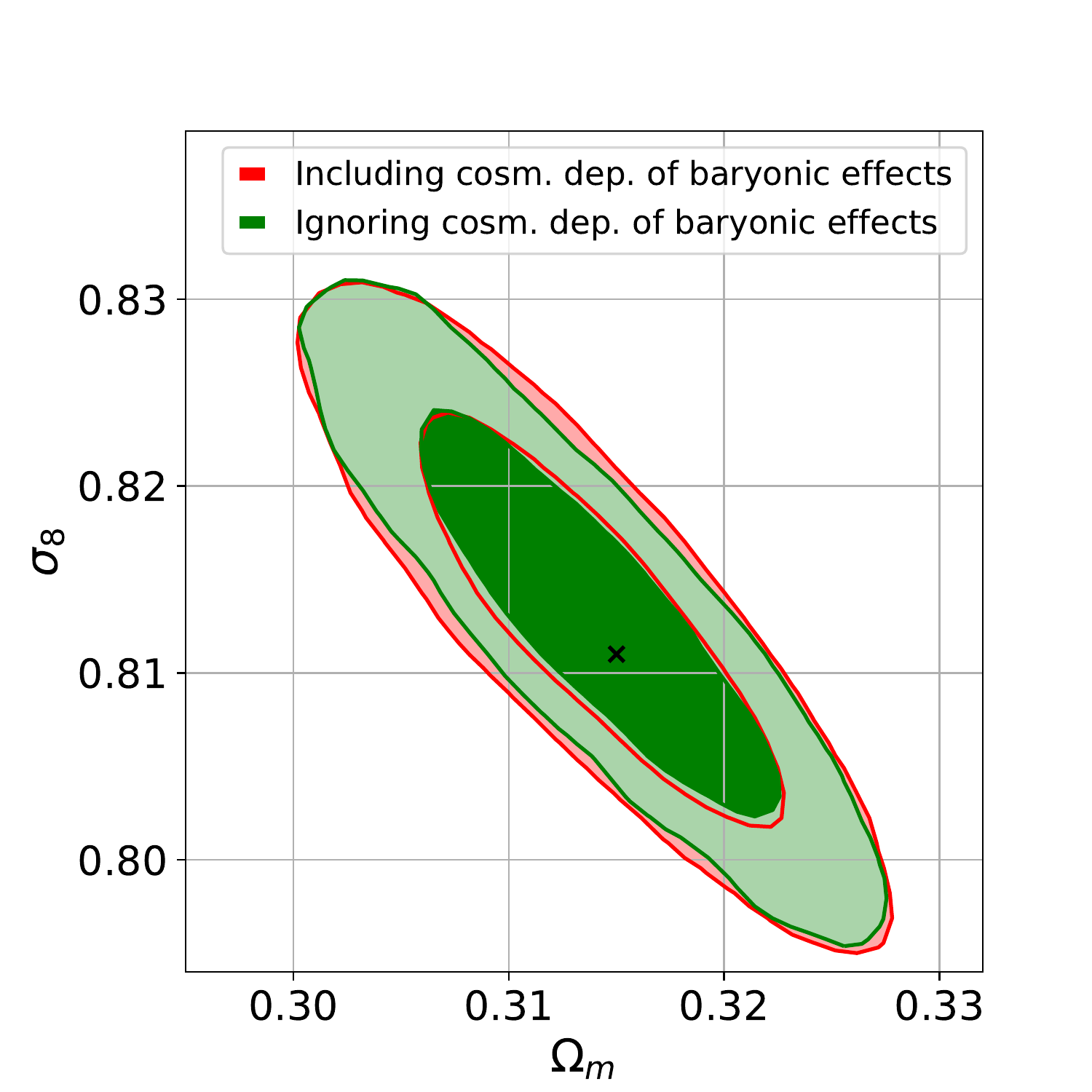}
\includegraphics[width=0.49\textwidth,trim=0.0cm 0.1cm 1.5cm 0.4cm,clip]{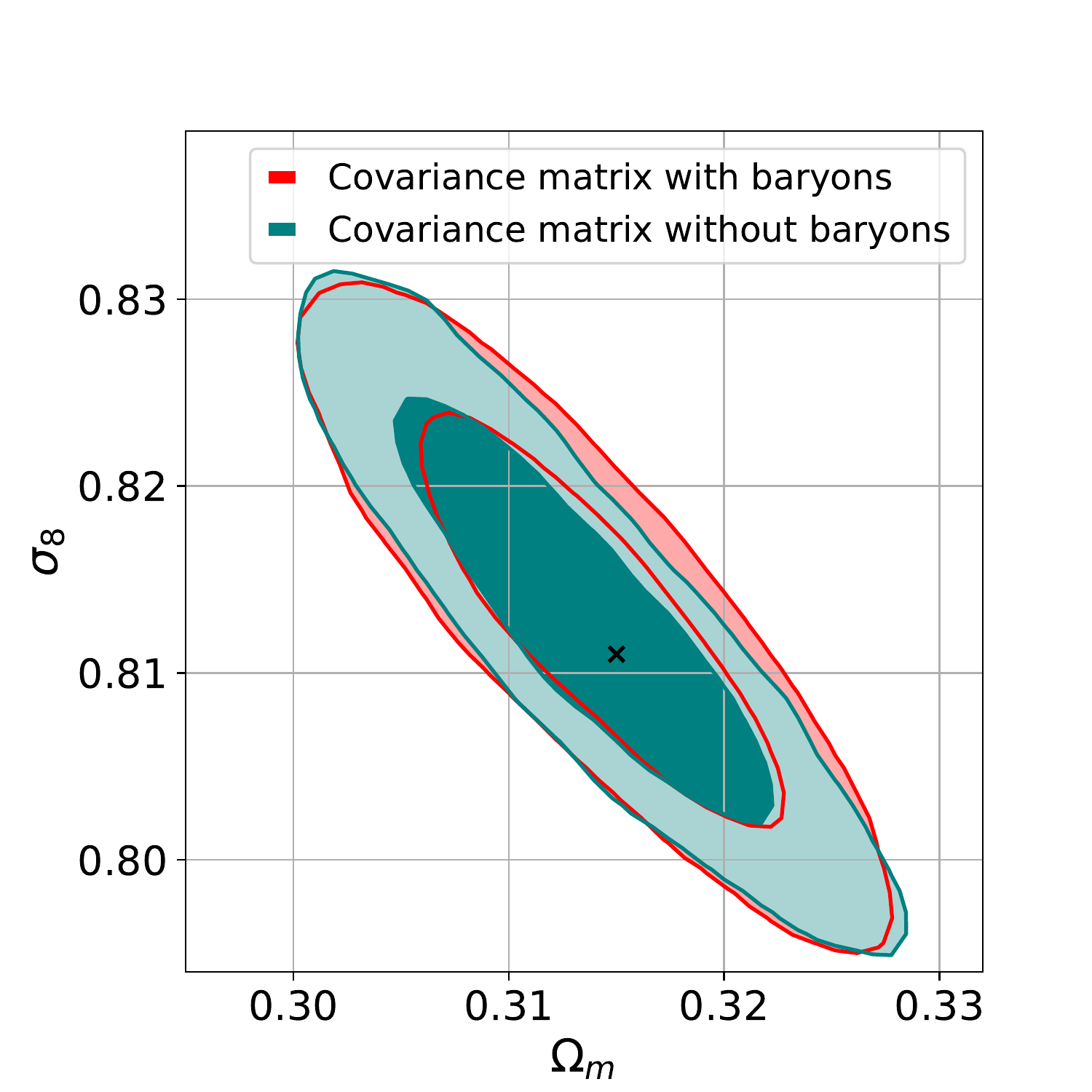}
\caption{Investigating the effects of simplifying assumptions of baryonic modelling on the cosmological parameters $\Omega_m$ and $\sigma_8$. \emph{Left:} Change of the posterior contours if the cosmology dependence between baryonic and cosmological parameters is ignored (green) instead of being properly included (red). The former is obtained by fixing the cosmic baryon fraction to $f_b=0.155$ in the BC model during the MCMC run. \emph{Right:} Change of the posterior contours if the baryons are ignored in the covariance matrix (steal-blue) instead of being included (red). We have checked that the level of agreement between the contours is similar for the other cosmological parameters.}
\label{fig:contourCOSCOV}
\end{figure}

The test is performed by comparing the posterior contours of the full analysis with a test-case where the baryonic fraction $f_b=\Omega_b/\Omega_m$ of Eq.~(\ref{fractions}) is fixed to the Planck-value of $f_b=0.155$. This means that while gravitational clustering remains sensitive to variations of $\Omega_b$ and $\Omega_m$, the baryonic suppression effect ($S_{\rm BCM}$) becomes completely decoupled from cosmology. In the left panel of Fig.~\ref{fig:contourCOSCOV} we compare the $\Omega_m$-$\sigma_8$ contours from a BC model with fixed $f_b$ (green) to the generic case with fully cosmology-dependent baryonic parameters (red). Both contours are very similar, showing no significant bias in the $\Omega_m$-$\sigma_8$ plane. We have checked that the same is true for the contours of all other parameters.

In order to investigate potential biases due to the selected value of $f_b$, we perform the same test assuming baryonic fractions of $f_b=0.13$ and $f_b=0.18$, respectively. Note that these values differ from the assumed cosmology of the mock data sample. As a result, we find a visible but very small shift of the $\Omega_m$-$\sigma_8$ contours towards the bottom-right corner for increasing $f_b$. However, the shift remains much smaller than the estimated posterior contours for reasonable values of $f_b$.

The results presented in the left panel of Fig.~\ref{fig:contourCOSCOV} retrospectively validate the assumptions made in weak-lensing analysis by KiDS \citep{Hildebrandt:2016iqg}, DES \citep{Abbott:2017wau}, and HSC \citep{Hikage:2018qbn}, where parametrised baryonic suppression functions were multiplied to the power spectrum without considering any cosmology dependence. Furthermore, the results show that ignoring cosmology dependence of baryonic effects also consist of an acceptable strategy for for future, stage-IV weak lensing surveys. It remains to be checked if these conclusions are also valid for other measures beyond the two-point statistic.

\subsection{Baryonic effects on the Covariance matrix}
In Sec.~\ref{sec:mockobs}, we have calculated the covariance matrix of the tomographic shear power spectrum with and without baryonic suppression effects. The goal of the current section is to establish, if it is sufficient to only account for baryonic effects at the level of the power spectrum or if they also have to be included in the covariance matrix. Note that we only compare the two cases of a dark-matter-only and a fixed baryonified covariance matrix (assuming the benchmark model B-avrg, see S19) and we do not vary the baryonic parameters during the parameter inference.

The right-hand panel of Fig.~\ref{fig:contourCOSCOV} illustrates the $\Omega_m$-$\sigma_8$ contours when baryons are ignored (steel blue) and included (red) in the covariance matrix. Both contour areas are very similar in size and they show no noticeable offset with respect to each other. Although Fig.~\ref{fig:contourCOSCOV} only illustrates the $\Omega_m$-$\sigma_8$ contours, we have checked that this is also true for all other parameters.

Based on the findings of the section, we conclude that for future stage-IV weak-lensing surveys, it will be sufficient to calculate the covariance matrix of the cosmic shear power spectrum using gravity-only $N$-body simulations. This considerably simplifies the analysis pipeline of stage-IV weak-lensing surveys.

\begin{figure}[tbp]
\centering
\includegraphics[width=0.99\textwidth,trim=0.3cm 0.2cm 0.7cm 0.3cm,clip]{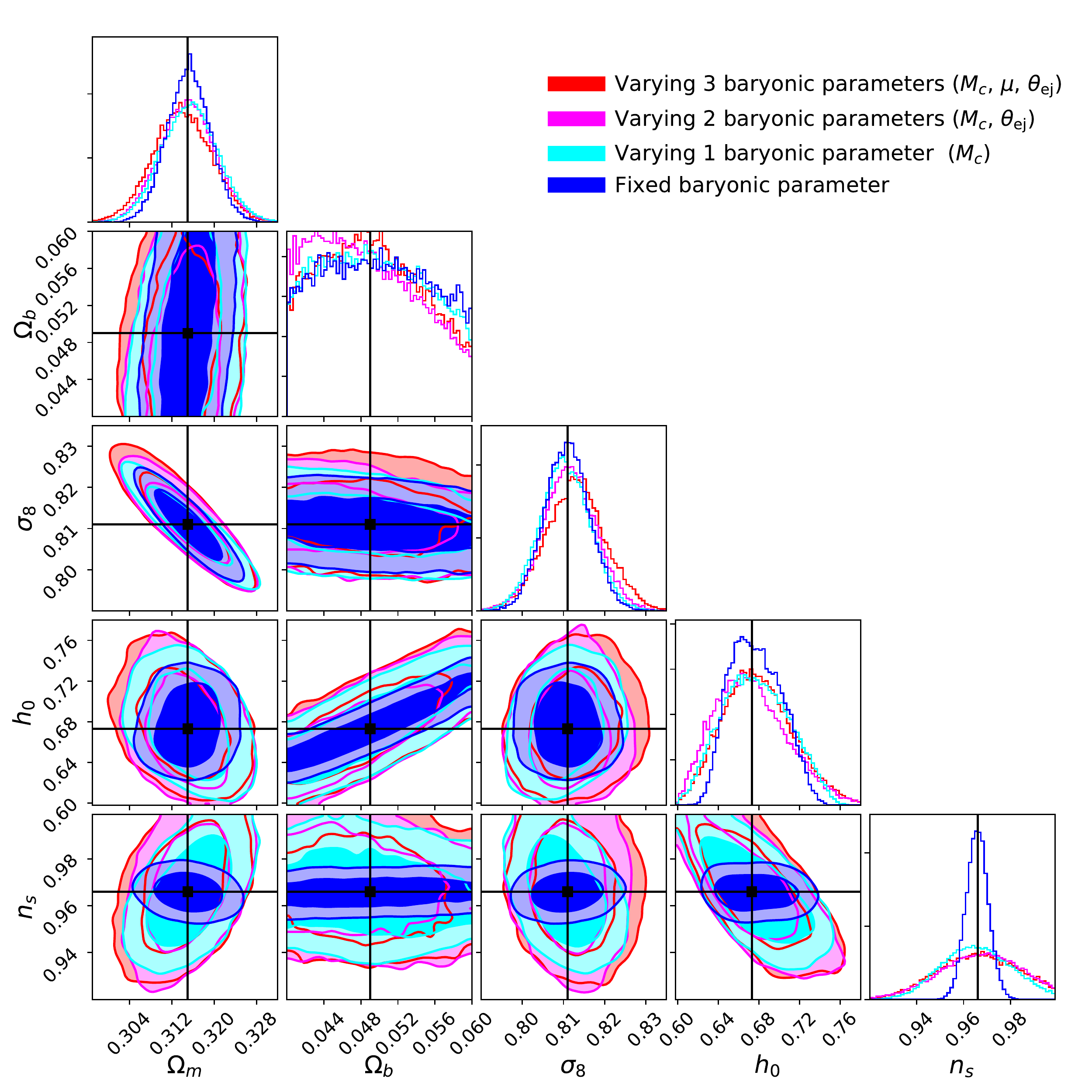}
\caption{Change of the cosmological parameter contours when the amount of free baryonic parameters are reduced from the original 3 (red) to 0 (blue). The red and blue contours have already been shown in Fig.~\ref{fig:contourDMOvsBCM}. The magenta contours correspond to the case where $M_c$ and $\theta_{\rm ej}$ are allowed to vary while $\mu$ is fixed. The cyan contours show the case where only $M_c$ can vary while $\mu$ and $\theta_{\rm ej}$ are fixed. The black lines indicate the true cosmology from the mock data.}
\label{fig:contour3210}
\end{figure}

\subsection{Reducing the number of baryonic parameters}
The baryonic correction model introduced in S19 consists of 5 free parameters, 3 describing to the gas distribution ($M_c$, $\mu$, $\theta_{\rm ej}$) and 2 related to the stellar component ($\eta_{\rm star}$, $\eta_{\rm cga}$). The same parameters (together with the cosmic baryon fraction $f_b$) were used to construct the baryonic emulator in Sec.~\ref{sec:emulator}. For the forecast analysis, however, we only varied the gas parameters, whereas the stellar parameters were kept constant. This can be justified by the fact that the stellar fractions are well known from observations and that they have a comparably small effect on the baryonic suppression signal. However, we explicitly show in Appendix~\ref{sec:53barpars} that adding stellar parameters to the inference sampling does not have any noticeable effect on the resulting cosmological parameter contours.

In the present section, we aim to investigate the role of the three baryonic parameters $M_c$, $\mu$, $\theta_{\rm ej}$ on the cosmological parameters contours. More specifically, we want to find out if further reducing the number of free baryonic parameters has a noticeable effect on the size of the error bars. We therefore run two additional MCMC chains, one where $\mu$ is fixed while $M_c$, and $\theta_{\rm ej}$ are allowed to vary, and another one where $\mu$ and $\theta_{\rm ej}$ are both fixed and only $M_c$ is allowed to vary.

The cosmological parameter contours resulting from this exercise are plotted in Fig.~\ref{fig:contour3210}. While the red and blue contours correspond to the cases of fully varying and fully fixed baryonic parameters (already shown in Fig.~\ref{fig:contourDMOvsBCM}), the magenta and cyan cases show the new results with one ($\mu$) and two ($\mu$, $\theta_{\rm ej}$) fixed parameters, respectively. As expected, the contours of all cosmological parameters become larger the more baryonic parameters are allowed to vary. The increase is largest when going from 0 to 1 baryonic parameter (blue to cyan). However, there is a further gradual increase of contours when going from 1 to 2 (cyan to magenta) and from 2 to 3 (magenta to red) baryonic parameters. This becomes evident, for example, in the parameter plane of $\Omega_m$ and $\sigma_8$, where for each added free parameter, the contour areas increase by about $\sim 30$ percent.

Based on the results of Fig.~\ref{fig:contour3210}, we conclude that it is necessary to allow all 3 baryonic model parameters to vary in order to get an accurate estimate of the total error budget. We therefore retain the same baryonic prescription for the analysis pipeline of Paper II.


\section{Conclusions and outlook}\label{sec:conclusions}
Future stage-IV weak lensing surveys will be affected by baryonic processes, more particularly the redistribution of gas due to high energetic feedback effects driven by active galactic nuclei (AGN). In this paper, we carried out a forecast study investigating the effects of baryons on cosmological parameter estimates for a Euclid-like weak-lensing survey. We run a suite of simulations to construct a mock data set consisting of a tomographic shear power spectrum and a corresponding covariance matrix covering the range $\ell=20-4000$. Based on this mock, we run a MCMC chain, simultaneously varying five cosmological ($\Omega_m$, $\Omega_b$, $\sigma_8$, $h_0$, $n_s$), one intrinsic alignment ($A_{\rm IA}$), and three baryonic ($M_c$, $\mu$, $\theta_{\rm ej}$) parameters. The resulting posterior contours are compared to other likelihood analyses, where baryonic effects are either ignored or fixed to the true value. In the following, we list the main conclusions obtained from this investigation:
\begin{itemize}
\item Ignoring the effects of baryons in the prediction pipeline leads to very wrong conclusions regarding cosmology. All cosmological parameters are biased by 5 standard deviations or more compared to the assumed cosmology of the mock data-set. Furthermore, ignoring baryons results in a severe underestimation of the theoretical uncertainties, leading to very tight contours and the false impression of high accuracy.
\item One straight-forward way to reduce the bias without properly modelling the baryonic effects is to only include the largest cosmological modes that remain unaffected by baryonic feedback. This strategy, however, requires a cut at $\ell\lesssim100$, which means that most of the data remains unused. While this approach indeed gets rid of the bias, it leads to strongly increased error-bars of the cosmological parameters. Regarding $\sigma_8$ and $\Omega_m$, the 95 percent confidence levels increase by more than a factor of 3 and 5, respectively. In terms of contour areas, this corresponds to an increase of a factor of 15. Hence, ignoring all scales affected by baryons strongly reduces the constraining power of stage-IV weak lensing surveys.
\item Including baryonic effects via the baryonic correction model (assuming 3 free baryonic parameters) results in tight (and unbiased) posterior contours of cosmological parameters. This is especially true for $\Omega_m$ and $\sigma_8$, whereas $h_0$ and $n_s$ are less well constrainable. We also obtain constraints on the baryonic parameters themselves, which means that future shear observations can be used to learn more about feedback processes. 
\item Next to the strategy of including baryonic parameters and marginalising over them, we have also investigated what happens to the cosmological parameters if the baryonic parameters are fixed to their true value. This corresponds to an idealistic case where baryonic effects are fully constrained by external data (such as X-ray or SZ observations, for example). Fixing baryonic parameters leads to a $\sim20-30$ percent reduction of errors for $\sigma_8$, $\Omega_m$, while the improvement is about 50 percent for $h_0$ and a factor of 3 for $n_s$. However, considering that the latter two parameters are already strongly constrained by CMB observations, we conclude that parametrising baryonic effects and marginalising over the baryonic parameters leads to surprisingly strong constraints on cosmology.
\item The baryonic suppression effect shows a weak dependence with the cosmological parameters $\Omega_b$ and $\Omega_m$ while other parameters are not affected. This is because the cosmic baryon fraction $f_b=\Omega_b/\Omega_m$ determines the amount of available gas that is involved in the feedback process. In terms of the baryon-induced suppression of the matter power spectrum, the amplitude of the suppression increases by about a factor of two if $f_b$ is changed by a factor of two. However, despite this obvious link between baryonic and cosmological effects, we have shown that fixing $f_b$ to a constant value in the BC model only leads to very small shifts of the posterior contours well below the estimated one sigma error. This means that correcting the nonlinear clustering signal in a cosmology independent way is an acceptable approximation to account for baryonic effects for a Euclid-like survey.
\item We have constructed two versions of the covariance matrix one with and one without baryonic correction effects, running MCMC chains for both. We found no noticeable difference of the posterior contours between the two cases. Therefore, we conclude that, while baryonic effects play an important role in the prediction of the angular power spectrum, it is safe to neglect them in the covariance matrix.
\item Finally, we have investigated what happens if the baryonic correction model with originally 3 free parameters ($M_c$, $\mu$, $\theta_{\rm ej}$) is reduced to 2 parameters ($M_c$, $\theta_{\rm ej}$) or 1 parameter ($M_c$) instead. We showed that each baryonic parameter contributes to the cosmological parameter contours, justifying the use of 3 parameters to describe the baryonic effects on the weak-lensing signal.
\end{itemize} 
The analysis presented here is based on the \emph{baryonic correction model} of S19 \citep{Schneider:2018pfw}, which consists of an empirical method to model baryonic effects on the large-scale density field of the universe. The model is consistent with full hydrodynamical simulations showing 2 percent agreement or better with respect to the power spectrum at redshift zero and up to $k\sim10$ h/Mpc. Note, however, that the model currently does not assume any explicit redshift dependence of the gas profiles. Current X-ray observation and some hydrodynamical simulations are consistent with this assumption, but rather large uncertainties remain. We have run a MCMC chain where we allowed for additional freedom in the redshift dependence of the baryonic power suppression. This additional redshift parameter does not lead to significantly different posterior contours, confirming that the above results are not driven by too restricted model assumptions regarding the redshift evolution.

In a companion paper \citep[Paper II]{Schneider:2019bbb} we will extend the present analysis to a more realistic cosmological model that includes massive neutrinos. Furthermore, we will investigate three straight-forward extensions to $\Lambda$CDM: a dark energy model with dynamical equation of state (wCDM), a modified gravity model based on the $f(R)$ extension of the Einstein-Hilbert action (fRCDM), and a model with a mixed dark matter sector ($\Lambda$MDM). Another goal of Paper II is to investigate how much can be gained in terms of cosmological parameter contours, if Euclid-like data from the weak-lensing shear power spectrum is combined with X-ray observations of the cluster gas fractions from the upcoming eROSITA survey.

\section*{Acknowledgments}
The authors would like to thank Uwe Schmidt for computational support as well as Tomasz Kacprzak, Adam Amara, Raphael Sgier, and Federica Tarsitano for many helpful scientific discussions. This work has been supported by the Swiss National Science Foundation via the project numbers PZ00P2\_161363 and PCEFP2\_181157. AR wants to thank the KIPAC institute at Stanford University where part of this work has been completed. Some of the results in this paper have been derived using the healpy and HEALPix software \citep{Zonca:2019aaa}. Some of the plots rely on the {\tt conrer.py} package \citep{Foreman-Mackey:2016aaa}.


\bibliographystyle{unsrtnat}
\bibliography{ASbib}

\begin{thebibliography}{81}
\providecommand{\natexlab}[1]{#1}
\providecommand{\url}[1]{\texttt{#1}}
\expandafter\ifx\csname urlstyle\endcsname\relax
  \providecommand{\doi}[1]{doi: #1}\else
  \providecommand{\doi}{doi: \begingroup \urlstyle{rm}\Url}\fi

\bibitem[Schneider and Teyssier(2015)]{Schneider:2015wta}
Aurel Schneider and Romain Teyssier.
\newblock {A new method to quantify the effects of baryons on the matter power
  spectrum}.
\newblock \emph{JCAP}, 1512\penalty0 (12):\penalty0 049, 2015.
\newblock \doi{10.1088/1475-7516/2015/12/049}.

\bibitem[Heitmann et~al.(2008)]{Heitmann:2007hr}
Katrin Heitmann et~al.
\newblock {The Cosmic Code Comparison Project}.
\newblock \emph{Comput. Sci. Dis.}, 1:\penalty0 015003, 2008.
\newblock \doi{10.1088/1749-4699/1/1/015003}.

\bibitem[Schneider et~al.(2016)Schneider, Teyssier, Potter, Stadel, Onions,
  Reed, Smith, Springel, Pearce, and Scoccimarro]{Schneider:2015yka}
Aurel Schneider, Romain Teyssier, Doug Potter, Joachim Stadel, Julian Onions,
  Darren~S. Reed, Robert~E. Smith, Volker Springel, Frazer~R. Pearce, and Roman
  Scoccimarro.
\newblock {Matter power spectrum and the challenge of percent accuracy}.
\newblock \emph{JCAP}, 1604\penalty0 (04):\penalty0 047, 2016.
\newblock \doi{10.1088/1475-7516/2016/04/047}.

\bibitem[Knabenhans et~al.(2018)]{Knabenhans:2018cng}
Mischa Knabenhans et~al.
\newblock {Euclid preparation: II. The EuclidEmulator -- A tool to compute the
  cosmology dependence of the nonlinear matter power spectrum}.
\newblock 2018.

\bibitem[Smith and Angulo(2018)]{Smith:2018zcj}
Robert~E. Smith and Raul~E. Angulo.
\newblock {Precision modelling of the matter power spectrum in a Planck-like
  Universe}.
\newblock 2018.

\bibitem[van Daalen et~al.(2011)van Daalen, Schaye, Booth, and
  Vecchia]{vanDaalen:2011xb}
Marcel~P. van Daalen, Joop Schaye, C.~M. Booth, and Claudio~Dalla Vecchia.
\newblock {The effects of galaxy formation on the matter power spectrum: A
  challenge for precision cosmology}.
\newblock \emph{Mon. Not. Roy. Astron. Soc.}, 415:\penalty0 3649--3665, 2011.
\newblock \doi{10.1111/j.1365-2966.2011.18981.x}.

\bibitem[{Semboloni} et~al.(2011){Semboloni}, {Hoekstra}, {Schaye}, {van
  Daalen}, and {McCarthy}]{Semboloni:2011aaa}
E.~{Semboloni}, H.~{Hoekstra}, J.~{Schaye}, M.~P. {van Daalen}, and I.~G.
  {McCarthy}.
\newblock {Quantifying the effect of baryon physics on weak lensing
  tomography}.
\newblock \emph{Mon. Not. Roy. Astron. Soc.}, 417:\penalty0 2020--2035,
  November 2011.
\newblock \doi{10.1111/j.1365-2966.2011.19385.x}.

\bibitem[Huang et~al.(2018)Huang, Eifler, Mandelbaum, and
  Dodelson]{Huang:2018wpy}
Hung-Jin Huang, Tim Eifler, Rachel Mandelbaum, and Scott Dodelson.
\newblock {Modeling baryonic physics in future weak lensing surveys}.
\newblock 2018.

\bibitem[Parimbelli et~al.(2018)Parimbelli, Viel, and
  Sefusatti]{Parimbelli:2018yzv}
Gabriele Parimbelli, Matteo Viel, and Emiliano Sefusatti.
\newblock {On the degeneracy between baryon feedback and massive neutrinos as
  probed by matter clustering and weak lensing}.
\newblock 2018.

\bibitem[Mummery et~al.(2017)Mummery, McCarthy, Bird, and
  Schaye]{Mummery:2017lcn}
Benjamin~O. Mummery, Ian~G. McCarthy, Simeon Bird, and Joop Schaye.
\newblock {The separate and combined effects of baryon physics and neutrino
  free-streaming on large-scale structure}.
\newblock \emph{Mon. Not. Roy. Astron. Soc.}, 471\penalty0 (1):\penalty0
  227--242, 2017.
\newblock \doi{10.1093/mnras/stx1469}.

\bibitem[Hellwing et~al.(2016)Hellwing, Schaller, Frenk, Theuns, Schaye, Bower,
  and Crain]{Hellwing:2016ucy}
Wojciech~A. Hellwing, Matthieu Schaller, Carlos~S. Frenk, Tom Theuns, Joop
  Schaye, Richard~G. Bower, and Robert~A. Crain.
\newblock {The effect of baryons on redshift space distortions and cosmic
  density and velocity fields in the EAGLE simulation}.
\newblock \emph{Mon. Not. Roy. Astron. Soc.}, 461\penalty0 (1):\penalty0
  L11--L15, 2016.
\newblock \doi{10.1093/mnrasl/slw081}.

\bibitem[Springel et~al.(2017)]{Springel:2017tpz}
Volker Springel et~al.
\newblock {First results from the IllustrisTNG simulations: matter and galaxy
  clustering}.
\newblock 2017.

\bibitem[Chisari et~al.(2018)Chisari, Richardson, Devriendt, Dubois, Schneider,
  Brun, Beckmann, Peirani, Slyz, and Pichon]{Chisari:2018prw}
Nora~Elisa Chisari, Mark L.~A. Richardson, Julien Devriendt, Yohan Dubois,
  Aurel Schneider, M.~C. Brun, Amandine~Le, Ricarda~S. Beckmann, Sebastien
  Peirani, Adrianne Slyz, and Christophe Pichon.
\newblock {The impact of baryons on the matter power spectrum from the
  Horizon-AGN cosmological hydrodynamical simulation}.
\newblock 2018.

\bibitem[Mohammed and Seljak(2014)]{Mohammed:2014lja}
Irshad Mohammed and Uros Seljak.
\newblock {Analytic model for the matter power spectrum, its covariance matrix,
  and baryonic effects}.
\newblock \emph{Mon. Not. Roy. Astron. Soc.}, 445\penalty0 (4):\penalty0
  3382--3400, 2014.
\newblock \doi{10.1093/mnras/stu1972}.

\bibitem[Dai et~al.(2018)Dai, Feng, and Seljak]{Dai:2018vvv}
Biwei Dai, Yu~Feng, and Uros Seljak.
\newblock {A gradient based method for modeling baryons and matter in halos of
  fast simulations}.
\newblock 2018.

\bibitem[Fedeli et~al.(2014)Fedeli, Semboloni, Velliscig, Van~Daalen, Schaye,
  and Hoekstra]{Fedeli:2014gja}
C.~Fedeli, E.~Semboloni, M.~Velliscig, M.~Van~Daalen, J.~Schaye, and
  H.~Hoekstra.
\newblock {The clustering of baryonic matter. II: halo model and hydrodynamic
  simulations}.
\newblock \emph{JCAP}, 1408:\penalty0 028, 2014.
\newblock \doi{10.1088/1475-7516/2014/08/028}.

\bibitem[Mohammed et~al.(2014)Mohammed, Martizzi, Teyssier, and
  Amara]{Mohammed:2014mba}
Irshad Mohammed, Davide Martizzi, Romain Teyssier, and Adam Amara.
\newblock {Baryonic effects on weak-lensing two-point statistics and its
  cosmological implications}.
\newblock 2014.

\bibitem[Mead et~al.(2015)Mead, Peacock, Heymans, Joudaki, and
  Heavens]{Mead:2015yca}
Alexander Mead, John Peacock, Catherine Heymans, Shahab Joudaki, and Alan
  Heavens.
\newblock {An accurate halo model for fitting non-linear cosmological power
  spectra and baryonic feedback models}.
\newblock \emph{Mon. Not. Roy. Astron. Soc.}, 454\penalty0 (2):\penalty0
  1958--1975, 2015.
\newblock \doi{10.1093/mnras/stv2036}.

\bibitem[{Schaye} et~al.(2010){Schaye}, {Dalla Vecchia}, {Booth}, {Wiersma},
  {Theuns}, {Haas}, {Bertone}, {Duffy}, {McCarthy}, and {van de
  Voort}]{Schaye:2010aaa}
J.~{Schaye}, C.~{Dalla Vecchia}, C.~M. {Booth}, R.~P.~C. {Wiersma},
  T.~{Theuns}, M.~R. {Haas}, S.~{Bertone}, A.~R. {Duffy}, I.~G. {McCarthy}, and
  F.~{van de Voort}.
\newblock {The physics driving the cosmic star formation history}.
\newblock \emph{Mon. Not. Roy. Astron. Soc.}, 402:\penalty0 1536--1560, March
  2010.
\newblock \doi{10.1111/j.1365-2966.2009.16029.x}.

\bibitem[Navarro et~al.(1996)Navarro, Frenk, and White]{Navarro:1995iw}
Julio~F. Navarro, Carlos~S. Frenk, and Simon D.~M. White.
\newblock {The Structure of cold dark matter halos}.
\newblock \emph{Astrophys. J.}, 462:\penalty0 563--575, 1996.
\newblock \doi{10.1086/177173}.

\bibitem[Baltz et~al.(2009)Baltz, Marshall, and Oguri]{Baltz:2007vq}
Edward~A. Baltz, Phil Marshall, and Masamune Oguri.
\newblock {Analytic models of plausible gravitational lens potentials}.
\newblock \emph{JCAP}, 0901:\penalty0 015, 2009.
\newblock \doi{10.1088/1475-7516/2009/01/015}.

\bibitem[Diemer and Kravtsov(2015)]{Diemer:2014gba}
Benedikt Diemer and Andrey~V. Kravtsov.
\newblock {A universal model for halo concentrations}.
\newblock \emph{Astrophys. J.}, 799\penalty0 (1):\penalty0 108, 2015.
\newblock \doi{10.1088/0004-637X/799/1/108}.

\bibitem[{Oguri} and {Hamana}(2011)]{Oguri:2011aaa}
Masamune {Oguri} and Takashi {Hamana}.
\newblock {Detailed cluster lensing profiles at large radii and the impact on
  cluster weak lensing studies}.
\newblock \emph{Mon. Not. Roy. Astron. Soc.}, 414:\penalty0 1851--1861, July
  2011.
\newblock \doi{10.1111/j.1365-2966.2011.18481.x}.

\bibitem[Sheth and Tormen(1999)]{Sheth:1999mn}
Ravi~K. Sheth and Giuseppe Tormen.
\newblock {Large scale bias and the peak background split}.
\newblock \emph{Mon. Not. Roy. Astron. Soc.}, 308:\penalty0 119, 1999.
\newblock \doi{10.1046/j.1365-8711.1999.02692.x}.

\bibitem[Hayashi and White(2008)]{Hayashi:2007uk}
E.~Hayashi and S.~D.~M. White.
\newblock {Understanding the shape of the halo-mass and galaxy-mass
  cross-correlation functions}.
\newblock \emph{Mon. Not. Roy. Astron. Soc.}, 388:\penalty0 2, 2008.
\newblock \doi{10.1111/j.1365-2966.2008.13371.x}.

\bibitem[Kravtsov et~al.(2018)Kravtsov, Vikhlinin, and
  Meshscheryakov]{Kravtsov:2014sra}
Andrey Kravtsov, Alexey Vikhlinin, and Alexander Meshscheryakov.
\newblock {Stellar mass -- halo mass relation and star formation efficiency in
  high-mass halos}.
\newblock \emph{Astron. Lett.}, 44\penalty0 (1):\penalty0 8--34, 2018.
\newblock \doi{10.1134/S1063773717120015}.

\bibitem[Moster et~al.(2013)Moster, Naab, and White]{Moster:2012fv}
Benjamin~P. Moster, Thorsten Naab, and Simon D.~M. White.
\newblock {Galactic star formation and accretion histories from matching
  galaxies to dark matter haloes}.
\newblock \emph{Mon. Not. Roy. Astron. Soc.}, 428:\penalty0 3121, 2013.
\newblock \doi{10.1093/mnras/sts261}.

\bibitem[Croston et~al.(2008)Croston, Pratt, Boehringer, Arnaud, Pointecouteau,
  Ponman, Sanderson, Temple, Bower, and Donahue]{Croston:2008yr}
J.~H. Croston, G.~W. Pratt, H.~Boehringer, M.~Arnaud, E.~Pointecouteau, T.~J.
  Ponman, A.~J.~R. Sanderson, R.~F. Temple, R.~G. Bower, and M.~Donahue.
\newblock {Galaxy-cluster gas-density distributions of the Representative
  XMM-Newton Cluster Structure Survey (REXCESS)}.
\newblock \emph{Astron. Astrophys.}, 487:\penalty0 431, 2008.
\newblock \doi{10.1051/0004-6361:20079154}.

\bibitem[Sanders et~al.(2018)Sanders, Fabian, Russell, and
  Walker]{Sanders:2017lce}
J.~S. Sanders, A.~C. Fabian, H.~R. Russell, and S.~A. Walker.
\newblock {Hydrostatic Chandra X-ray analysis of SPT-selected galaxy clusters ?
  I. Evolution of profiles and core properties}.
\newblock \emph{Mon. Not. Roy. Astron. Soc.}, 474\penalty0 (1):\penalty0
  1065--1098, 2018.
\newblock \doi{10.1093/mnras/stx2796}.

\bibitem[Eckert et~al.(2016)]{Eckert:2015rlr}
D.~Eckert et~al.
\newblock {The XXL Survey. XIII. Baryon content of the bright cluster sample}.
\newblock \emph{Astron. Astrophys.}, 592:\penalty0 A12, 2016.
\newblock \doi{10.1051/0004-6361/201527293}.

\bibitem[{Barnes} and {White}(1984)]{Barnes:1984aaa}
J.~{Barnes} and S.~D.~M. {White}.
\newblock {The response of a spheroid to a disc field or were bulges ever
  ellipticals?}
\newblock \emph{Mon. Not. Roy. Astron. Soc.}, 211:\penalty0 753--765, December
  1984.
\newblock \doi{10.1093/mnras/211.4.753}.

\bibitem[Blumenthal et~al.(1986)Blumenthal, Faber, Flores, and
  Primack]{Blumenthal:1985qy}
George~R. Blumenthal, S.~M. Faber, Ricardo Flores, and Joel~R. Primack.
\newblock {Contraction of Dark Matter Galactic Halos Due to Baryonic Infall}.
\newblock \emph{Astrophys. J.}, 301:\penalty0 27, 1986.
\newblock \doi{10.1086/163867}.

\bibitem[Abadi et~al.(2010)Abadi, Navarro, Fardal, Babul, and
  Steinmetz]{Abadi:2009ve}
Mario~G. Abadi, Julio~F. Navarro, Mark Fardal, Arif Babul, and Matthias
  Steinmetz.
\newblock {Galaxy-Induced Transformation of Dark Matter Halos}.
\newblock \emph{Mon. Not. Roy. Astron. Soc.}, 407:\penalty0 435--446, 2010.
\newblock \doi{10.1111/j.1365-2966.2010.16912.x}.

\bibitem[{Teyssier} et~al.(2011){Teyssier}, {Moore}, {Martizzi}, {Dubois}, and
  {Mayer}]{Teyssier:2011aaa}
R.~{Teyssier}, B.~{Moore}, D.~{Martizzi}, Y.~{Dubois}, and L.~{Mayer}.
\newblock {Mass distribution in galaxy clusters: the role of Active Galactic
  Nuclei feedback}.
\newblock \emph{Mon. Not. Roy. Astron. Soc.}, 414:\penalty0 195--208, June
  2011.
\newblock \doi{10.1111/j.1365-2966.2011.18399.x}.

\bibitem[{Stadel}(2001)]{Stadel:2001aaa}
J.~G. {Stadel}.
\newblock \emph{{Cosmological N-body simulations and their analysis}}.
\newblock PhD thesis, UNIVERSITY OF WASHINGTON, 2001.

\bibitem[Potter et~al.(2016)Potter, Stadel, and Teyssier]{Potter:2016ttn}
Douglas Potter, Joachim Stadel, and Romain Teyssier.
\newblock {PKDGRAV3: Beyond Trillion Particle Cosmological Simulations for the
  Next Era of Galaxy Surveys}.
\newblock 2016.

\bibitem[Ade et~al.(2015)]{Planck:2015xua}
P.~A.~R. Ade et~al.
\newblock {Planck 2015 results. XIII. Cosmological parameters}.
\newblock 2015.

\bibitem[{Knollmann} and {Knebe}(2009)]{Knollmann:2009aaa}
S.~R. {Knollmann} and A.~{Knebe}.
\newblock {AHF: Amiga's Halo Finder}.
\newblock \emph{Astrophys. J. Supp.}, 182:\penalty0 608--624, June 2009.
\newblock \doi{10.1088/0067-0049/182/2/608}.

\bibitem[LaRoque et~al.(2006)LaRoque, Bonamente, Carlstrom, Joy, Nagai, Reese,
  and Dawson]{LaRoque:2006te}
Samuel LaRoque, M.~Bonamente, J.~Carlstrom, M.~Joy, D.~Nagai, E.~Reese, and
  K.~Dawson.
\newblock {X-ray and Sunyaev-Zel'dovich Effect Measurements of the Gas Mass
  Fraction in Galaxy Clusters}.
\newblock \emph{Astrophys. J.}, 652:\penalty0 917--936, 2006.
\newblock \doi{10.1086/508139}.

\bibitem[Morandi et~al.(2015)Morandi, Sun, Forman, and Jones]{Morandi:2015pra}
Andrea Morandi, Ming Sun, William Forman, and Christine Jones.
\newblock {The galaxy cluster outskirts probed by Chandra}.
\newblock \emph{Mon. Not. Roy. Astron. Soc.}, 450\penalty0 (3):\penalty0
  2261--2278, 2015.
\newblock \doi{10.1093/mnras/stv660}.

\bibitem[Chiu et~al.(2018)]{Chiu:2017nwm}
I.~Chiu et~al.
\newblock {Baryon Content in a Sample of 91 Galaxy Clusters Selected by the
  South Pole Telescope at 0.2 < z < 1.25}.
\newblock \emph{Mon. Not. Roy. Astron. Soc.}, 478\penalty0 (3):\penalty0
  3072--3099, 2018.
\newblock \doi{10.1093/mnras/sty1284}.

\bibitem[{Leauthaud} et~al.(2012){Leauthaud}, {George}, {Behroozi}, {Bundy},
  {Tinker}, {Wechsler}, {Conroy}, {Finoguenov}, and
  {Tanaka}]{Leauthaud:2012aaa}
A.~{Leauthaud}, M.~R. {George}, P.~S. {Behroozi}, K.~{Bundy}, J.~{Tinker},
  R.~H. {Wechsler}, C.~{Conroy}, A.~{Finoguenov}, and M.~{Tanaka}.
\newblock {The Integrated Stellar Content of Dark Matter Halos}.
\newblock \emph{Ap.J.}, 746:\penalty0 95, February 2012.
\newblock \doi{10.1088/0004-637X/746/1/95}.

\bibitem[Behroozi et~al.(2013)Behroozi, Wechsler, and Conroy]{Behroozi:2012iw}
Peter~S. Behroozi, Risa~H. Wechsler, and Charlie Conroy.
\newblock {The Average Star Formation Histories of Galaxies in Dark Matter
  Halos from $z=$0-8}.
\newblock \emph{Astrophys. J.}, 770:\penalty0 57, 2013.
\newblock \doi{10.1088/0004-637X/770/1/57}.

\bibitem[Giodini et~al.(2009)]{Giodini:2009qf}
S.~Giodini et~al.
\newblock {Stellar and total baryon mass fractions in groups and clusters since
  redshift 1}.
\newblock \emph{Astrophys. J.}, 703:\penalty0 982--993, 2009.
\newblock \doi{10.1088/0004-637X/703/1/982}.

\bibitem[{McCarthy} et~al.(2010){McCarthy}, {Schaye}, {Ponman}, {Bower},
  {Booth}, {Dalla Vecchia}, {Crain}, {Springel}, {Theuns}, and
  {Wiersma}]{McCarthy:2010aaa}
I.~G. {McCarthy}, J.~{Schaye}, T.~J. {Ponman}, R.~G. {Bower}, C.~M. {Booth},
  C.~{Dalla Vecchia}, R.~A. {Crain}, V.~{Springel}, T.~{Theuns}, and R.~P.~C.
  {Wiersma}.
\newblock {The case for AGN feedback in galaxy groups}.
\newblock \emph{Mon. Not. Roy. Astron. Soc.}, 406:\penalty0 822--839, August
  2010.
\newblock \doi{10.1111/j.1365-2966.2010.16750.x}.

\bibitem[Springel(2005)]{Springel:2005mi}
Volker Springel.
\newblock {The Cosmological simulation code GADGET-2}.
\newblock \emph{Mon. Not. Roy. Astron. Soc.}, 364:\penalty0 1105--1134, 2005.
\newblock \doi{10.1111/j.1365-2966.2005.09655.x}.

\bibitem[Spergel et~al.(2007)]{Spergel:2006hy}
D.~N. Spergel et~al.
\newblock {Wilkinson Microwave Anisotropy Probe (WMAP) three year results:
  implications for cosmology}.
\newblock \emph{Astrophys. J. Suppl.}, 170:\penalty0 377, 2007.
\newblock \doi{10.1086/513700}.

\bibitem[{Eckert} et~al.(2012){Eckert}, {Vazza}, {Ettori}, {Molendi}, {Nagai},
  {Lau}, {Roncarelli}, {Rossetti}, {Snowden}, and
  {Gastaldello}]{Eckert:2012aaa}
D.~{Eckert}, F.~{Vazza}, S.~{Ettori}, S.~{Molendi}, D.~{Nagai}, E.~T. {Lau},
  M.~{Roncarelli}, M.~{Rossetti}, S.~L. {Snowden}, and F.~{Gastaldello}.
\newblock {The gas distribution in the outer regions of galaxy clusters}.
\newblock \emph{A\&A}, 541:\penalty0 A57, May 2012.
\newblock \doi{10.1051/0004-6361/201118281}.

\bibitem[Pierre et~al.(2016)]{Pierre:2015cqe}
M.~Pierre et~al.
\newblock {The XXL Survey - I. Scientific motivations ? XMM-Newton observing
  plan ? Follow-up observations and simulation programme}.
\newblock \emph{Astron. Astrophys.}, 592:\penalty0 A1, 2016.
\newblock \doi{10.1051/0004-6361/201526766}.

\bibitem[Arnaud et~al.(2005)Arnaud, Pointecouteau, and Pratt]{Arnaud:2005ur}
Monique Arnaud, E.~Pointecouteau, and G.~W. Pratt.
\newblock {The Structural and scaling properties of nearby galaxy clusters. 2.
  The M-T relation}.
\newblock \emph{Astron. Astrophys.}, 441:\penalty0 893--903, 2005.
\newblock \doi{10.1051/0004-6361:20052856}.

\bibitem[Eckert et~al.(2018)]{Eckert:2018mlz}
D.~Eckert et~al.
\newblock {Non-thermal pressure support in X-COP galaxy clusters}.
\newblock 2018.

\bibitem[Ettori et~al.(2018)Ettori, Ghirardini, Eckert, Pointecouteau,
  Gastaldello, Sereno, Gaspari, Ghizzardi, Roncarelli, and
  Rossetti]{Ettori:2018tus}
S.~Ettori, V.~Ghirardini, D.~Eckert, E.~Pointecouteau, F.~Gastaldello,
  M.~Sereno, M.~Gaspari, S.~Ghizzardi, M.~Roncarelli, and M.~Rossetti.
\newblock {Hydrostatic mass profiles in X-COP galaxy clusters}.
\newblock 2018.

\bibitem[Nagai et~al.(2007)Nagai, Vikhlinin, and Kravtsov]{Nagai:2006sz}
Daisuke Nagai, Alexey Vikhlinin, and Andrey~V. Kravtsov.
\newblock {Testing X-ray Measurements of Galaxy Clusters with Cosmological
  Simulations}.
\newblock \emph{Astrophys. J.}, 655:\penalty0 98--108, 2007.
\newblock \doi{10.1086/509868}.

\bibitem[Brun et~al.(2014)Brun, McCarthy, Schaye, and Ponman]{Brun:2013yva}
Amandine M. C.~Le Brun, Ian~G. McCarthy, Joop Schaye, and Trevor~J. Ponman.
\newblock {Towards a realistic population of simulated galaxy groups and
  clusters}.
\newblock \emph{Mon. Not. Roy. Astron. Soc.}, 441\penalty0 (2):\penalty0
  1270--1290, 2014.
\newblock \doi{10.1093/mnras/stu608}.

\bibitem[Lieu et~al.(2016)]{Lieu:2015pit}
Maggie Lieu et~al.
\newblock {The XXL Survey IV. Mass-temperature relation of the bright cluster
  sample}.
\newblock \emph{Astron. Astrophys.}, 592:\penalty0 A4, 2016.
\newblock \doi{10.1051/0004-6361/201526883}.

\bibitem[Sereno and Ettori(2015)]{Sereno:2014pfa}
Mauro Sereno and Stefano Ettori.
\newblock {Comparing masses in literature (CoMaLit) ? I. Bias and scatter in
  weak lensing and X-ray mass estimates of clusters}.
\newblock \emph{Mon. Not. Roy. Astron. Soc.}, 450\penalty0 (4):\penalty0
  3633--3648, 2015.
\newblock \doi{10.1038/ncomms8173, 10.1093/mnras/stv810}.

\bibitem[Sun et~al.(2009)Sun, Voit, Donahue, Jones, and Forman]{Sun:2008eh}
M.~Sun, G.~M. Voit, M.~Donahue, C.~Jones, and W.~Forman.
\newblock {Chandra studies of the X-ray gas properties of galaxy groups}.
\newblock \emph{Astrophys. J.}, 693:\penalty0 1142--1172, 2009.
\newblock \doi{10.1088/0004-637X/693/2/1142}.

\bibitem[Vikhlinin et~al.(2009)]{Vikhlinin:2008cd}
A.~Vikhlinin et~al.
\newblock {Chandra Cluster Cosmology Project II: Samples and X-ray Data
  Reduction}.
\newblock \emph{Astrophys. J.}, 692:\penalty0 1033--1059, 2009.
\newblock \doi{10.1088/0004-637X/692/2/1033}.

\bibitem[Gonzalez et~al.(2013)Gonzalez, Sivanandam, Zabludoff, and
  Zaritsky]{Gonzalez:2013awy}
Anthony~H. Gonzalez, Suresh Sivanandam, Ann~I. Zabludoff, and Dennis Zaritsky.
\newblock {Galaxy Cluster Baryon Fractions Revisited}.
\newblock \emph{Astrophys. J.}, 778:\penalty0 14, 2013.
\newblock \doi{10.1088/0004-637X/778/1/14}.

\bibitem[Vogelsberger et~al.(2014)Vogelsberger, Genel, Springel, Torrey,
  Sijacki, Xu, Snyder, Nelson, and Hernquist]{Vogelsberger:2014dza}
Mark Vogelsberger, Shy Genel, Volker Springel, Paul Torrey, Debora Sijacki,
  Dandan Xu, Gregory~F. Snyder, Dylan Nelson, and Lars Hernquist.
\newblock {Introducing the Illustris Project: Simulating the coevolution of
  dark and visible matter in the Universe}.
\newblock \emph{Mon. Not. Roy. Astron. Soc.}, 444\penalty0 (2):\penalty0
  1518--1547, 2014.
\newblock \doi{10.1093/mnras/stu1536}.

\bibitem[van Daalen and Schaye(2015)]{vanDaalen:2015msa}
Marcel~P. van Daalen and Joop Schaye.
\newblock {The contributions of matter inside and outside of haloes to the
  matter power spectrum}.
\newblock \emph{Mon. Not. Roy. Astron. Soc.}, 452\penalty0 (3):\penalty0
  2247--2257, 2015.
\newblock \doi{10.1093/mnras/stv1456}.

\bibitem[Schneider et~al.(2012)Schneider, Smith, Maccio, and
  Moore]{Schneider:2011yu}
Aurel Schneider, Robert~E. Smith, Andrea~V. Maccio, and Ben Moore.
\newblock {Nonlinear Evolution of Cosmological Structures in Warm Dark Matter
  Models}.
\newblock \emph{Mon. Not. Roy. Astron. Soc.}, 424:\penalty0 684, 2012.
\newblock \doi{10.1111/j.1365-2966.2012.21252.x}.

\bibitem[Kilbinger et~al.(2013)]{Kilbinger:2012qz}
Martin Kilbinger et~al.
\newblock {CFHTLenS: Combined probe cosmological model comparison using 2D weak
  gravitational lensing}.
\newblock \emph{Mon. Not. Roy. Astron. Soc.}, 430:\penalty0 2200--2220, 2013.
\newblock \doi{10.1093/mnras/stt041}.

\bibitem[Heymans et~al.(2013)]{Heymans:2013fya}
Catherine Heymans et~al.
\newblock {CFHTLenS tomographic weak lensing cosmological parameter
  constraints: Mitigating the impact of intrinsic galaxy alignments}.
\newblock \emph{Mon. Not. Roy. Astron. Soc.}, 432:\penalty0 2433, 2013.
\newblock \doi{10.1093/mnras/stt601}.

\bibitem[{Kaiser}(1992)]{Kaiser:1992aaa}
N.~{Kaiser}.
\newblock {Weak gravitational lensing of distant galaxies}.
\newblock \emph{Astrophys. J.}, 388:\penalty0 272--286, April 1992.
\newblock \doi{10.1086/171151}.

\bibitem[Ade et~al.(2016)]{Ade:2015xua}
P.~A.~R. Ade et~al.
\newblock {Planck 2015 results. XIII. Cosmological parameters}.
\newblock \emph{Astron. Astrophys.}, 594:\penalty0 A13, 2016.
\newblock \doi{10.1051/0004-6361/201525830}.

\bibitem[Hildebrandt et~al.(2017)]{Hildebrandt:2016iqg}
H.~Hildebrandt et~al.
\newblock {KiDS-450: Cosmological parameter constraints from tomographic weak
  gravitational lensing}.
\newblock \emph{Mon. Not. Roy. Astron. Soc.}, 465:\penalty0 1454, 2017.
\newblock \doi{10.1093/mnras/stw2805}.

\bibitem[Mccarthy et~al.(2018)Mccarthy, Bird, Schaye, Harnois-Deraps, Font, and
  Van~Waerbeke]{McCarthy:2017csu}
Ian~G. Mccarthy, Simeon Bird, Joop Schaye, Joachim Harnois-Deraps, Andreea~S.
  Font, and Ludovic Van~Waerbeke.
\newblock {The BAHAMAS project: the CMB large-scale structure tension and the
  roles of massive neutrinos and galaxy formation}.
\newblock \emph{Mon. Not. Roy. Astron. Soc.}, 476\penalty0 (3):\penalty0
  2999--3030, 2018.
\newblock \doi{10.1093/mnras/sty377}.

\bibitem[Joudaki et~al.(2017{\natexlab{a}})]{Joudaki:2016mvz}
Shahab Joudaki et~al.
\newblock {CFHTLenS revisited: assessing concordance with Planck including
  astrophysical systematics}.
\newblock \emph{Mon. Not. Roy. Astron. Soc.}, 465\penalty0 (2):\penalty0
  2033--2052, 2017{\natexlab{a}}.
\newblock \doi{10.1093/mnras/stw2665}.

\bibitem[Joudaki et~al.(2017{\natexlab{b}})]{Joudaki:2016kym}
Shahab Joudaki et~al.
\newblock {KiDS-450: Testing extensions to the standard cosmological model}.
\newblock \emph{Mon. Not. Roy. Astron. Soc.}, 471\penalty0 (2):\penalty0
  1259--1279, 2017{\natexlab{b}}.
\newblock \doi{10.1093/mnras/stx998}.

\bibitem[Hikage et~al.(2018)]{Hikage:2018qbn}
Chiaki Hikage et~al.
\newblock {Cosmology from cosmic shear power spectra with Subaru Hyper
  Suprime-Cam first-year data}.
\newblock 2018.

\bibitem[Battaglia et~al.(2017)Battaglia, Ferraro, Schaan, and
  Spergel]{Battaglia:2017neq}
Nicholas Battaglia, Simone Ferraro, Emmanuel Schaan, and David Spergel.
\newblock {Future constraints on halo thermodynamics from combined
  Sunyaev-Zel'dovich measurements}.
\newblock \emph{JCAP}, 1711\penalty0 (11):\penalty0 040, 2017.
\newblock \doi{10.1088/1475-7516/2017/11/040}.

\bibitem[Aguirre et~al.(2018)]{Ade:2018sbj}
James Aguirre et~al.
\newblock {The Simons Observatory: Science goals and forecasts}.
\newblock 2018.

\bibitem[Komatsu et~al.(2011)]{Komatsu:2010fb}
E.~Komatsu et~al.
\newblock {Seven-Year Wilkinson Microwave Anisotropy Probe (WMAP) Observations:
  Cosmological Interpretation}.
\newblock \emph{Astrophys. J. Suppl.}, 192:\penalty0 18, 2011.
\newblock \doi{10.1088/0067-0049/192/2/18}.

\bibitem[Dubois et~al.(2014)]{Dubois:2014lxa}
Y.~Dubois et~al.
\newblock {Dancing in the dark: galactic properties trace spin swings along the
  cosmic web}.
\newblock \emph{Mon. Not. Roy. Astron. Soc.}, 444\penalty0 (2):\penalty0
  1453--1468, 2014.
\newblock \doi{10.1093/mnras/stu1227}.

\bibitem[{Dubois} et~al.(2016){Dubois}, {Peirani}, {Pichon}, {Devriendt},
  {Gavazzi}, {Welker}, and {Volonteri}]{Dubois:2016aaa}
Y.~{Dubois}, S.~{Peirani}, C.~{Pichon}, J.~{Devriendt}, R.~{Gavazzi},
  C.~{Welker}, and M.~{Volonteri}.
\newblock {The HORIZON-AGN simulation: morphological diversity of galaxies
  promoted by AGN feedback}, December 2016.

\bibitem[Teyssier(2002)]{Teyssier:2001cp}
Romain Teyssier.
\newblock {Cosmological hydrodynamics with adaptive mesh refinement: a new high
  resolution code called ramses}.
\newblock \emph{Astron. Astrophys.}, 385:\penalty0 337--364, 2002.
\newblock \doi{10.1051/0004-6361:20011817}.

\bibitem[Pillepich et~al.(2018{\natexlab{a}})]{Pillepich:2017fcc}
Annalisa Pillepich et~al.
\newblock {First results from the IllustrisTNG simulations: the stellar mass
  content of groups and clusters of galaxies}.
\newblock \emph{Mon. Not. Roy. Astron. Soc.}, 475:\penalty0 648,
  2018{\natexlab{a}}.
\newblock \doi{10.1093/mnras/stx3112}.

\bibitem[Pillepich et~al.(2018{\natexlab{b}})]{Pillepich:2017jle}
Annalisa Pillepich et~al.
\newblock {Simulating Galaxy Formation with the IllustrisTNG Model}.
\newblock \emph{Mon. Not. Roy. Astron. Soc.}, 473\penalty0 (3):\penalty0
  4077--4106, 2018{\natexlab{b}}.
\newblock \doi{10.1093/mnras/stx2656}.

\bibitem[Springel(2010)]{Springel:2010aaa}
Volker Springel.
\newblock {E pur si muove: Galilean-invariant cosmological hydrodynamical
  simulations on a moving mesh}.
\newblock \emph{Mon. Not. Roy. Astron. Soc.}, 401:\penalty0 791--851, January
  2010.
\newblock \doi{10.1111/j.1365-2966.2009.15715.x}.

\bibitem[{Budzynski} et~al.(2012){Budzynski}, {Koposov}, {McCarthy}, {McGee},
  and {Belokurov}]{Budzynski:2012aaa}
J.~M. {Budzynski}, S.~E. {Koposov}, I.~G. {McCarthy}, S.~L. {McGee}, and
  V.~{Belokurov}.
\newblock {The radial distribution of galaxies in groups and clusters}.
\newblock \emph{Mon. Not. Roy. Astron. Soc.}, 423:\penalty0 104--121, June
  2012.
\newblock \doi{10.1111/j.1365-2966.2012.20663.x}.

\end{thebibliography}


\begin{thebibliography}{68}
\providecommand{\natexlab}[1]{#1}
\providecommand{\url}[1]{\texttt{#1}}
\expandafter\ifx\csname urlstyle\endcsname\relax
  \providecommand{\doi}[1]{doi: #1}\else
  \providecommand{\doi}{doi: \begingroup \urlstyle{rm}\Url}\fi

\bibitem[Schneider et~al.(2019{\natexlab{a}})Schneider, Teyssier, Stadel,
  Chisari, Le~Brun, Amara, and Refregier]{Schneider:2018pfw}
Aurel Schneider, Romain Teyssier, Joachim Stadel, Nora~Elisa Chisari, Amandine
  M.~C. Le~Brun, Adam Amara, and Alexandre Refregier.
\newblock {Quantifying baryon effects on the matter power spectrum and the weak
  lensing shear correlation}.
\newblock \emph{JCAP}, 1903\penalty0 (03):\penalty0 020, 2019{\natexlab{a}}.
\newblock \doi{10.1088/1475-7516/2019/03/020}.

\bibitem[Heitmann et~al.(2008)]{Heitmann:2007hr}
Katrin Heitmann et~al.
\newblock {The Cosmic Code Comparison Project}.
\newblock \emph{Comput. Sci. Dis.}, 1:\penalty0 015003, 2008.
\newblock \doi{10.1088/1749-4699/1/1/015003}.

\bibitem[Schneider et~al.(2016)Schneider, Teyssier, Potter, Stadel, Onions,
  Reed, Smith, Springel, Pearce, and Scoccimarro]{Schneider:2015yka}
Aurel Schneider, Romain Teyssier, Doug Potter, Joachim Stadel, Julian Onions,
  Darren~S. Reed, Robert~E. Smith, Volker Springel, Frazer~R. Pearce, and Roman
  Scoccimarro.
\newblock {Matter power spectrum and the challenge of percent accuracy}.
\newblock \emph{JCAP}, 1604\penalty0 (04):\penalty0 047, 2016.
\newblock \doi{10.1088/1475-7516/2016/04/047}.

\bibitem[Garrison et~al.(2019)Garrison, Eisenstein, and
  Pinto]{Garrison:2018juw}
Lehman~H. Garrison, Daniel~J. Eisenstein, and Philip~A. Pinto.
\newblock {A High-Fidelity Realization of the Euclid Code Comparison $N$-body
  Simulation with Abacus}.
\newblock \emph{Mon. Not. Roy. Astron. Soc.}, 485\penalty0 (3):\penalty0
  3370--3377, 2019.
\newblock \doi{10.1093/mnras/stz634}.

\bibitem[Takahashi et~al.(2012)Takahashi, Sato, Nishimichi, Taruya, and
  Oguri]{Takahashi:2012em}
Ryuichi Takahashi, Masanori Sato, Takahiro Nishimichi, Atsushi Taruya, and
  Masamune Oguri.
\newblock {Revising the Halofit Model for the Nonlinear Matter Power Spectrum}.
\newblock \emph{Astrophys. J.}, 761:\penalty0 152, 2012.
\newblock \doi{10.1088/0004-637X/761/2/152}.

\bibitem[Mead et~al.(2015)Mead, Peacock, Heymans, Joudaki, and
  Heavens]{Mead:2015yca}
Alexander Mead, John Peacock, Catherine Heymans, Shahab Joudaki, and Alan
  Heavens.
\newblock {An accurate halo model for fitting non-linear cosmological power
  spectra and baryonic feedback models}.
\newblock \emph{Mon. Not. Roy. Astron. Soc.}, 454\penalty0 (2):\penalty0
  1958--1975, 2015.
\newblock \doi{10.1093/mnras/stv2036}.

\bibitem[Smith and Angulo(2018)]{Smith:2018zcj}
Robert~E. Smith and Raul~E. Angulo.
\newblock {Precision modelling of the matter power spectrum in a Planck-like
  Universe}.
\newblock 2018.

\bibitem[Cataneo et~al.(2019)Cataneo, Lombriser, Heymans, Mead, Barreira, Bose,
  and Li]{Cataneo:2018cic}
Matteo Cataneo, Lucas Lombriser, Catherine Heymans, Alexander Mead, Alexandre
  Barreira, Sownak Bose, and Baojiu Li.
\newblock {On the road to percent accuracy: non-linear reaction of the matter
  power spectrum to dark energy and modified gravity}.
\newblock \emph{Mon. Not. Roy. Astron. Soc.}, 488\penalty0 (2):\penalty0
  2121--2142, 2019.
\newblock \doi{10.1093/mnras/stz1836}.

\bibitem[Heitmann et~al.(2014)Heitmann, Lawrence, Kwan, Habib, and
  Higdon]{Heitmann:2013bra}
Katrin Heitmann, Earl Lawrence, Juliana Kwan, Salman Habib, and David Higdon.
\newblock {The Coyote Universe Extended: Precision Emulation of the Matter
  Power Spectrum}.
\newblock \emph{Astrophys. J.}, 780:\penalty0 111, 2014.
\newblock \doi{10.1088/0004-637X/780/1/111}.

\bibitem[Knabenhans et~al.(2019)]{Knabenhans:2018cng}
Mischa Knabenhans et~al.
\newblock {Euclid preparation: II. The EuclidEmulator -- A tool to compute the
  cosmology dependence of the nonlinear matter power spectrum}.
\newblock \emph{Mon. Not. Roy. Astron. Soc.}, 484:\penalty0 5509--5529, 2019.
\newblock \doi{10.1093/mnras/stz197}.

\bibitem[DeRose et~al.(2019)DeRose, Wechsler, Tinker, Becker, Mao, McClintock,
  McLaughlin, Rozo, and Zhai]{DeRose:2018xdj}
Joseph DeRose, Risa~H. Wechsler, Jeremy~L. Tinker, Matthew~R. Becker, Yao-Yuan
  Mao, Thomas McClintock, Sean McLaughlin, Eduardo Rozo, and Zhongxu Zhai.
\newblock {The Aemulus Project I: Numerical Simulations for Precision
  Cosmology}.
\newblock \emph{Astrophys. J.}, 875\penalty0 (1):\penalty0 69, 2019.
\newblock \doi{10.3847/1538-4357/ab1085}.

\bibitem[Ravanbakhsh et~al.(2017)Ravanbakhsh, Oliva, Fromenteau, Price, Ho,
  Schneider, and Poczos]{Ravanbakhsh:2017bbi}
Siamak Ravanbakhsh, Junier Oliva, Sebastien Fromenteau, Layne~C. Price, Shirley
  Ho, Jeff Schneider, and Barnabas Poczos.
\newblock {Estimating Cosmological Parameters from the Dark Matter
  Distribution}.
\newblock 2017.

\bibitem[Fluri et~al.(2018)Fluri, Kacprzak, Refregier, Amara, Lucchi, and
  Hofmann]{Fluri:2018hoy}
Janis Fluri, Tomasz Kacprzak, Alexandre Refregier, Adam Amara, Aurelien Lucchi,
  and Thomas Hofmann.
\newblock {Cosmological constraints from noisy convergence maps through deep
  learning}.
\newblock \emph{Phys. Rev.}, D98\penalty0 (12):\penalty0 123518, 2018.
\newblock \doi{10.1103/PhysRevD.98.123518}.

\bibitem[Alsing et~al.(2019)Alsing, Charnock, Feeney, and
  Wandelt]{Alsing:2019xrx}
Justin Alsing, Tom Charnock, Stephen Feeney, and Benjamin Wandelt.
\newblock {Fast likelihood-free cosmology with neural density estimators and
  active learning}.
\newblock \emph{Mon. Not. Roy. Astron. Soc.}, 488\penalty0 (3):\penalty0
  4440--4458, 2019.
\newblock \doi{10.1093/mnras/stz1960}.

\bibitem[Manrique-Yus and Sellentin(2019)]{Manrique-Yus:2019hqc}
Andrea Manrique-Yus and Elena Sellentin.
\newblock {Euclid-era cosmology for everyone: Neural net assisted MCMC sampling
  for the joint 3x2 likelihood}.
\newblock 2019.

\bibitem[van Daalen et~al.(2011)van Daalen, Schaye, Booth, and
  Vecchia]{vanDaalen:2011xb}
Marcel~P. van Daalen, Joop Schaye, C.~M. Booth, and Claudio~Dalla Vecchia.
\newblock {The effects of galaxy formation on the matter power spectrum: A
  challenge for precision cosmology}.
\newblock \emph{Mon. Not. Roy. Astron. Soc.}, 415:\penalty0 3649--3665, 2011.
\newblock \doi{10.1111/j.1365-2966.2011.18981.x}.

\bibitem[Hellwing et~al.(2016)Hellwing, Schaller, Frenk, Theuns, Schaye, Bower,
  and Crain]{Hellwing:2016ucy}
Wojciech~A. Hellwing, Matthieu Schaller, Carlos~S. Frenk, Tom Theuns, Joop
  Schaye, Richard~G. Bower, and Robert~A. Crain.
\newblock {The effect of baryons on redshift space distortions and cosmic
  density and velocity fields in the EAGLE simulation}.
\newblock \emph{Mon. Not. Roy. Astron. Soc.}, 461\penalty0 (1):\penalty0
  L11--L15, 2016.
\newblock \doi{10.1093/mnrasl/slw081}.

\bibitem[Mummery et~al.(2017)Mummery, McCarthy, Bird, and
  Schaye]{Mummery:2017lcn}
Benjamin~O. Mummery, Ian~G. McCarthy, Simeon Bird, and Joop Schaye.
\newblock {The separate and combined effects of baryon physics and neutrino
  free-streaming on large-scale structure}.
\newblock \emph{Mon. Not. Roy. Astron. Soc.}, 471\penalty0 (1):\penalty0
  227--242, 2017.
\newblock \doi{10.1093/mnras/stx1469}.

\bibitem[Springel et~al.(2017)]{Springel:2017tpz}
Volker Springel et~al.
\newblock {First results from the IllustrisTNG simulations: matter and galaxy
  clustering}.
\newblock 2017.

\bibitem[Chisari et~al.(2018)Chisari, Richardson, Devriendt, Dubois, Schneider,
  Brun, Beckmann, Peirani, Slyz, and Pichon]{Chisari:2018prw}
Nora~Elisa Chisari, Mark L.~A. Richardson, Julien Devriendt, Yohan Dubois,
  Aurel Schneider, M.~C. Brun, Amandine~Le, Ricarda~S. Beckmann, Sebastien
  Peirani, Adrianne Slyz, and Christophe Pichon.
\newblock {The impact of baryons on the matter power spectrum from the
  Horizon-AGN cosmological hydrodynamical simulation}.
\newblock 2018.

\bibitem[van Daalen et~al.(2019)van Daalen, McCarthy, and
  Schaye]{vanDaalen:2019pst}
Marcel~P. van Daalen, Ian~G. McCarthy, and Joop Schaye.
\newblock {Exploring the effects of galaxy formation on matter clustering
  through a library of simulation power spectra}.
\newblock 2019.

\bibitem[Foreman et~al.(2019)Foreman, Coulton, Villaescusa-Navarro, and
  Barreira]{Foreman:2019ahr}
Simon Foreman, William Coulton, Francisco Villaescusa-Navarro, and Alexandre
  Barreira.
\newblock {Baryonic effects on the matter bispectrum}.
\newblock 2019.

\bibitem[Barreira et~al.(2019)Barreira, Nelson, Pillepich, Springel, Schmidt,
  Pakmor, Hernquist, and Vogelsberger]{Barreira:2019ckp}
Alexandre Barreira, Dylan Nelson, Annalisa Pillepich, Volker Springel, Fabian
  Schmidt, Ruediger Pakmor, Lars Hernquist, and Mark Vogelsberger.
\newblock {Separate Universe Simulations with IllustrisTNG: baryonic effects on
  power spectrum responses and higher-order statistics}.
\newblock \emph{Mon. Not. Roy. Astron. Soc.}, 488\penalty0 (2):\penalty0
  2079--2092, 2019.
\newblock \doi{10.1093/mnras/stz1807}.

\bibitem[{Semboloni} et~al.(2011){Semboloni}, {Hoekstra}, {Schaye}, {van
  Daalen}, and {McCarthy}]{Semboloni:2011aaa}
E.~{Semboloni}, H.~{Hoekstra}, J.~{Schaye}, M.~P. {van Daalen}, and I.~G.
  {McCarthy}.
\newblock {Quantifying the effect of baryon physics on weak lensing
  tomography}.
\newblock \emph{Mon. Not. Roy. Astron. Soc.}, 417:\penalty0 2020--2035,
  November 2011.
\newblock \doi{10.1111/j.1365-2966.2011.19385.x}.

\bibitem[Mohammed et~al.(2014)Mohammed, Martizzi, Teyssier, and
  Amara]{Mohammed:2014mba}
Irshad Mohammed, Davide Martizzi, Romain Teyssier, and Adam Amara.
\newblock {Baryonic effects on weak-lensing two-point statistics and its
  cosmological implications}.
\newblock 2014.

\bibitem[Fedeli et~al.(2014)Fedeli, Semboloni, Velliscig, Van~Daalen, Schaye,
  and Hoekstra]{Fedeli:2014gja}
C.~Fedeli, E.~Semboloni, M.~Velliscig, M.~Van~Daalen, J.~Schaye, and
  H.~Hoekstra.
\newblock {The clustering of baryonic matter. II: halo model and hydrodynamic
  simulations}.
\newblock \emph{JCAP}, 1408:\penalty0 028, 2014.
\newblock \doi{10.1088/1475-7516/2014/08/028}.

\bibitem[Debackere et~al.(2019)Debackere, Schaye, and
  Hoekstra]{Debackere:2019cec}
Stijn N.~B. Debackere, Joop Schaye, and Henk Hoekstra.
\newblock {The impact of the observed baryon distribution in haloes on the
  total matter power spectrum}.
\newblock 2019.

\bibitem[Chisari et~al.(2019)]{Chisari:2019tus}
Nora~Elisa Chisari et~al.
\newblock {Modelling baryonic feedback for survey cosmology}.
\newblock 2019.
\newblock \doi{10.21105/astro.1905.06082}.

\bibitem[Schneider and Teyssier(2015)]{Schneider:2015wta}
Aurel Schneider and Romain Teyssier.
\newblock {A new method to quantify the effects of baryons on the matter power
  spectrum}.
\newblock \emph{JCAP}, 1512\penalty0 (12):\penalty0 049, 2015.
\newblock \doi{10.1088/1475-7516/2015/12/049}.

\bibitem[Weiss et~al.(2019)Weiss, Schneider, Sgier, Kacprzak, Amara, and
  Refregier]{Weiss:2019jfx}
Andreas~J. Weiss, Aurel Schneider, Raphael Sgier, Tomasz Kacprzak, Adam Amara,
  and Alexandre Refregier.
\newblock {Effects of baryons on weak lensing peak statistics}.
\newblock 2019.

\bibitem[Fluri et~al.(2019)Fluri, Kacprzak, Lucchi, Refregier, Amara, Hofmann,
  and Schneider]{Fluri:2019qtp}
Janis Fluri, Tomasz Kacprzak, Aurelien Lucchi, Alexandre Refregier, Adam Amara,
  Thomas Hofmann, and Aurel Schneider.
\newblock {Cosmological constraints with deep learning from KiDS-450 weak
  lensing maps}.
\newblock \emph{Phys. Rev.}, D100\penalty0 (6):\penalty0 063514, 2019.
\newblock \doi{10.1103/PhysRevD.100.063514}.

\bibitem[Schneider et~al.(2019{\natexlab{b}})Schneider, Refregier, Stoira,
  Knabenhans, Grandis, Eckert, Stadel, and Teyssier]{Schneider:2019bbb}
Aurel Schneider, Alexandre Refregier, Nicola Stoira, Mischa Knabenhans,
  Sebastian Grandis, Dominique Eckert, Joachim Stadel, and Romain Teyssier.
\newblock {Baryonic effects for weak lensing. Part II. Combination with X-ray
  data and extended cosmologies}.
\newblock 2019{\natexlab{b}}.

\bibitem[Moster et~al.(2013)Moster, Naab, and White]{Moster:2012fv}
Benjamin~P. Moster, Thorsten Naab, and Simon D.~M. White.
\newblock {Galactic star formation and accretion histories from matching
  galaxies to dark matter haloes}.
\newblock \emph{Mon. Not. Roy. Astron. Soc.}, 428:\penalty0 3121, 2013.
\newblock \doi{10.1093/mnras/sts261}.

\bibitem[Abadi et~al.(2010)Abadi, Navarro, Fardal, Babul, and
  Steinmetz]{Abadi:2009ve}
Mario~G. Abadi, Julio~F. Navarro, Mark Fardal, Arif Babul, and Matthias
  Steinmetz.
\newblock {Galaxy-Induced Transformation of Dark Matter Halos}.
\newblock \emph{Mon. Not. Roy. Astron. Soc.}, 407:\penalty0 435--446, 2010.
\newblock \doi{10.1111/j.1365-2966.2010.16912.x}.

\bibitem[Joudaki et~al.(2017{\natexlab{a}})]{Joudaki:2016mvz}
Shahab Joudaki et~al.
\newblock {CFHTLenS revisited: assessing concordance with Planck including
  astrophysical systematics}.
\newblock \emph{Mon. Not. Roy. Astron. Soc.}, 465\penalty0 (2):\penalty0
  2033--2052, 2017{\natexlab{a}}.
\newblock \doi{10.1093/mnras/stw2665}.

\bibitem[Joudaki et~al.(2017{\natexlab{b}})]{Joudaki:2016kym}
Shahab Joudaki et~al.
\newblock {KiDS-450: Testing extensions to the standard cosmological model}.
\newblock \emph{Mon. Not. Roy. Astron. Soc.}, 471\penalty0 (2):\penalty0
  1259--1279, 2017{\natexlab{b}}.
\newblock \doi{10.1093/mnras/stx998}.

\bibitem[Hildebrandt et~al.(2017)]{Hildebrandt:2016iqg}
H.~Hildebrandt et~al.
\newblock {KiDS-450: Cosmological parameter constraints from tomographic weak
  gravitational lensing}.
\newblock \emph{Mon. Not. Roy. Astron. Soc.}, 465:\penalty0 1454, 2017.
\newblock \doi{10.1093/mnras/stw2805}.

\bibitem[Abbott et~al.(2017)]{Abbott:2017wau}
T.~M.~C. Abbott et~al.
\newblock {Dark Energy Survey Year 1 Results: Cosmological Constraints from
  Galaxy Clustering and Weak Lensing}.
\newblock 2017.

\bibitem[Hikage et~al.(2018)]{Hikage:2018qbn}
Chiaki Hikage et~al.
\newblock {Cosmology from cosmic shear power spectra with Subaru Hyper
  Suprime-Cam first-year data}.
\newblock 2018.

\bibitem[Mccarthy et~al.(2018)Mccarthy, Bird, Schaye, Harnois-Deraps, Font, and
  Van~Waerbeke]{McCarthy:2017csu}
Ian~G. Mccarthy, Simeon Bird, Joop Schaye, Joachim Harnois-Deraps, Andreea~S.
  Font, and Ludovic Van~Waerbeke.
\newblock {The BAHAMAS project: the CMB large-scale structure tension and the
  roles of massive neutrinos and galaxy formation}.
\newblock \emph{Mon. Not. Roy. Astron. Soc.}, 476\penalty0 (3):\penalty0
  2999--3030, 2018.
\newblock \doi{10.1093/mnras/sty377}.

\bibitem[Hinshaw et~al.(2013)]{Hinshaw:2012aka}
G.~Hinshaw et~al.
\newblock {Nine-Year Wilkinson Microwave Anisotropy Probe (WMAP) Observations:
  Cosmological Parameter Results}.
\newblock \emph{Astrophys. J. Suppl.}, 208:\penalty0 19, 2013.
\newblock \doi{10.1088/0067-0049/208/2/19}.

\bibitem[Ade et~al.(2014)]{Ade:2013zuv}
P.~A.~R. Ade et~al.
\newblock {Planck 2013 results. XVI. Cosmological parameters}.
\newblock \emph{Astron. Astrophys.}, 571:\penalty0 A16, 2014.
\newblock \doi{10.1051/0004-6361/201321591}.

\bibitem[Spergel et~al.(2007)]{Spergel:2006hy}
D.~N. Spergel et~al.
\newblock {Wilkinson Microwave Anisotropy Probe (WMAP) three year results:
  implications for cosmology}.
\newblock \emph{Astrophys. J. Suppl.}, 170:\penalty0 377, 2007.
\newblock \doi{10.1086/513700}.

\bibitem[Komatsu et~al.(2011)]{Komatsu:2010fb}
E.~Komatsu et~al.
\newblock {Seven-Year Wilkinson Microwave Anisotropy Probe (WMAP) Observations:
  Cosmological Interpretation}.
\newblock \emph{Astrophys. J. Suppl.}, 192:\penalty0 18, 2011.
\newblock \doi{10.1088/0067-0049/192/2/18}.

\bibitem[{Stadel}(2001)]{Stadel:2001aaa}
J.~G. {Stadel}.
\newblock \emph{{Cosmological N-body simulations and their analysis}}.
\newblock PhD thesis, UNIVERSITY OF WASHINGTON, 2001.

\bibitem[Potter et~al.(2016)Potter, Stadel, and Teyssier]{Potter:2016ttn}
Douglas Potter, Joachim Stadel, and Romain Teyssier.
\newblock {PKDGRAV3: Beyond Trillion Particle Cosmological Simulations for the
  Next Era of Galaxy Surveys}.
\newblock 2016.

\bibitem[{Knollmann} and {Knebe}(2009)]{Knollmann:2009aaa}
S.~R. {Knollmann} and A.~{Knebe}.
\newblock {AHF: Amiga's Halo Finder}.
\newblock \emph{Astrophys. J. Supp.}, 182:\penalty0 608--624, June 2009.
\newblock \doi{10.1088/0067-0049/182/2/608}.

\bibitem[Marelli and Sudret()]{Marelli:2014aaa}
Stefano Marelli and Bruno Sudret.
\newblock \emph{UQLab: A Framework for Uncertainty Quantification in Matlab},
  pages 2554--2563.
\newblock \doi{10.1061/9780784413609.257}.
\newblock URL \url{https://ascelibrary.org/doi/abs/10.1061/9780784413609.257}.

\bibitem[{Limber}(1953)]{Limber:1953aaa}
D.~N. {Limber}.
\newblock {The Analysis of Counts of the Extragalactic Nebulae in Terms of a
  Fluctuating Density Field.}
\newblock \emph{Astrophys. J.}, 117:\penalty0 134, January 1953.
\newblock \doi{10.1086/145672}.

\bibitem[{Hirata} and {Seljak}(2004)]{Hirata:2004aaa}
Christopher~M. {Hirata} and Uro{\v{s}} {Seljak}.
\newblock {Intrinsic alignment-lensing interference as a contaminant of cosmic
  shear}.
\newblock \emph{Phys. Rev. D}, 70\penalty0 (6):\penalty0 063526, Sep 2004.
\newblock \doi{10.1103/PhysRevD.70.063526}.

\bibitem[Bridle and King(2007)]{Bridle:2007ft}
Sarah Bridle and Lindsay King.
\newblock {Dark energy constraints from cosmic shear power spectra: impact of
  intrinsic alignments on photometric redshift requirements}.
\newblock \emph{New J. Phys.}, 9:\penalty0 444, 2007.
\newblock \doi{10.1088/1367-2630/9/12/444}.

\bibitem[Kacprzak et~al.(2019)]{Kacprzak:2019tzh}
T.~Kacprzak et~al.
\newblock {Monte Carlo Control Loops for cosmic shear cosmology with DES Year
  1}.
\newblock 2019.

\bibitem[{Eisenstein} and {Hu}(1998)]{Eisenstein:1998aaa}
Daniel~J. {Eisenstein} and Wayne {Hu}.
\newblock {Baryonic Features in the Matter Transfer Function}.
\newblock \emph{Astrophysical Journal}, 496:\penalty0 605--614, March 1998.
\newblock \doi{10.1086/305424}.

\bibitem[Refregier et~al.(2018)Refregier, Gamper, Amara, and
  Heisenberg]{Refregier:2017seh}
Alexandre Refregier, Lukas Gamper, Adam Amara, and Lavinia Heisenberg.
\newblock {PyCosmo: An Integrated Cosmological Boltzmann Solver}.
\newblock \emph{Astron. Comput.}, 25:\penalty0 38--43, 2018.
\newblock \doi{10.1016/j.ascom.2018.08.001}.

\bibitem[{Hahn} and {Abel}(2011)]{Hahn:2011aaa}
Oliver {Hahn} and Tom {Abel}.
\newblock {Multi-scale initial conditions for cosmological simulations}.
\newblock \emph{Monthly Notices of the RAS}, 415:\penalty0 2101--2121, August
  2011.
\newblock \doi{10.1111/j.1365-2966.2011.18820.x}.

\bibitem[Ade et~al.(2016)]{Ade:2015fva}
P.~A.~R. Ade et~al.
\newblock {Planck 2015 results. XXIV. Cosmology from Sunyaev-Zeldovich cluster
  counts}.
\newblock \emph{Astron. Astrophys.}, 594:\penalty0 A24, 2016.
\newblock \doi{10.1051/0004-6361/201525833}.

\bibitem[Aghanim et~al.(2018)]{Aghanim:2018eyx}
N.~Aghanim et~al.
\newblock {Planck 2018 results. VI. Cosmological parameters}.
\newblock 2018.

\bibitem[Steigman(2007)]{Steigman:2007xt}
Gary Steigman.
\newblock {Primordial Nucleosynthesis in the Precision Cosmology Era}.
\newblock \emph{Ann. Rev. Nucl. Part. Sci.}, 57:\penalty0 463--491, 2007.
\newblock \doi{10.1146/annurev.nucl.56.080805.140437}.

\bibitem[{Akeret} et~al.(2012){Akeret}, {Seehars}, {Amara}, {Refregier}, and
  {Csillaghy}]{Akeret:2012aaa}
Jo{\"e}l {Akeret}, Sebastian {Seehars}, Adam {Amara}, Alexandre {Refregier},
  and Andr{\'e} {Csillaghy}.
\newblock {CosmoHammer: Cosmological parameter estimation with the MCMC
  Hammer}.
\newblock \emph{arXiv e-prints}, art. arXiv:1212.1721, Dec 2012.

\bibitem[{Foreman-Mackey} et~al.(2013){Foreman-Mackey}, {Hogg}, {Lang}, and
  {Goodman}]{Foreman-Mackey:2013aaa}
Daniel {Foreman-Mackey}, David~W. {Hogg}, Dustin {Lang}, and Jonathan
  {Goodman}.
\newblock {emcee: The MCMC Hammer}.
\newblock \emph{PASP}, 125\penalty0 (925):\penalty0 306, Mar 2013.
\newblock \doi{10.1086/670067}.

\bibitem[Jing et~al.(2006)Jing, Zhang, Lin, Gao, and Springel]{Jing:2005gm}
Y.~P. Jing, Pengjie Zhang, W.~P. Lin, L.~Gao, and V.~Springel.
\newblock {The influence of baryons on the clustering of matter and weak
  lensing surveys}.
\newblock \emph{Astrophys. J.}, 640:\penalty0 L119--L122, 2006.
\newblock \doi{10.1086/503547}.

\bibitem[Rudd et~al.(2008)Rudd, Zentner, and Kravtsov]{Rudd:2007zx}
Douglas~H. Rudd, Andrew~R. Zentner, and Andrey~V. Kravtsov.
\newblock {Effects of Baryons and Dissipation on the Matter Power Spectrum}.
\newblock \emph{Astrophys. J.}, 672:\penalty0 19--32, 2008.
\newblock \doi{10.1086/523836}.

\bibitem[Huang et~al.(2018)Huang, Eifler, Mandelbaum, and
  Dodelson]{Huang:2018wpy}
Hung-Jin Huang, Tim Eifler, Rachel Mandelbaum, and Scott Dodelson.
\newblock {Modeling baryonic physics in future weak lensing surveys}.
\newblock 2018.

\bibitem[Hartlap et~al.(2006)Hartlap, Simon, and Schneider]{Hartlap:2006kj}
J.~Hartlap, Patrick Simon, and P.~Schneider.
\newblock {Why your model parameter confidences might be too optimistic:
  Unbiased estimation of the inverse covariance matrix}.
\newblock \emph{Astron. Astrophys.}, 2006.
\newblock \doi{10.1051/0004-6361:20066170}.
\newblock [Astron. Astrophys.464,399(2007)].

\bibitem[Sellentin and Heavens(2016)]{Sellentin:2015waz}
Elena Sellentin and Alan~F. Heavens.
\newblock {Parameter inference with estimated covariance matrices}.
\newblock \emph{Mon. Not. Roy. Astron. Soc.}, 456\penalty0 (1):\penalty0
  L132--L136, 2016.
\newblock \doi{10.1093/mnrasl/slv190}.

\bibitem[Zonca et~al.(2019)Zonca, Singer, Lenz, Reinecke, Rosset, Hivon, and
  Gorski]{Zonca:2019aaa}
Andrea Zonca, Leo Singer, Daniel Lenz, Martin Reinecke, Cyrille Rosset, Eric
  Hivon, and Krzysztof Gorski.
\newblock healpy: equal area pixelization and spherical harmonics transforms
  for data on the sphere in python.
\newblock \emph{Journal of Open Source Software}, 4\penalty0 (35):\penalty0
  1298, March 2019.
\newblock \doi{10.21105/joss.01298}.
\newblock URL \url{https://doi.org/10.21105/joss.01298}.

\bibitem[{Foreman-Mackey} et~al.(2016){Foreman-Mackey}, {Vousden},
  {Price-Whelan}, {Pitkin}, {Zabalza}, {Ryan}, {Emily}, {Smith}, {Ashton},
  {Cruz}, {Kerzendorf}, {Caswell}, {Hoyer}, {Barbary}, {Czekala}, {Rein},
  {Gentry}, {Brewer}, and {Hogg}]{Foreman-Mackey:2016aaa}
Dan {Foreman-Mackey}, Will {Vousden}, Adrian {Price-Whelan}, Matt {Pitkin},
  Victor {Zabalza}, Geoffrey {Ryan}, {Emily}, Michael {Smith}, Gregory
  {Ashton}, Kelle {Cruz}, Wolfgang {Kerzendorf}, Thomas~A {Caswell}, Stephan
  {Hoyer}, Kyle {Barbary}, Ian {Czekala}, Hanno {Rein}, Eric {Gentry},
  Brendon~J {Brewer}, and David~W {Hogg}.
\newblock {Corner.Py: Corner.Py V2.0.0}.
\newblock art. 10.5281/zenodo.53155, May 2016.
\newblock \doi{10.5281/zenodo.53155}.

\bibitem[Behroozi et~al.(2013)Behroozi, Wechsler, and Conroy]{Behroozi:2012iw}
Peter~S. Behroozi, Risa~H. Wechsler, and Charlie Conroy.
\newblock {The Average Star Formation Histories of Galaxies in Dark Matter
  Halos from $z=$0-8}.
\newblock \emph{Astrophys. J.}, 770:\penalty0 57, 2013.
\newblock \doi{10.1088/0004-637X/770/1/57}.

\end{thebibliography}


\appendix

\section{More about the baryonic emulator}
In this appendix we give some more details about the construction and the testing of the \emph{baryonic emulator} presented in Sec.~\ref{sec:BCM}. 

\begin{figure}[tbp]
\centering
\includegraphics[width=0.6\textwidth]{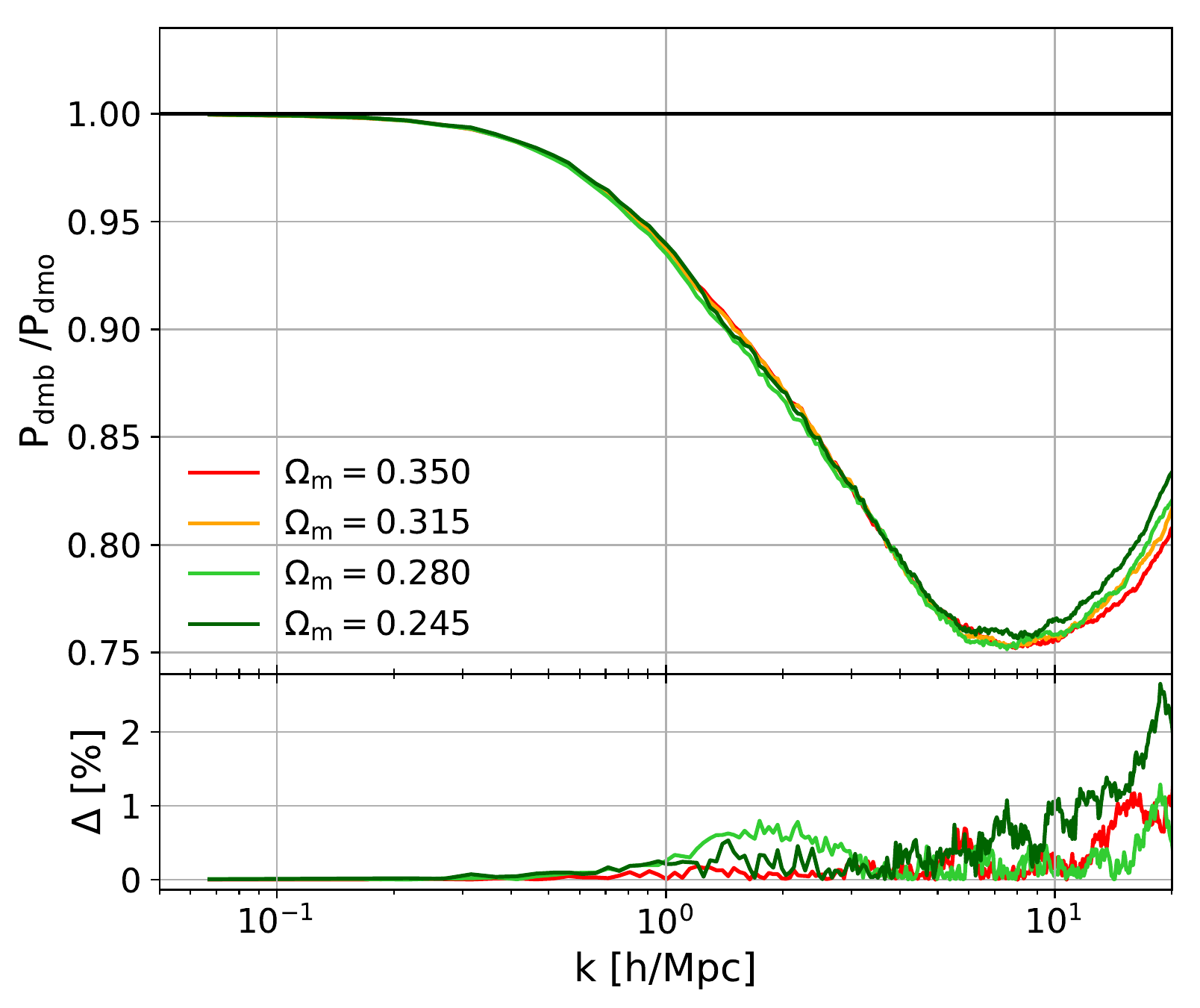}
\caption{Changes in the baryonic power suppression if we use $N$-body simulations with varying $\Omega_m$ while keeping $\Omega_m$ constant in the formalism of the baryonic correction model (i.e. Eq.~\ref{fractions}). All changes stay below one percent for $k<10$ h/Mpc (see lower panel).}
\label{fig:reference_simulation}
\end{figure}

\subsection{Reference $N$-body simulation for the emulator}\label{sec:ref_sim}
In Section \ref{cosmo_dependence} of the main text we have argued that, for building the baryonic emulator, it is sufficient to only vary the parameters of the baryonic correction model while keeping a fixed background $N$-body simulation. This assumption allowed us to increase the experimental design to $N_{\rm ED}=1000$, thereby improving the accuracy of the emulator to a higher level. In this Appendix, we provide further evidence as to why this assumption is valid.

Fig.~\ref{fig:cosmo_dependence} of the main text shows that the baryonic suppression signal does not depend on the cosmological parameter $h_0$, $\sigma_8$, and $n_s$ while it is affected by the variation of the cosmic baryon fraction $f_b=\Omega_b/\Omega_m$. This is not surprising, since the baryonic correction model directly depends on $f_b$, whereas all other cosmological parameter only come in through changes of the underlying $N$-body simulation. However, since we are only emulating the ratio of two power spectra (with the same underlying cosmology), these effects cancel out.

In Fig. \ref{fig:reference_simulation} we show that if $\Omega_m$ (and therefore $f_b$) is varied in the underlying $N$-body simulation without changing $f_b$ in the formalism of the baryonic correction model, then the baryonic suppression signal is only affected at the sub-percent level below $k\sim 10$ h/Mpc. Conversely, this means that any cosmology dependence of the baryonic suppression effect comes from Eq.~(\ref{fractions}) (that explicitly depends on $f_b$) and not from the underlying $N$-body simulation. As a consequence, we can use the same simulation to construct our emulator without compromising its accuracy at a significant level.


\subsection{Emulator performance}\label{sec:emu_perform}
In Sec.~\ref{sec:emulator} of the main text, we have presented the \emph{baryonic emulator}, quantified its overall precision (see Fig. \ref{fig:emu_error_std}), and we have shown a comparison between emulator and baryonic correction model for two selected sample points in the parameter space (Fig.~\ref{fig:Samples0_1}). In the present Appendix, we give more details about the performance of the emulator, also showing the results from the remaining four sample points that make up the test sample of our emulator.

The test sample has been randomly selected using the optimised latin hypercube sampling (LHS) method. Note that the test sample does not coincide with the experimental design (ED) on which the emulator has been constructed. The parameter values of the six sample points are given in Table~\ref{tab:emutestsample}.
 
\begin{table}[tbp]
\centering
\begin{tabular}{c c c c c c c}
Sample & $f_b$ & $\log M_c$ & $\mu$ & $\theta_{\rm ej}$ & $\eta_{\rm star} $ & $\eta_{\rm cga}$  \\
\hline
0 & 0.207 & 16.601 & 0.173 & 3.308 & 0.314 & 0.569 \\
1 & 0.172 & 15.197 & 0.939 & 5.102 & 0.370 & 0.556 \\
2 & 0.132 & 13.529 & 0.339 & 6.605 & 0.360 & 0.625 \\
3 & 0.185 & 14.419 & 0.498 & 2.007 & 0.267 & 0.680 \\
4 & 0.150 & 15.985 & 0.797 & 7.354 & 0.209 & 0.638 \\
5 & 0.162 & 12.801 & 0.583 & 4.136 & 0.236 & 0.532 \\
\hline
\end{tabular}
\caption{Parameter values of the sample on which the \emph{baryonic emulator} is tested. The test sample is randomly selected based on the LHC sampling method and does not coincide with the sample points from the experimental design (which is used to construct the emulator).}
\label{tab:emutestsample}
\end{table}

\begin{figure}[tbp]
\centering
\includegraphics[height=0.235\textwidth,trim=0.2cm 0.0cm 0.35cm 0.0cm,clip]{Figs/std_Sample0.pdf}
\includegraphics[height=0.235\textwidth,trim=1.88cm 0.0cm 0.35cm 0.0cm,clip]{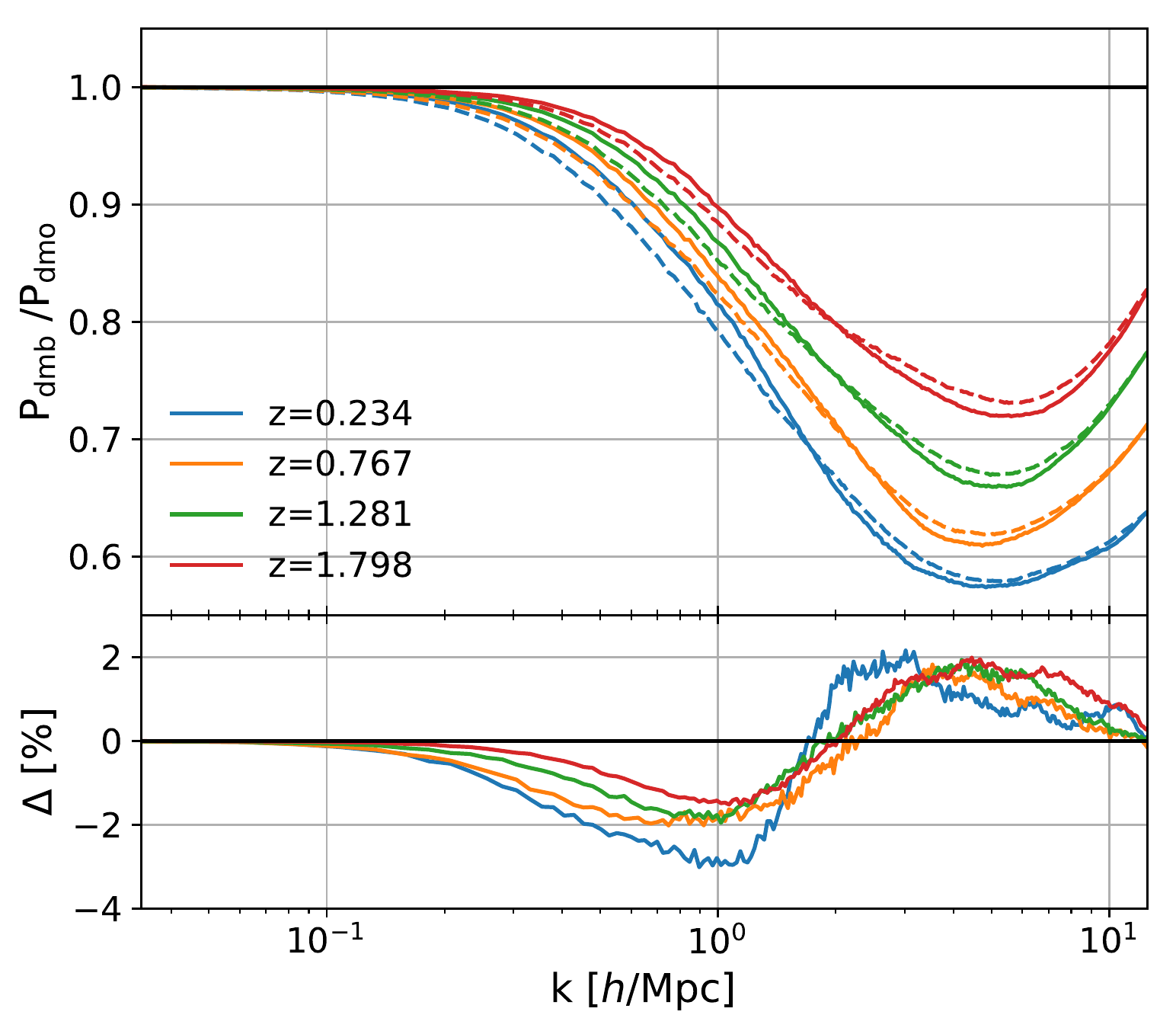}
\includegraphics[height=0.235\textwidth,trim=0.3cm 0.0cm 0.35cm 0.0cm,clip]{Figs/std_Sample1.pdf}
\includegraphics[height=0.235\textwidth,trim=1.88cm 0.0cm 0.1cm 0.0cm,clip]{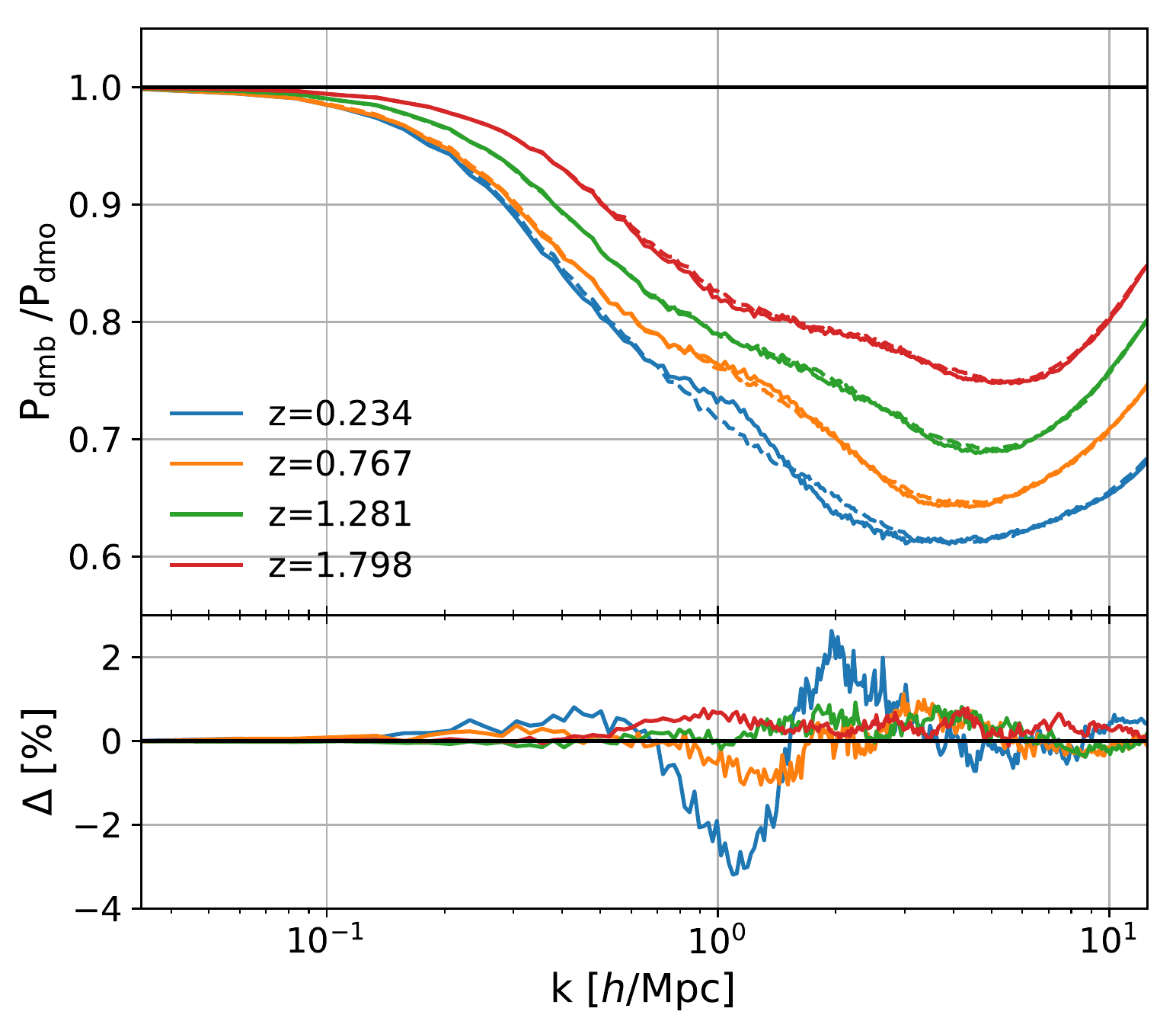}\\
\includegraphics[height=0.235\textwidth,trim=0.2cm 0.0cm 0.35cm 0.0cm,clip]{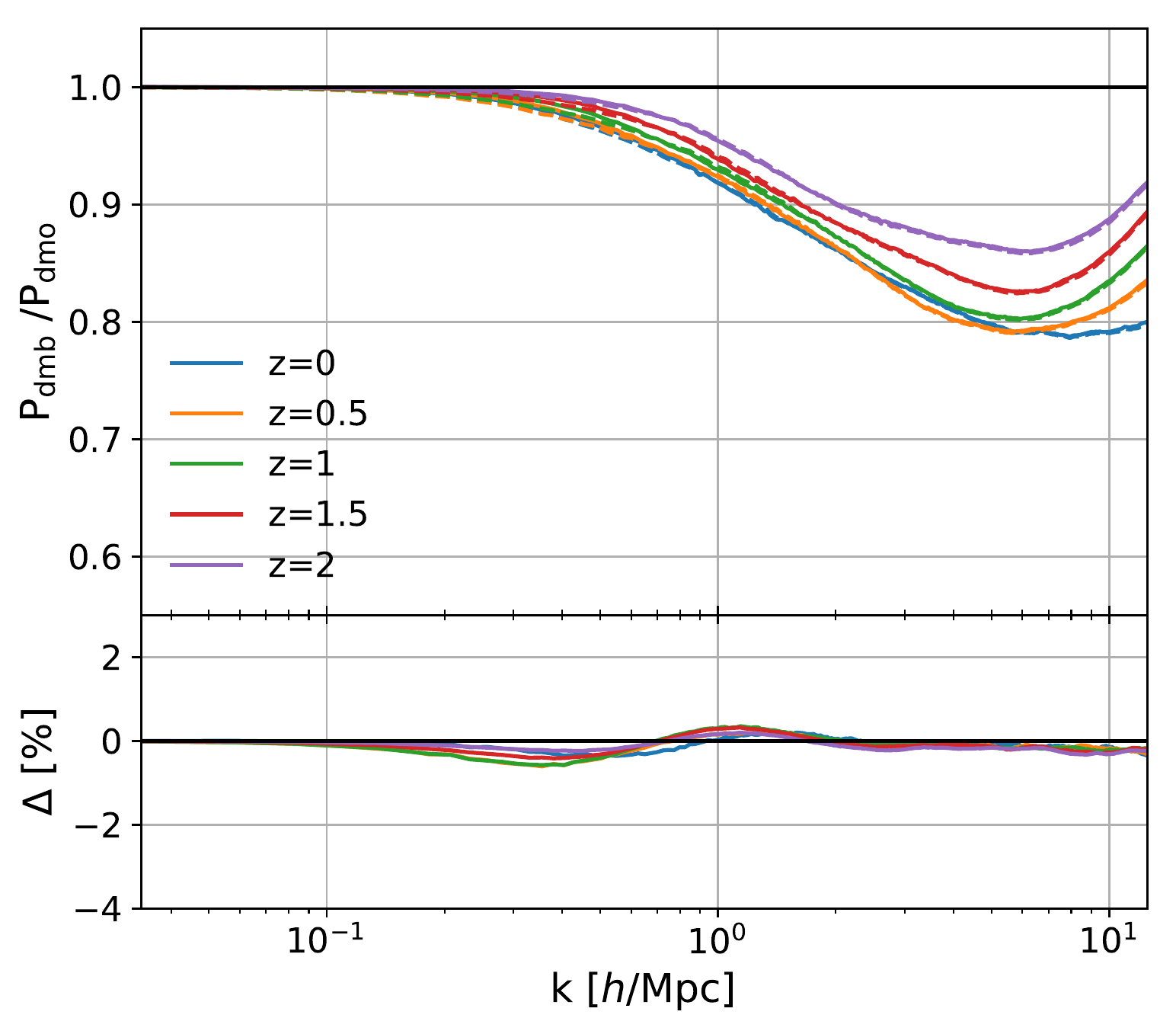}
\includegraphics[height=0.235\textwidth,trim=1.88cm 0.0cm 0.35cm 0.0cm,clip]{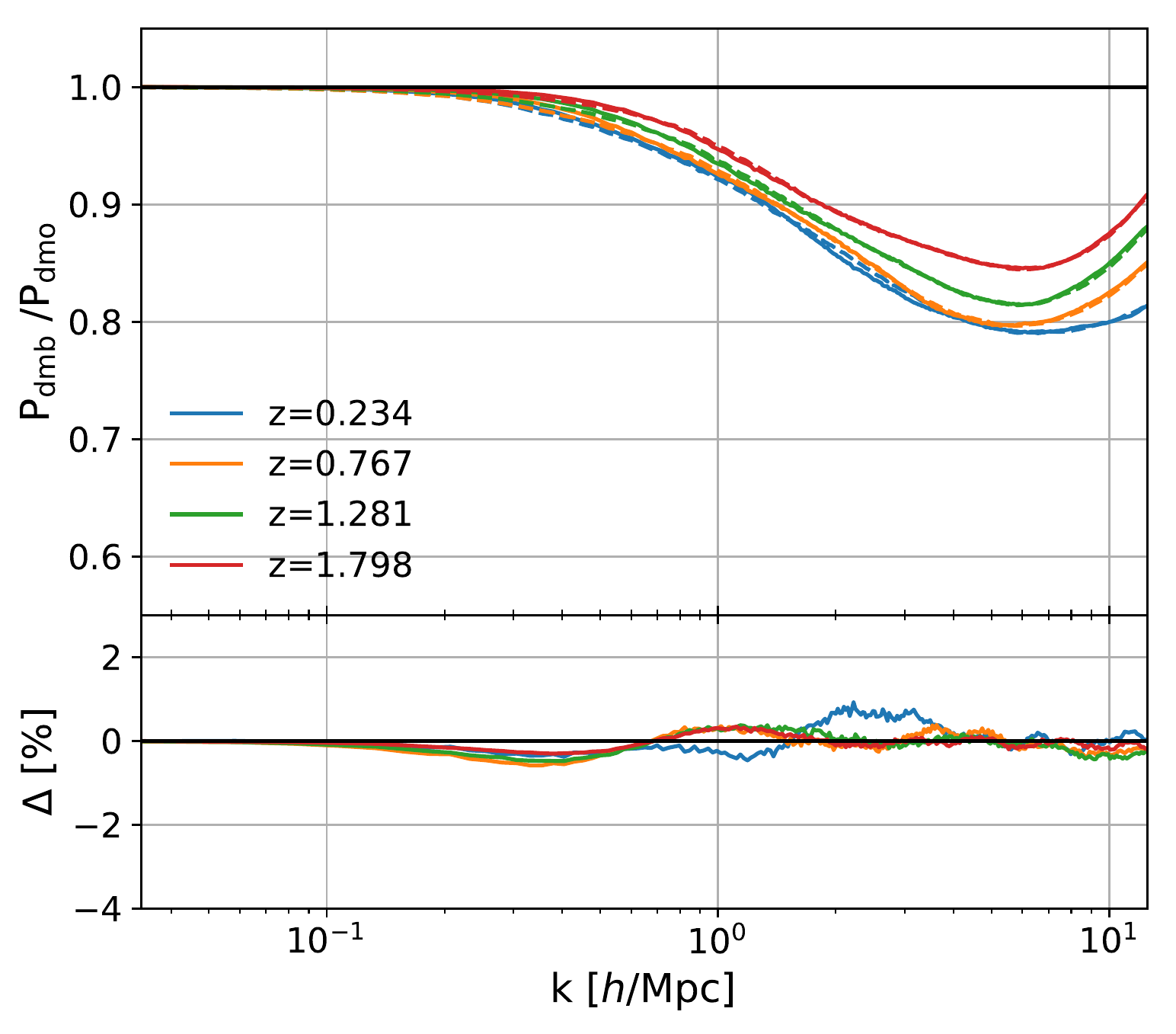}
\includegraphics[height=0.235\textwidth,trim=0.3cm 0.0cm 0.35cm 0.0cm,clip]{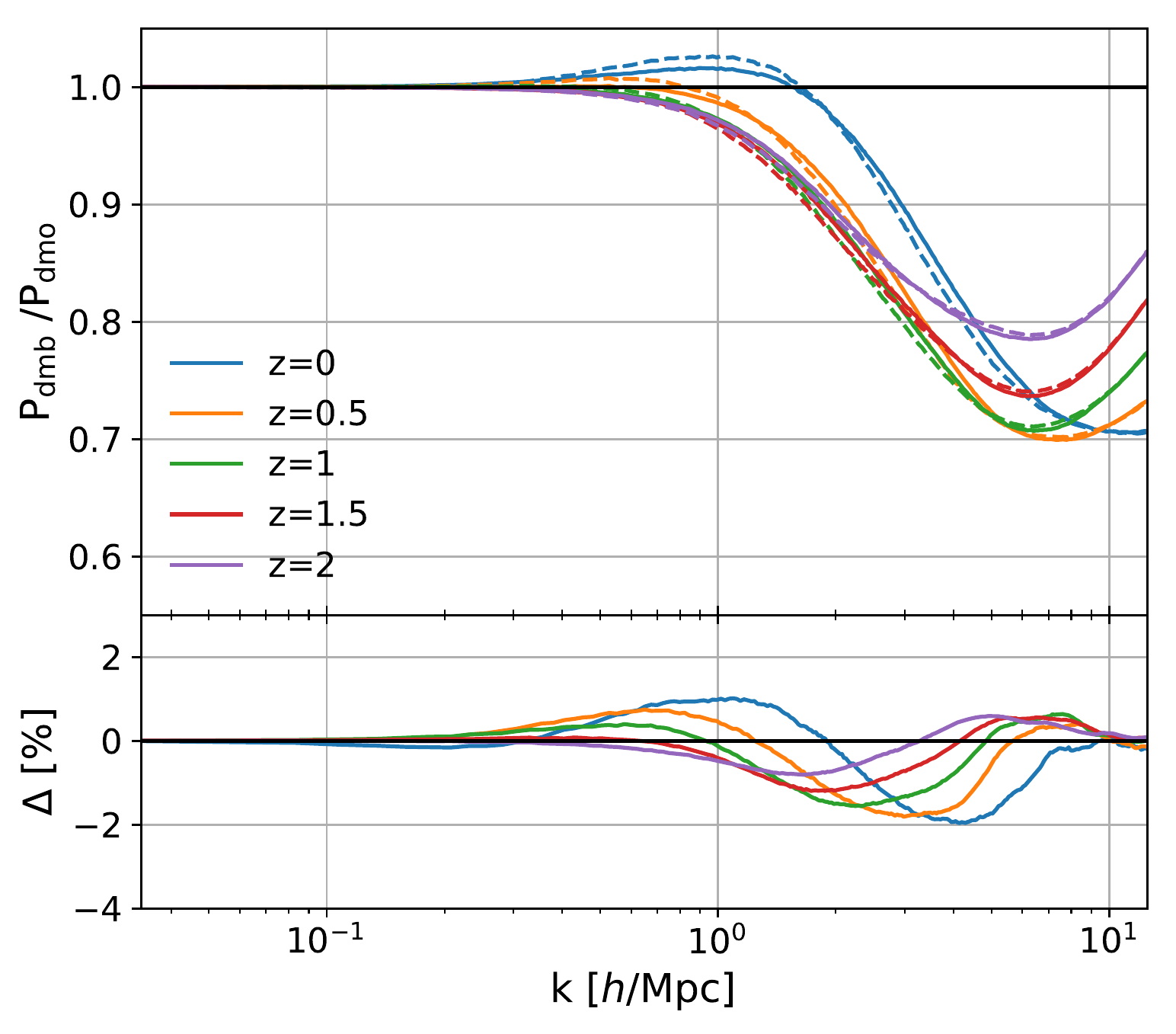}
\includegraphics[height=0.235\textwidth,trim=1.88cm 0.0cm 0.1cm 0.0cm,clip]{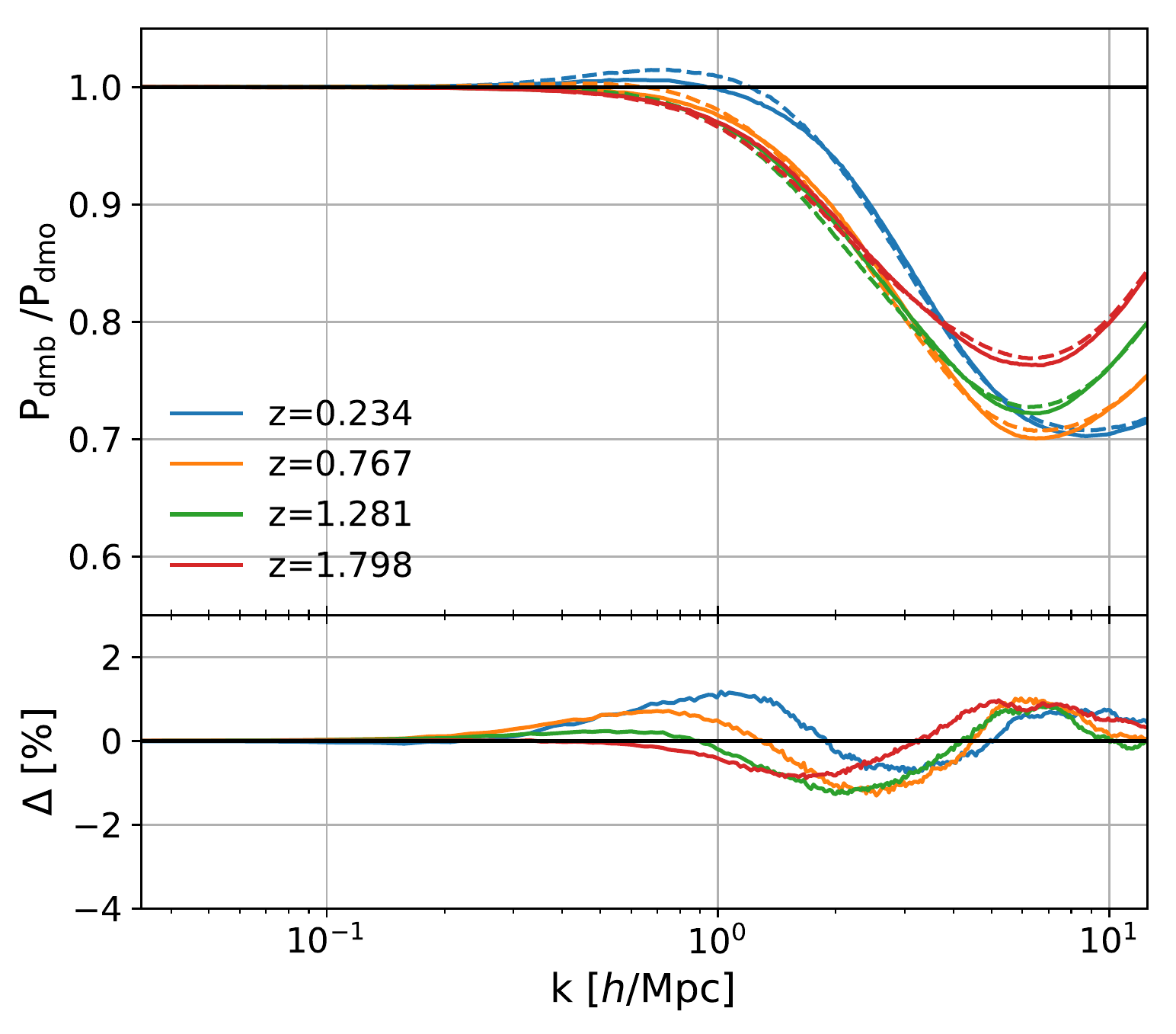}\\
\includegraphics[height=0.235\textwidth,trim=0.2cm 0.0cm 0.35cm 0.0cm,clip]{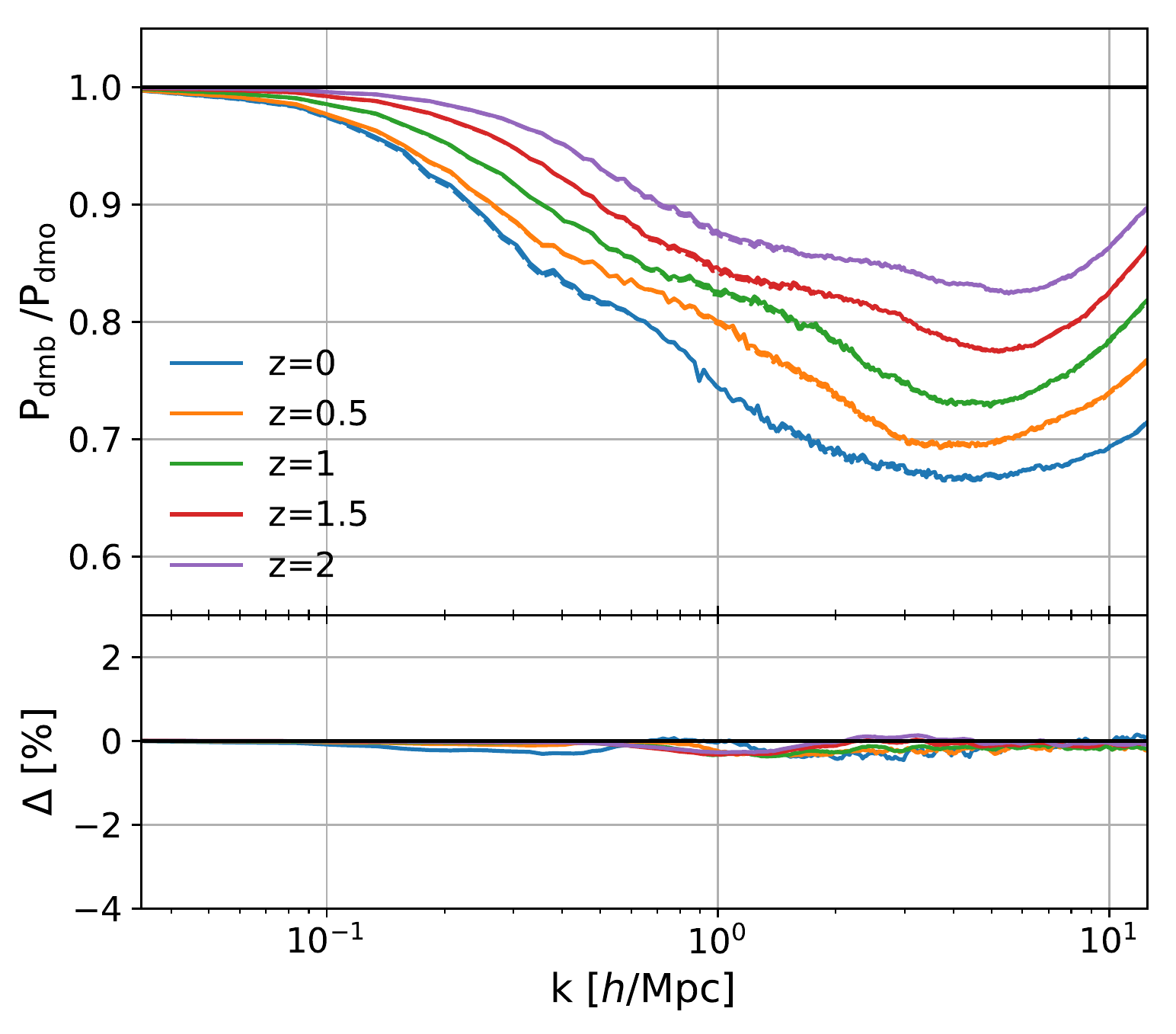}
\includegraphics[height=0.235\textwidth,trim=1.88cm 0.0cm 0.35cm 0.0cm,clip]{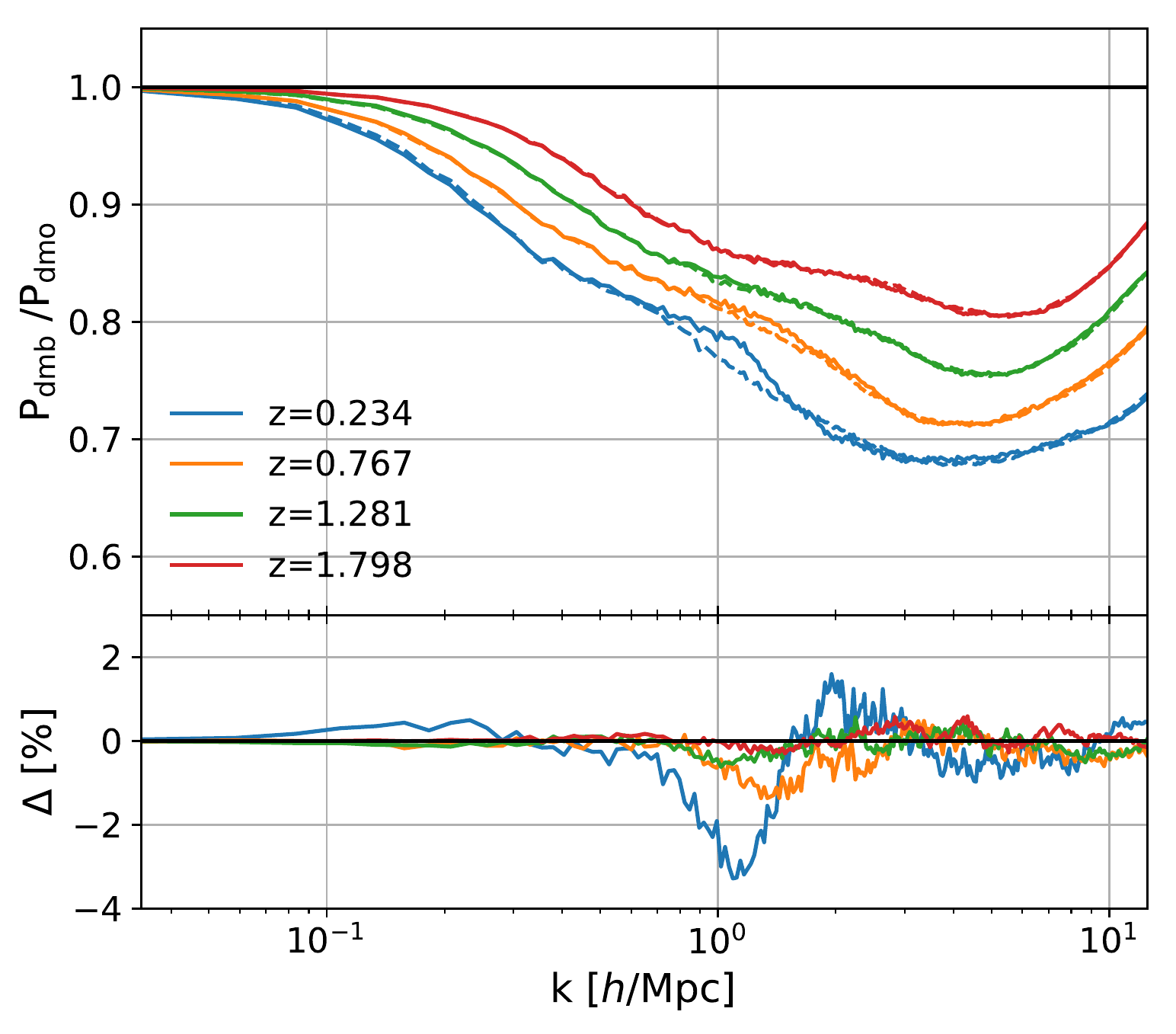}
\includegraphics[height=0.235\textwidth,trim=0.3cm 0.0cm 0.35cm 0.0cm,clip]{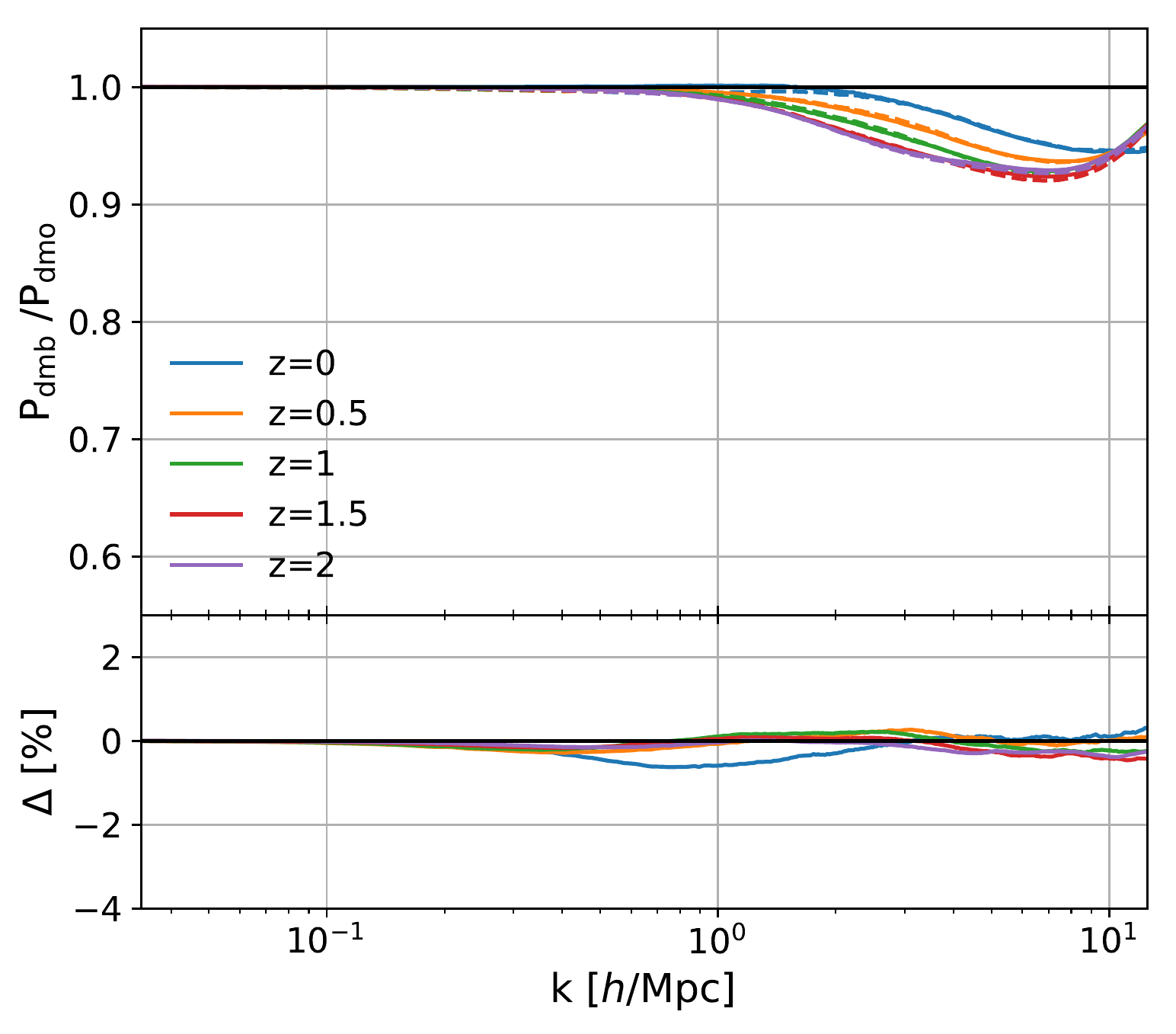}
\includegraphics[height=0.235\textwidth,trim=1.88cm 0.0cm 0.1cm 0.0cm,clip]{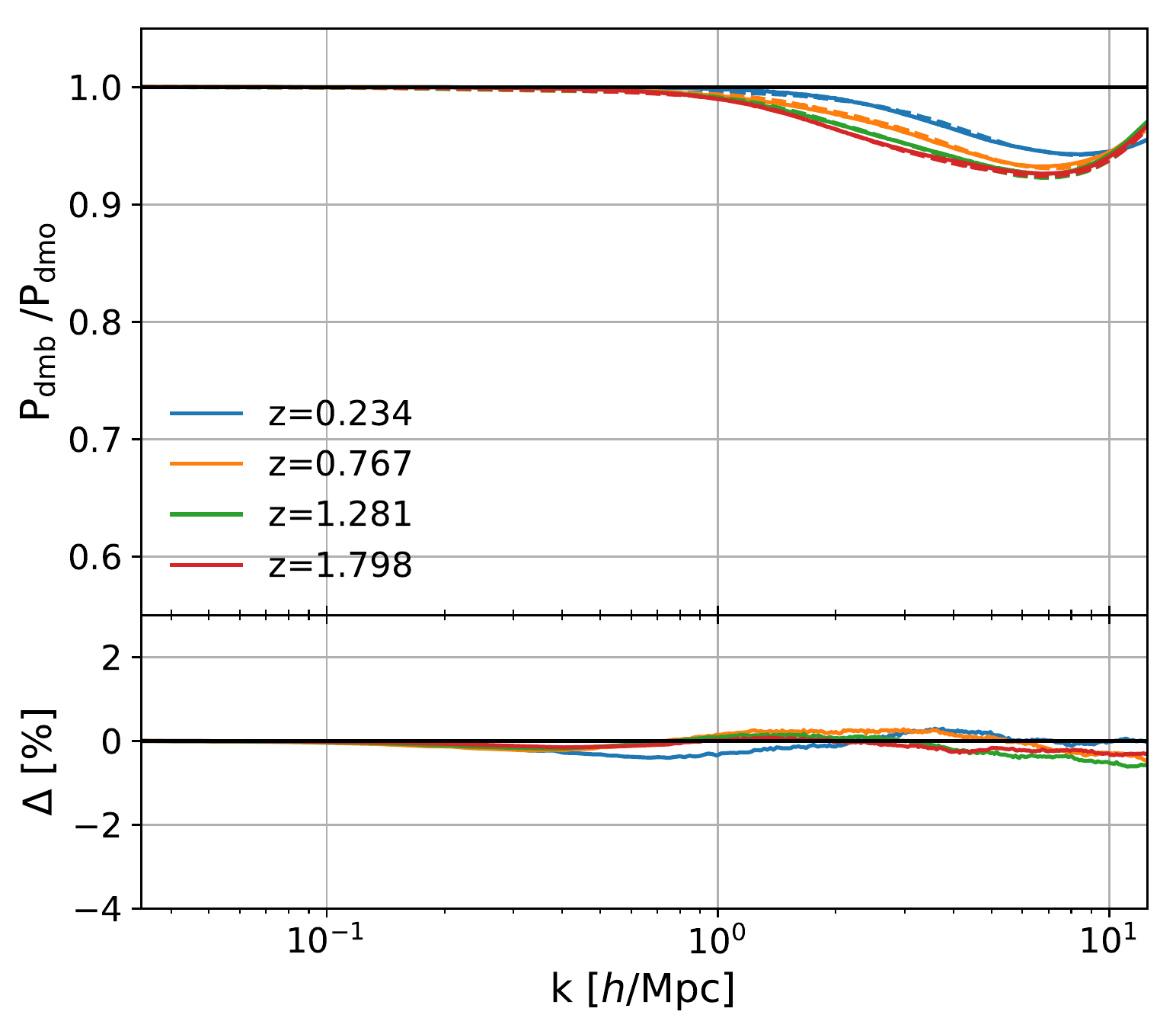}\\
\caption{Comparison between the \emph{baryonic correction model} and the \emph{baryonic emulator} for the six sample points from the emulator test-set. Sample one two six are plotted from top left to bottom right (always two panels per sample showing different redshifts). The first and the third pair of panels referring to sample 1 and 3 have been shown in the main text but are plotted again to allow for a better comparison. The solid lines show the baryonic power suppression from the BC model, the dashed lines are the results from the emulator. All parameter values for each sample are summarised in Table~\ref{tab:emutestsample}.}
\label{fig:SampleALL}
\end{figure}

In Fig.~\ref{fig:SampleALL} we show a comparison of the relative power spectra obtained from the baryonic emulator (dashed) and directly from the baryonic correction model (solid) for each of the six test-sample points. In order to provide a good overview, we again include the power spectra illustrated in Fig.~\ref{fig:Samples0_1} of the main text. However, this time we not only show the redshifts ($z_{\rm emu}$) on which the emulator has been constructed ($z_{\rm emu}=0$, 0.5, 1.0, 1.5, 2.0), but also intermediate redshifts  ($z_{\rm interp}=0.234$, 0.767, 1,281, 1.798), where the emulation code interpolates between the emulated values. The results from the sample points 0-5 are arranged from top-left to bottom-right (aways one panel showing the power spectra at $z_{\rm emu}$ and one at $z_{\rm interp}$). The first and the third panel of the top row are identical to Fig.~\ref{fig:Samples0_1}.

From Fig.~\ref{fig:SampleALL} we conclude that the emulation is well performed within a $1-2$ percent error margin in most of the considered cases. However, there are some outliers at some specific redshifts and $k$-modes, where the errors of the emulator increases to the $2 -3$ percent level. This qualitative conclusion is in agreement with the more quantitative error analysis of Sec.~\ref{sec:emulator}.

\begin{figure}[tpb]
\centering
\includegraphics[width=\textwidth]{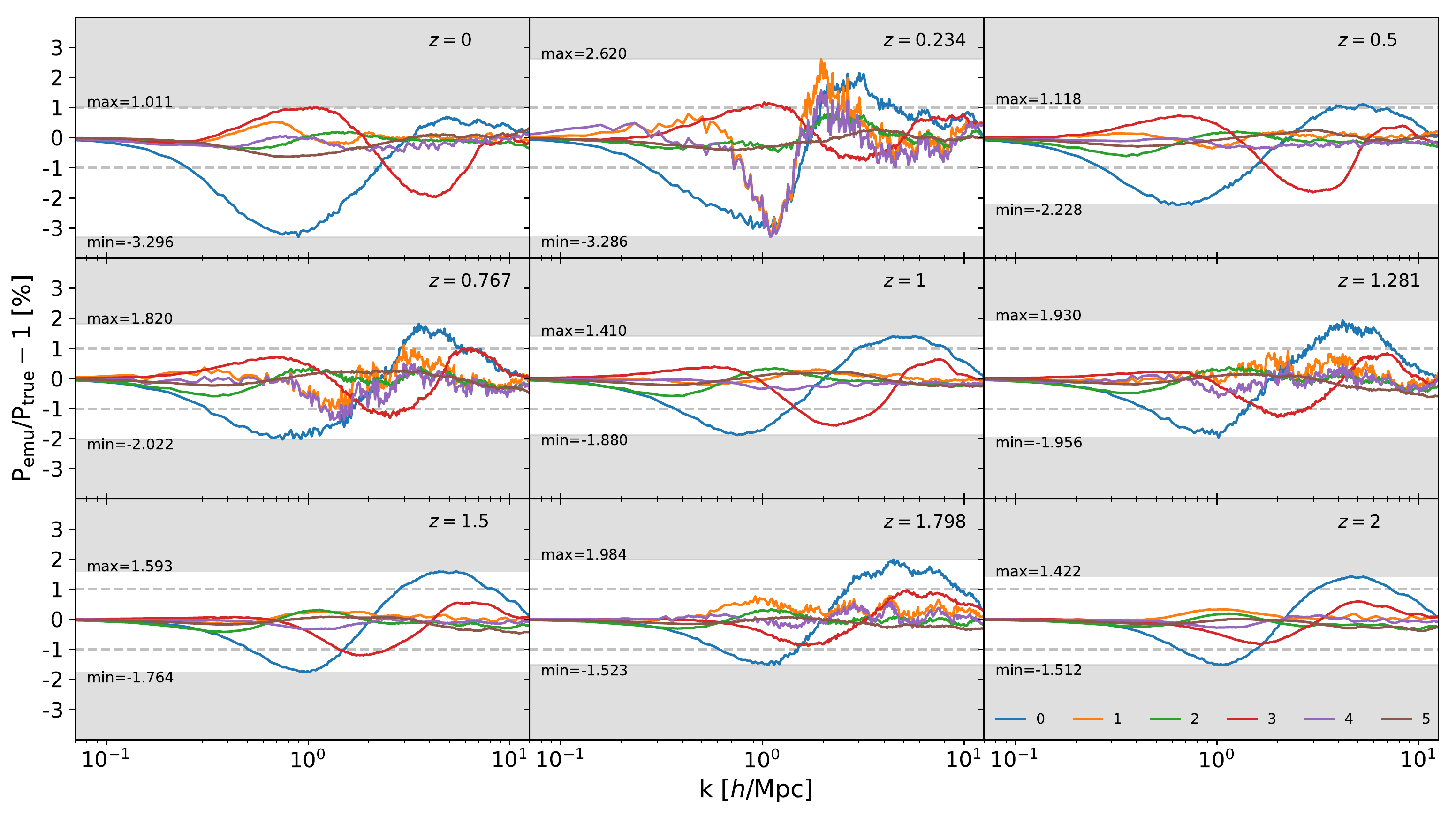}
\caption{Relative emulation error ($\mathrm{P}_{\rm emu}/\mathrm{P}_{\rm true}-1$) at different redshifts (from top-left to bottom-right) for the 6 points of the test sample (different colours). The dashed lines in each plot represent the $\pm1\%$ error range. The shaded region highlight the maximum and the minimum relative error at each $k$-mode and for each redshift.}
\label{fig:emu_error_per_z}
\end{figure}

Yet another way to illustrate the performance of the \emph{baryonic emulator} is to visualise the errors per redshift $z$ and not per sample point. The results are presented in Fig. \ref{fig:emu_error_per_z}. In terms of maximum absolute error, the emulation performance is worst at redshift 0 (with max(P$_{\rm emu}$/P$_{\rm true}-1) = 3.296$) and best at redshift 2 (with max(P$_{\rm emu}$/P$_{\rm true}-1) = 1.512$).

\section{Cosmological inference: why 3 (and not 5) baryonic parameters?}\label{sec:53barpars}
Both the \emph{baryonic emulator} and the underlying \emph{baryonic correction model} are based on 5 baryonic parameters ($M_c$, $\mu$, $\theta_{\rm ej}$, $\eta_{\rm star}$, $\eta_{\rm cga}$). However, the forecast analysis presented in Sec.~\ref{sec:likelihoodanalysis} is limited to the variation of the gas parameters ($M_c$, $\mu$, $\theta_{\rm ej}$), whereas the stellar parameters ($\eta_{\rm star}$, $\eta_{\rm cga}$) are fixed to their benchmark values.

In the main text, we have justified the restriction to 3 baryonic parameters with the fact that the stellar parameters ($\eta_{\rm star}$, $\eta_{\rm cga}$) have a comparably small effect on the matter power spectrum that only becomes apparent at relatively small scales (see for example Fig.~2 of S19). Furthermore, the parameters are already well constrained with the observed luminosity function coupled to dark matter theory via the abundance matching techniques \citep[e.g. Refs][]{Moster:2012fv,Behroozi:2012iw}. Note that abundance matching galaxies to haloes is not independent of cosmology. However, we do not expect substantial changes of $\eta_{\rm star}$, $\eta_{\rm cga}$ for reasonable modifications of cosmological parameter values (for example within the prior range of our analysis).

In this Appendix, we test the effects of varying $\eta_{\rm star}$ and $\eta_{\rm cga}$ (together with the other baryonic cosmological and intrinsic-alignment parameters) on the posterior contours of our forecast analysis. We thereby assume the flat priors $\eta_{\rm star}\in [0.2,0.4]$ and $\eta_{\rm cga}\in [0.5,0.7]$. All other priors are given in Table~\ref{tab:prior0} of the main text.

\begin{figure}[tbp]
\centering
\includegraphics[width=0.99\textwidth,trim=0.3cm 0.2cm 0.7cm 0.3cm,clip]{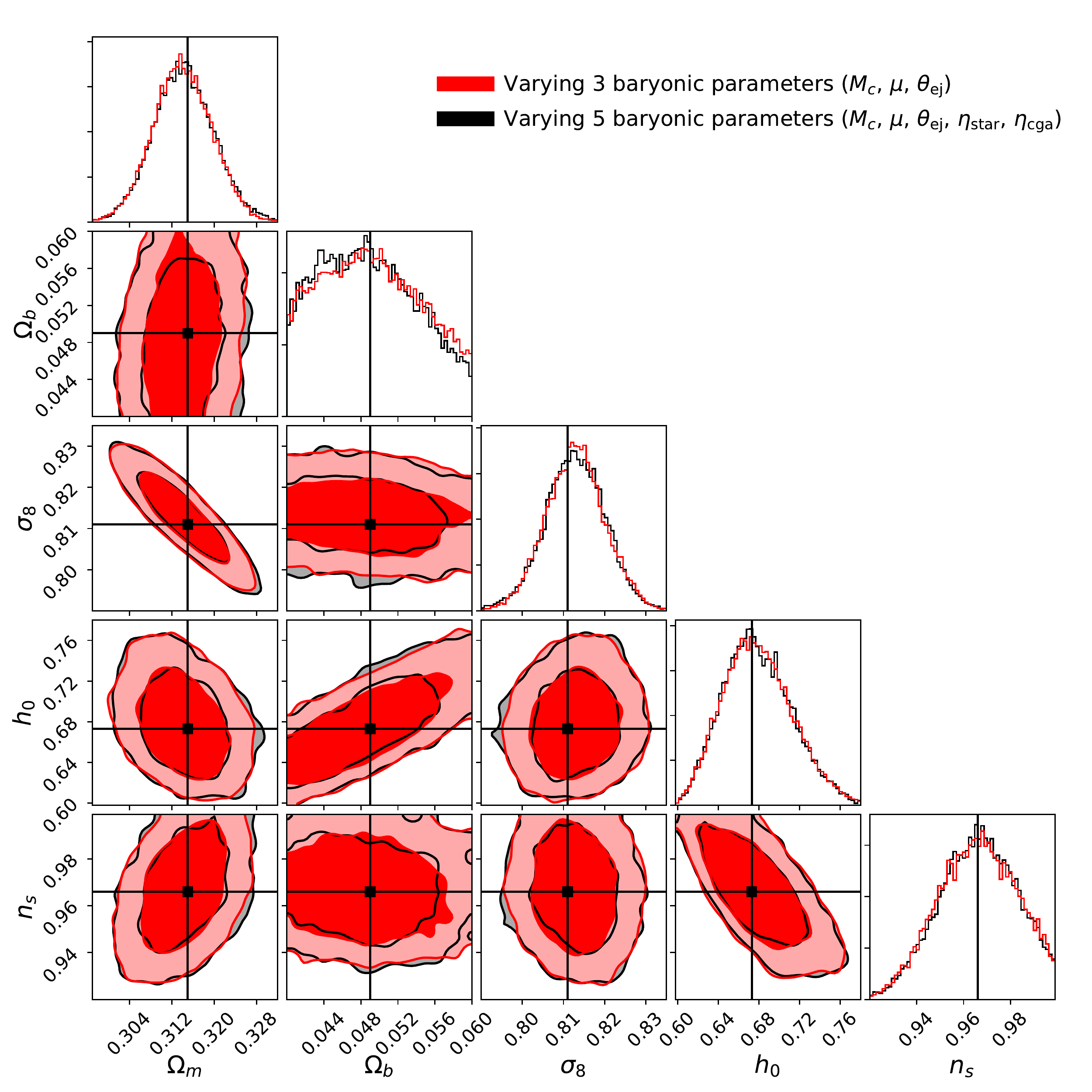}
\caption{Comparison of the posterior contours when we extend the number of free baryonic parameters from 3 (red) to 5 (black). The former corresponds to the default setup applied in our forecast analysis of Paper I and II. The latter includes all free parameters originally proposed in the S19. The black lines indicate the true cosmology of the mock sample.}
\label{fig:contour53}
\end{figure}

Fig.~\ref{fig:contour53} shows the resulting posterior contours for the five cosmological parameters ($\Omega_m$, $\Omega_b$, $\sigma_8$, $h_0$, $n_s$). The black lines indicate the true cosmology assumed when for the construction of the mock observations. The red contours illustrate the result from the original forecast analysis with 3 baryonic parameters ($M_c$, $\mu$, $\theta_{\rm ej}$), already shown in Fig.~\ref{fig:contourDMOvsBCM} and \ref{fig:contour3210} from the main text. The black contours, on the other hand, correspond to the extended cosmological inference analysis including all 5 baryonic parameters ($M_c$, $\mu$, $\theta_{\rm ej}$, $\eta_{\rm star}$, $\eta_{\rm cga}$). The red and black contour areas are very similar. Although we added two new parameters, the black contours have not become larger than the red ones. This means that varying stellar parameters ($\eta_{\rm star}$, $\eta_{\rm cga}$) has no noticeable effect on the cosmological parameter estimates. We therefore conclude that fixing the stellar parameters to their default values is a valid approximation for future stage-IV weak-lensing surveys.

\end{document}